\documentclass[11pt,preprint]{aastex}
\begin{document}

\shorttitle{GRB jet-star interaction.}
\shortauthors{Morsony et al.}

\title{Temporal and Angular Properties of GRB Jets Emerging from
Massive Stars}

\author{Brian J. Morsony, Davide Lazzati and 
Mitchell C. Begelman\altaffilmark{1}}
\affil{JILA, 440 UCB, University of Colorado, 
Boulder, CO 80309-0440, USA}
\email{morsony@quixote.colorado.edu; lazzati@colorado.edu; 
mitchb@jilau1.colorado.edu}
\altaffiltext{1}{Department of Astrophysical and Planetary Sciences,
University of Colorado at Boulder}

\begin{abstract}
We study the long-term evolution of relativistic jets in collapsars
and examine the effects of viewing angle on the subsequent gamma ray
bursts.  We carry out a series of high-resolution simulations of a jet
propagating through a stellar envelope in 2D cylindrical coordinates
using the FLASH relativistic hydrodynamics module.  For the first
time, simulations are carried out using an adaptive mesh that allows
for a large dynamic range inside the star while still being efficient
enough to follow the evolution of the jet long after it breaks out
from the star.  Our simulations allow us to single out three phases in
the jet evolution: a precursor phase in which relativistic material
turbulently shed from the head of the jet first emerges from the star,
a shocked jet phase where a fully shocked jet of material is emerging,
and an unshocked jet phase where the jet consists of a free-streaming,
unshocked core surrounded by a thin boundary layer of shocked jet
material.  The appearance of these phases will be different to
observers at different angles.  The precursor has a wide opening angle
and would be visible far off axis. The shocked phase has a relatively
narrow opening angle that is constant in time.  During the unshocked
jet phase the opening angle increases logarithmically with time. As a
consequence, some observers see prolonged dead times of emission even
for constant properties of the jet injected in the stellar core. We
also present an analytic model that is able to reproduce the overall
properties of the jet and its evolution.  We finally discuss the
observational implications of our results, emphasizing the possible
ways to test progenitor models through the effects of jet propagation
in the star.  In an appendix, we present 1D and 2D tests of the FLASH
relativistic hydrodynamics module.
\end{abstract}

\section{Introduction}
The association of gamma-ray bursts (GRBs) with supernova explosions
creates an apparent contradiction. On the one hand, the observed
gamma-ray spectra require the emitting material to be outflowing at
highly relativistic speed (Lithwick \& Sari 2001), limiting its rest
mass to a fraction of a solar mass. On the other hand, the jet expands
through a dense stellar core, potentially picking up several solar
masses in its way. Pioneering work by Woosley and collaborators
(MacFadyen \& Woosley 1999; MacFadyen et al. 2001) and by Aloy and
collaborators (Aloy et al. 2000) showed that a light jet can reach the
surface of a star virtually unpolluted if its head travels
sub-relativistically inside the star. In this way, the cold and dense
stellar material can be pushed aside. Once the stellar surface is
reached, the light jet accelerates and reaches the high Lorentz
factors required by $\gamma$-ray observations (see also Matzner 2003
for a simplified analytic treatment).

Once this basic question is answered, two main issues remain
open. First, what is the jet launching mechanism at the base of the
star and what are the properties of the injected outflow? Second, how
does the propagation of the jet influence its properties in the
radiative phase, which takes place when the jet has traveled far from
its birth site? The first question has been addressed with MHD
numerical simulations (Proga et al. 2003; Mizuno et al. 2004; De
Villiers et al. 2006; Nagataki et al. 2006). Due to the complexity of
dealing with 3D magnetic phenomena in the general relativistic regime,
these studies are still in their infancy. This paper deals with the
second of these questions. Zhang et al. (2003) performed fixed grid 2D
special relativistic simulations of jets in the cores of massive
stars. They observed a diverse phenomenology of time dependent
properties in the emerging jets, including variations in the opening
angle, but their resolution was not large enough to draw robust
conclusions at large radii, after the jet breaks out of the star.
Lazzati \& Begelman (2005, hereafter LB05) showed, with an analytic
model, that the star-jet interaction can be strong and can create
observable consequences. They argued that even if a steady jet is
injected in the stellar core, its properties at the surface of the
star will be strongly time-dependent, so that even the concept of a
jet opening angle becomes hard to define in a general sense. They
concluded that the time-integrated jet structure would match that of
the ``Universal Jet'' proposed by Rossi et al. (2002). Even though -
as we will show in this paper - their approximations were inaccurate
in places, and the jet structure they predicted is not observed in the
simulations, we confirm the basic concept that the propagation of the
jet through the star will strongly influence both, generating strong
temporal evolution in their properties. The influence of the jet's
initial conditions has been studied recently with a set of 2D
simulations (Mizuta et al. 2006). Their jets, however, are cylindrical
at the base, different from any previous work and from our
simulations presented here, where the outflow is set up to be conical
in the core of the star.

In this paper, we present the results of a set of high-resolution
2-dimensional simulations of a jet propagating through a stellar
progenitor. The jet is assumed to be already developed and mildly
relativistic at $10^9$~cm inside the star, even though it is still hot
and entropy-rich. The energy release is constant and the engine is
left to run for a time longer than the breakout time, i.e., the time
at which the jet breaks through the stellar surface. We show that the
jet-star interaction creates three well-defined phases. The first one
is characterized by turbulence. During this phase the jet propagates
through and eventually breaks out of the star. The second phase is a
transition phase, during which a heavily shocked jet flows through the
opened channel. Finally, a more stable configuration is attained, with
a freely flowing jet surrounded by a thin boundary layer that
progressively widens its opening angle. Some of these phases can be
dealt with analytically or semi-analytically with sufficient accuracy,
allowing us to understand the origin of the observed behavior.  The
observational implications of such non-steady flows are numerous,
including long dead times during the prompt emission and bumpy,
complex afterglows that do not obey the rules of the simple external
shock model.

This paper is organized as follows. In \S~2 we describe the numerical
code used. In \S~3 we describe the initial conditions for both the
stellar progenitor and the jet in its center while in \S~4 we detail
our results. In \S~5 we develop an analytic modeling of the jet and
its propagation, in \S~6 we present afterglow calculations based on
the numerical results of \S~4. We finally summarize our results in
\S~7.

\section{Numerical Methods}

The simulations presented here were performed using a modified version
of the FLASH adaptive mesh refinement (AMR) code (version 2.5) in 2D
cylindrical coordinates. The FLASH special relativistic hydrodynamics
module utilizes piecewise-parabolic interpolation for computing fluxes
at cell interfaces and uses a two-shock Riemann solver for solving the
fluid equations. See Mignone et al. (2005) for a full description of
the methods used in the relativistic portion of the FLASH code. An
adiabatic equation of state (EoS) with a fixed adiabatic index is
used. The choice of a single EoS for both the relativistic and non
relativistic material is dictated by the need to keep the running
times reasonable and by the fact that we will concentrate on the
properties of the relativistic material. Future work, that will
analyze also the behavior of the star material in detail, will adopt
a more realistic EoS. Block refinement is increased if the normalized
second derivative of the density or pressure is greater than $0.8$
anywhere in the block.  Refinement is decreased if the normalized
second derivative of the density and pressure is less than $0.2$
everywhere in the block.  Appendix \ref{codetesting} contains results
from a number of test problems using this code.

In order to achieve higher resolution near the center of the star and
close to the axis of the jet, we have modified the FLASH code so that
the maximum level of refinement allowed can be varied over the
simulation grid. See \S~\ref{gridsetup} for details of the grid
setup used for the simulations presented here. In order to improve the
efficiency of the simulations, the FLASH code was modified so that
quantities in a given cell are only computed and updated if
mod$(n_{step},2^{(l_{max} - l_i)}) = 0$, where $n_{step}$ is the time
step number, $l_{max}$ is the maximum level of refinement anywhere on
the grid, and $l_i$ is the refinement level of a particular cell. The
size of the time step used for computing new values in a cell is
correspondingly increased by a factor of $2^{(l_{max} - l_i)}$. In
other words, larger time steps are taken less often for cells at a
lower level of refinement. This adjustment is equivalent to having a
fixed Courant-Friedrichs-Levy (CFL) number at all levels of
refinement, rather than a lower CFL number in less resolved cells, as
is the default in FLASH. The minimum time step is determined by the
maximum velocity anywhere on the grid, ensuring that the CFL condition
is never violated. Additionally, in the relativistic case the maximum
velocity is always $\sim c$ (1 in relativistic units), so the minimum
time step is approximately constant. This improvement in efficiency is
generally applicable to the entire FLASH code and not just the
relativistic portion. Although some operations are still performed on
all cells at all time steps, this scheme is always more efficient than
the default in FLASH. For the simulations and grid setup described
here, this scheme resulted in approximately a factor of 10 decrease in
running time.  The tests of the adaptive mesh in Appendix
\ref{codetesting} include this change and do not show any adverse
effects.

\section{Setup}

\subsection{Stellar Model}

For the simulations presented here, two different stellar models are
used.  The first is a realistic stellar model, based on model 16TI
from Woosley \& Heger (2006) of a Wolf-Rayet star with an initial mass
of 16 M$_\sun$, metallicity of $1\%$ solar and a large angular
momentum of $3.3 \times 10^{52}$~erg~s$^{-1}$. The model has been
evolved to core collapse, and with a final mass and radius of
13.95~M$_\sun$ and $4.077 \times 10^{10}$~cm, respectively. (see
\url{http://www.ucolick.org/$\sim$alex/GRB2/}).  Simulations using
this model have names beginning with 16TI.  The second model is a
simple power-law model for a star with a mass of 15~M$_\sun$ and a
radius of $10^{11}$~cm. The density profile is modeled as a power law
$\rho \propto r^{-2.5}$ and the pressure is set such that $p = p_0
\rho^{4/3}$, where $p_0$ is computed based on $p = 1.8 \times
10^{18}$~erg~cm$^{-3}$ at $r = 10^{10}$~cm. This value of pressure at
$10^{10}$~cm is reasonable based on numerically modeled values for
stars of similar size and mass (Heger et al. 2005).  Since the stellar
pressure is small compared to the jet and cocoon pressure, the actual
value of the stellar pressure has little impact on the
simulations. All simulations use an ultra-relativistic equation of
state with $\Gamma = 4/3$, so setting $p = p_0 \rho^{4/3}$ ensures
that pressure is always small compared to $\rho\,c^2$ inside the star.
For both models, the density and pressure exterior to the star are set
to $10^{-9}$~g~cm$^{-3}$ and $9 \times 10^7$~erg~cm$^{-3}$,
respectively. The mass exterior to the star does not have a
significant impact on the dynamics of the simulations presented here,
and the exterior density and pressure are set to small, non-zero
values for numerical reasons.  A CFL number of $0.4$ is used for all
simulations.

The stars in our simulations are not stable objects since gravity is
not included. However, the timescale for the star to expand is large
compared to the length of our simulations. After 50 seconds, the end
time of our simulations, the material from the edge of the power-law
star has moved outward by $8\times10^9$~cm, which is smaller than the
distance scale for any realistic drop-off in density at the edge of
the star.

\subsection{Grid Setup \label{gridsetup}}

\begin{figure}
\epsscale{1.0}
%\plotone{grid_setup.eps}
\plotone{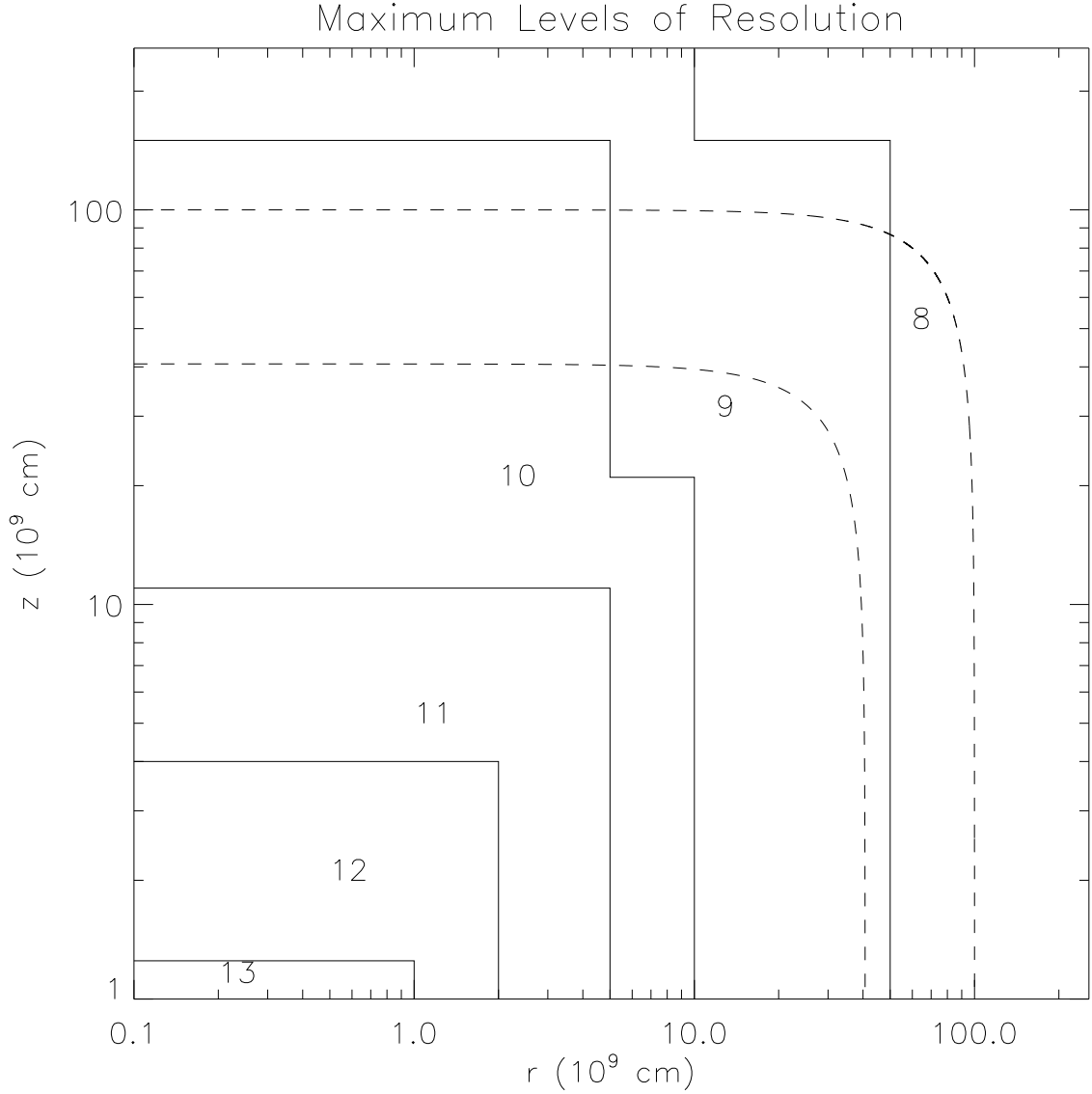}
\caption{\label{gridsetupfig} Maximum level of allowed resolution in
different grid regions. Each level down represents a factor of 2
decrease in maximum resolution. Level 8 resolution corresponds to a
pixel size of $2.5\times10^8$~cm. The two dotted lines are at a radius
of $10^{11}$~cm and $4.077\times10^{10}$~cm, the radii of the two
stellar models used. Note that the axes, as plotted, are logarithmic
in order to emphasize the regions close to the center of the star, and
that the horizontal axis does extend down to 0, rather than $10^8$~cm
as plotted, while the vertical axis begins at $10^9$~cm above the
equator of the star.}
\end{figure}

All simulations were run on an identical grid setup. The total grid
size is $2.56 \times 10^{11}$~cm by $2.56 \times 10^{11}$~cm. The
maximum allowed resolution varies from $7.8125 \times 10^6$~cm (13
levels of refinement) near the jet input to $2.5 \times 10^8$~cm (8
levels of refinement) far from the coordinate axis outside the
star. Figure~\ref{gridsetupfig} shows the maximum allowed refinement
in different regions. With this setup, the grid has a resolution of
$6.25 \times 10^7$~cm at $10^{11}$~cm, the surface of the star for the
power-law model. This corresponds to an angular resolution of
$0.0358\degr$ at this radius. This is a significant improvement over
Zhang et al. (2003), who used a grid in spherical coordinates with a
resolution of $0.25\degr$, and is comparable to Zhang et al. (2004),
who had a resolution of $10^8$~cm on a cylindrical grid at
$10^{11}$~cm. The variable maximum resolution allows the lower
boundary of the grid to be set at $10^9$~cm above the equator of the
star. Zhang et al. (2004) used a fixed grid inside the star, and their
lower boundary was placed at $10^{10}$~cm above the stellar equator.
Zhang et al. (2003) was able to place the lower boundary at a radius
of $2 \times 10^8$~cm in radial coordinates, but at the cost of
increasingly poor resolution across the jet at large radii. Similarly,
Aloy et al. (2000) could place the inner boundary at $2\times10^7$~cm
but the resolution was severely degraded at the star surface.

\subsection{Jet Parameters and Boundary Conditions}

The jet injection is modeled as a boundary condition on the lower edge
of the grid at $10^9$~cm. The opening angle and Lorentz factor of the
incoming jet are varied between simulations, but are constant at all
times in each run. The terminal Lorentz factor, or Lorentz factor at
infinity, $\gamma_\infty$, is defined as the Lorentz factor that the
material would achieve if all its internal energy were converted to
kinetic energy. It is calculated as $\gamma_\infty = (1 + 4p/ \rho
c^2)\gamma$, where $\gamma$ is the local bulk Lorentz
factor. $\gamma_\infty$ for the jet material is set to 400 for all
simulations and the luminosity of the central engine is set to $5.32
\times 10^{50}$~erg~s$^{-1}$. Table~\ref{modelparameters} lists the
parameters used for each simulation. Outside the jet injection region,
where the boundary values are fixed, the boundary conditions on the
lower edge of the grid are reflective, which is appropriate assuming
that symmetric jets emerge along both axes of the star. The outer
boundaries of the grid use outflow boundary conditions and the
symmetry axis is reflecting. Each simulation is run for 50 seconds,
giving an energy input of $2.66 \times 10^{52}$~ergs per jet, or a
total energy of $5.32 \times 10^{52}$~ergs assuming symmetric
jets.  These total energies are comparable to those assumed in previous
works (see Zhang et al. 2003 for a discussion).

\section{Results}

The early evolution of each of our models is generally similar to that
found in previous simulations (Zhang et al. 2003, 2004). The jet ram
pressure generates a bow shock that propagates sub-relativistically
through the star. Cold and dense stellar material is pushed to the
sides, partly mixing with shocked jet material to create a hot cocoon
that surrounds the jet. This allows the younger jet material to
propagate relativistically and unpolluted through an evacuated channel
along the symmetry axis. Once this material reaches the stellar
surface, it accelerates to its final Lorentz factor and can, once it
has become optically thin, generate the observed GRB spectra.

Figure~\ref{timesequence} shows a time sequence of the evolution of
model t10g5. The three upper panels are very similar to figures
obtained by other groups with previous simulations, even though our
AMR code can capture finer details. The main novelty of our work is
shown in the three bottom panels, where the jet is still powered
several tens of second after breakout and is evolving inside the
star. As we will explain in more detail in the following sections,
three phases can be identified. Initially the jet is confined, and hot
turbulent material is stored in a cocoon (first two panels). When the
jet head reaches the surface, the cocoon is released as a wide angle
outflow (third panel). Immediately afterward, a heavily shocked jet
flows outside the star (third and fourth panels). Eventually, a more
stable configuration emerges (fifth panel) in which the jet is
internally free-flowing, and is bounded by a shear layer at the
contact discontinuity with the star. Figure~\ref{rhopg} shows a
snapshot of simulation t10g2 at 30 seconds as an example of the
structure present shortly after breakout.

The three phases are identified through the behavior of the on-axis
energy flow, as shown in Fig.~\ref{fig:phasedef}. The transition
between the precursor and the shocked jet is defined as the moment at
which the energy flow along the axis becomes continuous, albeit
variable. The transition between the shocked and unshocked jet is
defined as the time at which the on-axis energy flow drops and becomes
steady, without prominent variations.  Table~\ref{breakout} lists the
times at which the cocoon, shocked jet, and unshocked jet reach the
initial surface of the star.  From these data we can see some expected
trends as different parameters are varied. Models with an initial
Lorentz factor of 2 (t10g2 and 16TIg2) have longer breakout times than
corresponding models with an initial Lorentz factor of 5 (t10g5 and
16TIg5). This is to be expected because a lower Lorentz factor at the
same energy implies less momentum in the direction of propagation and
a higher pressure (similar conclusions in a different geometry were
obtained by Mizuta et al. 2006). This means that the jet will be less
tightly collimated and will propagate more slowly.

Model t5g2, with a $5\degr$ injection opening angle, has a slightly
earlier cocoon breakout time than the corresponding model with a
$10\degr$ opening angle (t10g2).  A smaller opening angle means that
the same amount of energy and momentum is initially spread over a
smaller area, making the jet more penetrating.  However, as the jet is
being collimated, the initial opening angle is only important at the
beginning of the simulation, hence the overall difference between the
two simulations is small.  The shocked jet breakout time is earlier
for model t10g2 by 0.8s, and the unshocked jet breakout is earlier for
model t5g2 by 2s. These differences appear to depend more on the
details of turbulence near the jet head than on the initial opening
angle of the jet.

Simulations using the realistic stellar model (16TIg5 and 16TIg2) have
earlier breakout times than simulations using the power-law stellar
model and identical jet (t10g5 and t10g2, respectively).  This is
because the realistic stellar model is more compact, giving the star a
smaller radius and higher density.  The higher density leads to a more
tightly collimated jet, which is therefore more penetrating.  As a
result, the jet is able to travel through about the same amount of
total mass in a shorter time. It should be noted, however, that the
average speed of the bow shock is smaller in the denser, realistic,
stellar progenitor.

We also ran a simulation where a wide jet ($\theta_0=20\degr$) is
injected at the center of the star.  Model t20g5 appears to be
morphologically different from the other simulations. It has the
latest cocoon breakout time of any of the simulations, and the jet
material along the propagation axis does not reach the surface of the
star until the end of the simulation. Before this, relativistic
material is deflected around a dense wedge of stellar material and can
escape the star at large angles, but it does not appear that this
would produce a classical gamma ray burst.  This is in part a 2D
effect.  In three dimensions, the jet could be deflected sideways to
move around the wedge of material, but in our 2D simulations it is
forced to be axially symmetric and therefore stalls (see Zhang et
al. 2004).  Figure~\ref{t20g5compare} compares models t10g5 and
t20g5. The reason why model t20g5 is different is that the initial
opening angle is too large to produce a tightly collimated jet. The
pressure of the incoming jet falls as $\theta^{-2}_0$, meaning that a
wider jet is less penetrating. At 20 degrees, model t20g5 is not
penetrating enough to pierce the star in the manner necessary to
create a GRB, despite the large energy input of $5.32 \times 10^{52}$
ergs. This simulation may represent a different class of jet-powered
supernova or failed gamma ray burst, but it will not be considered in
latter portions of this paper that compare our simulations to
classical GRB properties. Analogous conclusions were drawn from
previous simulations (MacFadyen et al. 2001; Zhang et al. 2003) and
for 3D precessing jets (Zhang et al. 2004). It should be remembered,
however, that assuming a wide jet at $10^9$~cm in the star may be
unphysical. Even if a wide jet is launched at the black-hole scale,
recollimation is likely to take place at radii $10^8-10^9$~cm (see
model JA in Zhang et al. 2003). As a consequence, an initial condition
with a narrow, entropy rich, jet (model t5g2) is more realistic than a
wide fast outflow (model t20g5) at the inner boundary of our
simulations.

\subsection{Angular Distribution of Energy} \label{angulardistributionofenergy}

A snapshot file containing all the simulation data is produced every
1/15th of a second in simulation time.  The energy flux is determined
as a function of angle and time by taking a simulation snapshot and
finding all the points that will cross a fixed radius in the next
1/15th of a second.  The energy at each point is spread equally over
an angle of $\pm1/\gamma_\infty$ from the direction of motion of the
fluid at that point.  This approximately accounts for the hydrodynamic
spreading (or sideways expansion) of that element of the jet and for
the relativistic beaming of radiation eventually emitted by that
material.  The energy is then placed into angular bins.  The total
energy in each angular bin can then be calculated by adding the energy
from all points, and correcting for geometrical effects to give the
energy flux in each bin.  The energy in material above a given
terminal Lorentz factor (Lorentz factor at infinity) can be found by
excluding points with material below that $\gamma_\infty$.  The
results from each snapshot can be added over time to find the total
energy seen at a fixed radius at different angles for the entire
length of the simulation, or for shorter time intervals.  With this
data, we can examine the amount of energy crossing a fixed radius as a
function of time, angle, and minimum Lorentz factor. Bear in mind
that, given the definition of the angle adopted, the energy
distribution may be modified by hydrodynamic interactions during the
propagation to larger radii\footnote{An alternative definition of the
angle is provided by the positional angle of the point with respect to
the origin of the coordinates. This definition is also prone to change
with the expansion of the flow if the velocity and position vectors
are not aligned.}. For the results presented here, data from our
simulations have been taken at $2.4 \times 10^{11}$~cm and placed into
45 angular bins arranged in the following manner: 14 bins spaced every
$0.25\degr$ with centers ranging from $0.125\degr$ to $3.375\degr$, 17
bins spaced every $1.0\degr$ with centers ranging from $4.0\degr$ to
$20.0\degr$, and 14 bins spaced every $5.0\degr$ with centers ranging
from $23.0\degr$ to $88.0\degr$.

\subsubsection{Energy vs. Angle}

In order to determine the observational characteristics of a gamma ray
burst produced from our simulations, we examine the amount of energy
directed toward observers at different angles.
Figure~\ref{total_energy_angle_16TIg5} shows the total energy crossing
a radius of $2.4 \times 10^{11}$~cm during the simulation.  Minimum
terminal Lorentz factors of 1.01, 2, 10, 50, and 200 have been used
for the different lines in the figure.  The energy has been converted
to units of isotropic equivalent energy available at different angles.
The observed isotropic energy would then be the available energy times
the efficiency of converting this energy to gamma rays.  Different
minimum Lorentz factors are presented because the efficiency of gamma
ray production is likely to vary with Lorentz factor.

For each simulation, the total isotropic equivalent energy
$dE/d\Omega$ is constant or slowly decreasing from the jet axis out to
some cutoff.  Inside the cutoff, energy goes as $\sim$constant to
$\theta^{-1}$. Beyond this, the energy decreases rapidly.  In some
cases there is a significant contribution from precursor energy at
large angle, but when this is removed
(Fig.~\ref{no_pre_energy_angle_16TIg5}), the cutoff in energy becomes
even sharper.  For large $\gamma_\infty$, this decrease in energy can
be fit by an exponential or a steep power law ($dE/d\Omega \sim
\theta^{-4}$ or more).  If low-energy particles are included ($1.01
\le \gamma_\infty \le 2$) then energy falls off as $\sim \theta^{-2}$
at large angles.  These particles should not contribute to the prompt
burst, as they do not have a Lorentz factor large enough to produce
gamma rays, but they can produce X-rays and can be important to the
afterglow energetics.  This differs from the Zhang et al. (2004)
finding that the energy is distributed as $dE/d\Omega \sim
\theta^{-3}$. The difference is due to a combination of several
factors. First, since their simulations are performed on a fixed grid,
the resolution is poorer at the stellar surface compared to our AMR
simulations. Second, Zhang et al. (2004) consider a small range of
angles for their $\theta^{-3}$ fit ($3\degr<\theta<15\degr$). As a
matter of fact, some of our models (especially 16TIg2 and t10g5) could
be fit by $\theta^{-3}$ power-laws in small angle intervals. Finally,
their simulation domain inside the star is only from $10^{10}$~cm to
$8.8\times10^{10}$~cm, a volume that may be too small to allow the
development of turbulence that would result in mixing of the jet
material with the stellar material. From our simulations and the
comparison with previous work, it can be concluded that the energy
distribution with angle is not a robust property of the collapsar, but
depends on the stellar structure and the jet injection parameters.

As discussed above, the passage of jet material at a fixed radius
consists of two phases: the passage of the shocked and unshocked jet.
Figure~\ref{shock_unshock_energy_angle_16TIg5} shows the total energy
from each of these phases in model 16TIg5 for $\gamma_\infty \ge 50$ 
at a radius of $1.2 \times 10^{11}$~cm.
The shocked phase of the jet has an approximately uniform energy
distribution with a sharp cutoff at the edge of the jet.  During the
unshocked phase, however, mass and energy are concentrated in the
shocked boundary layer that surrounds the unshocked core of the jet.
This means the unshocked jet does not contribute much energy near the
jet axis.  However, at angles outside where the shocked phase of the
jet is visible, the unshocked phase is the dominant energy source. In
addition, while the energy in the shocked jet phase is fixed by the
stellar properties, the unshocked jet can extend in time and become
more and more prominent. Only self-consistent simulations of the jet
feeding, launching, and propagation can pin down any absolute
normalization between the energies of the two phases.  At any angle,
the unshocked phase only contributes significant energy while the thin
shocked boundary layer covers that angle.  As the opening angle of the
jet is increasing during this phase, an observer will only see the
boundary layer for a relatively short time.  Increasing the duration
of energy injection will therefore increase the final opening angle of
the jet, but it will not significantly increase the isotropic
equivalent energy seen by observers already within the jet.

\subsubsection{Phases of Jet Evolution}

Figure~\ref{energy_time_angle_16TIg5} shows logarithmic energy
(brightness) contours as a function of time (vertical axis) and angle
(horizontal axis) for model 16TIg5. Different panels show different minimum
$\gamma_\infty$ cutoffs. The figure shows that there are three phases
in the energy flux seen at a fixed radius.  First there is a precursor
phase, preceding significant on-axis energy flux, followed by a
shocked jet phase, and an unshocked jet phase.

The precursor phase can be divided in two subsequent events.  First a
thin, nearly isotropic shell of mildly relativistic material, visible
in the lower energy bands of Fig.~\ref{energy_time_angle_16TIg5}. This
phase is associated to the shock preceding the jet expanding into a
density gradient and carries very little energy. For this reason it
has observational consequences only if the other phases of the GRB are
not visible due to viewing angle constraints (see \S~\ref{secpr}).
This initial precursor is followed by a mixture of jet and stellar
material which is peaked at several degrees off axis with less energy
on axis.  This is the material that accumulates around the jet in the
form of a cocoon (Ramirez-Ruiz et al. 2002) during the initial phases
when the jet is confined inside the star. This material generally has
a lower Lorentz factor than the jet and is most clearly visible in the
lower energy plots.  A further discussion of this material is given in
\S~\ref{precursors}.

After the cocoon material, the shocked jet emerges. Differently from
the cocoon material, the shocked jet has a large Lorentz factor at the
stellar surface. During the shocked phase, jet material between the
jet head and reverse shock is passing by.  The shocked portion of the
jet has a highly structured interior, including turbulent structures
and internal shocks.  Although this makes the energy flux and width of
the jet variable during this period, there is no clear trend with time
in the jet properties.

Finally, the jet settles into a quasi-stationary configuration, which
we call the unshocked phase. The reverse shock has passed the radius
we are looking at, and the jet consists of unshocked material
surrounded by a narrow boundary layer of shocked material.  During
this phase the jet does not have the variable structure seen in the
precursor and shocked phases, but the width of the jet clearly
increases with time (see
\S~\ref{openingangle}). Figure~\ref{fig:frexp} shows the Lorentz
factor and pressure along the jet axis of model 16TIg5 at the time at
which the shocked-unshocked jet boundary crosses the star surface. The
figure confirms that the core of the unshocked jet, inside the
boundary layer, is free streaming. Dashed lines show how pressure and
Lorentz factor behave in a free-streaming, pressure dominated jet; the
simulation result is in excellent agreement with this behavior. Note
that the Lorentz factor deviates from the asymptotic solution at large
$\gamma$. This is due to the fact that at such highly relativistic
speeds, the pressure-dominated approximation does not hold any more.
The comparison of Fig.~\ref{fig:frexp} with previous results
(e.g. Fig.~2 in Aloy et al. 2000, Fig.~4, 5, and 6 in Zhang et
al. 2003, or Fig.~6 and 7 in Mizuta et al. 2006) reveals much sharper
features in our results. This is likely due to the increased
resolution of the AMR code with respect to fixed grid codes.

It is interesting to speculate on the temporal properties of the three
phases. It is usually assumed that the light curve of the GRB prompt
emission is due to internal shocks driven by Lorentz factor
inhomogeneities in the flow. These inhomogeneities are supposed to be
imprinted by the central engine. The propagation of the outflow
through the star, however, is likely to modify these structures,
erasing some and amplifying others. Which ones are erased and which
amplified will depend on the phase of the jet evolution. The precursor
material is made of jet and stellar material that has been completely
reshuffled and shocked in the bow shock and turbulent eddies
surrounding the jets. As a consequence, the ejection history of the
central engine has been forgotten. A similar conclusion holds for the
shocked jet, even though some trace of the engine properties may be
retained. In the unshocked jet, however, and especially in its freely
expanding core, all variability imprinted by the engine will be frozen
and advected to the radiative phase. The properties of the inner
engine will therefore be more clearly seen in the tail of the GRB
emission. Unluckily, this is the faintest phase.

\subsubsection{Opening Angle} \label{openingangle}

As noted from previous simulations (Zhang et al. 2003) and discussed
theoretically in LB05, the opening angle is not a constant property of
the outflow emerging from the star, nor is it an easily measurable
quantity. We define the opening angle as follows. We first select all
points that will cross a fixed radius within 1/15th of a second (the
same points used to find energy flux) and with a minimum value of
$\gamma_\infty$. We then compute the angles associated with all these
points and define the jet opening angle as the largest. In this case,
we define the angle as the geometrical angle of the point with respect
to the origin and polar axis of the coordinates, since adopting the
velocity vector angle introduces substantial noise (see footnote 1).

Figure~\ref{opening_angle_16TIg5} shows opening angle vs. time for
different minimum values of the terminal Lorentz factor
($\gamma_\infty$) at a radius of $1.2 \times 10^{11}$~cm in our
simulations. This figure emphasizes again the three phases of jet
evolution. As material first reaches this radius, the opening angle is
very wide for all but the highest values of $\gamma_\infty$.  This is
due to the precursor material discussed in \S~\ref{precursors}.
Following this is the shocked portion of the jet, characterized by a
fairly constant opening angle and no consistent evolution with time.
Using different values of $\gamma_\infty$ gives widely different
opening angles, ranging, for example, from $\sim 2\degr$ for
$\gamma_\infty = 200$ to $\sim 8\degr$ for $\gamma_\infty = 2$ in
model 16TIg5.  This is because material at the edge of the jet is
partially mixed with stellar material, lowering its terminal Lorentz
factor.  After the shocked portion of the jet has passed, the jet
consists of an unshocked core with a shocked boundary layer along the
edge.  As the unshocked jet passes, the opening angle of the jet is
slowly but consistently increasing.  The rate of increase is always
less than linear and can be well fit by a logarithmic increase.  The
rate of increase is obviously much slower than the exponential
increase predicted in LB05 but is in reasonable agreement with the
semi-analytic results presented in this paper (\S~\ref{analytic}).
The thickness of the boundary layer, measured as the difference
between the opening angles for $\gamma_{min} = 2$ and $\gamma_{min} =
200$, decreases with time.  For model 16TIg5, the thickness goes from
$\sim 6\degr$ in the shocked phase down to $\sim 4\degr$ by the end of
the simulation (see Fig.~\ref{opening_angle_16TIg5}).

Although the intrinsic opening angle of the jet is as small as
$\theta_0 = 5\degr$, at late times the jet will be over-pressured at
the base and expand to a larger opening angle.  After cocoon breakout,
the pressure in the cocoon decreases exponentially, while the incoming
jet pressure remains fixed.  When the cocoon pressure drops below the
incoming jet pressure, the jet will expand near its base by an angle
of up to $1/\gamma_0$, where $\gamma_0$ is the Lorentz factor of the
incoming jet.  For $\gamma_0 = 5$ this is about $11\degr$.  Therefore,
a jet with $\theta_0 = 5\degr$ and $\gamma_0 = 5$ can have a maximum
opening angle of about $16\degr$.  Pressure in the unshocked jet drops
as $r^{-4}$, so at large radii the jet is still being collimated by
the cocoon pressure.  In all simulations presented here, the opening
angle of the jet at large radii is always less than the maximum
opening angle for that simulation.

\subsection{Precursors}
\label{precursors}

As the jet propagates through the dense material of the star, a high
density wedge of stellar material develops at the head of the jet.
The jet material does not penetrate this wedge, but instead moves to
the sides.  Eventually this material will curl back on itself,
creating large vortexes in advance of the narrowly collimated jet
(Fig.~\ref{timesequence}).  As the jet continues to propagate, the
vortexes will eventually detach from the jet and be swept backward,
relative to jet propagation, in a phenomenon known as vortex shedding
(Scheck et al. 2002; Mizuta et al. 2004).  New vortexes will then
develop at the head of the jet in a repeating cycle.  However, when
the cocoon of material surrounding the jet breaks out of the star and
into the low density material surrounding it, the cocoon material is
released and these vortexes are no longer swept backward.  Whatever
material is being shed from the head of the jet at this time is then
free to expand ballistically.  The result is a significant amount of
relativistic material escaping over a large angle at close to the
breakout time. Even though the details of the process require a 3D
simulation for a deep investigation, the formation of a cocoon of high
pressure material is unavoidable (see Ramirez-Ruiz et al. 2002 for
discussion of cocoon properties).  Because the opening angle of the
jet increases slowly, for off-axis observers there can be a long delay
between when the cocoon material is seen and when the jet is seen.
This vortex material could therefore be responsible for precursor
events seen 10s of seconds before the main gamma ray burst (Lazzati
2005; Lazzati et al. 2005).  In our simulations, lasting 50 seconds,
delays of up to $\sim20$ seconds were seen between the precursor
material and jet material at certain angles
(Fig.~\ref{pre_angle_16TIg5}).  At larger angles, the jet never comes
into view, but the precursor is still visible. This could account for
the observed soft, low-energy gamma ray bursts such as GRB980425
(associated with SN 1998bw, Galama et al. 1998). Whether this faint
GRB had a relativistic jet associated with it is still a matter of
lively debate (Waxman 2004).

\subsubsection{Precursor Energetics}

To estimate the total energy contained in the precursor, we first find
the isotropic energy vs. angle of the precursor at $2.4 \times
10^{11}$~cm, as described above
(\S~\ref{angulardistributionofenergy}).  This energy can then be added
over angle to give the total energy in the precursor. The mode and
full width at one tenth of the maximum of the energy-weighted
direction of motion can also be calculated to give an idea of the
angle over which the precursor is visible.  All those quantities are
given in Tab.~\ref{precursorenergetics} for $\gamma_\infty>10$ and
$\gamma_\infty>50$.

For the five simulations, the total precursor energy with
$\gamma_\infty \ge 10$ ranges from about $3 \times 10^{50}$ to $5
\times 10^{51}$~ergs, with $45\%$ to $75\%$ of that energy carried by
material with $\gamma_\infty \ge 50$. This means that $\ga1\%$ of the
total input energy of our simulation, $2.66 \times 10^{52}$~ergs per
jet, ends up as relativistic material in the precursors.  This
material is spread over a wide opening angle ($\sim 40\degr$), so the
isotropic equivalent energy of the precursor emission should be within
an order of magnitude of the total precursor energy.  Models with a
later breakout time typically have more energy in the precursor
material, which is expected since a longer propagation time allows
more relativistic material to be shed from the jet and because
simulations with a lower initial Lorentz factor have later breakout
times and have less momentum in the jet, making it easier to deflect.

\subsubsection{Precursor Energy vs. Angle}

Figure~\ref{pre_energy_angle_16TIg5} shows isotropic equivalent energy
of the precursor vs. angle for our simulations.  The isotropic
equivalent precursor energy can be up to $2 \times 10^{52}$ ergs,
$\sim 2\%$ of the $10^{54}$ ergs typically seen in the core of the
jet.  Note in Fig.~\ref{pre_energy_angle_16TIg5} that near the axis
there is usually less precursor emission.  This is because vortex
material is being deflected away from the head of the jet.  This is in
part a 2D effect.  In 3D, the jet would be able to wobble and the
precursor material would not have to be rotationally symmetric.  This
should increase the amount of precursor material on axis.  However,
the qualitative effects would be the same.  Material would still be
shed from the head of the jet, creating a region of relativistic
material spread over a wide opening angle near the head of the jet.

The precursor energies found in our simulations are sufficient to
account for observed precursor emissions, which typically produce
$\sim$ 1/1000th as many photons as the complete GRB (Lazzati 2005).
As shown in Fig.~\ref{pre_energy_angle_16TIg5}, the isotropic
equivalent precursor energy typically peaks at $10^{52}$~ergs.  The
absolute energies of observed precursors are not known because the
redshifts of GRBs with precursors have not been measured. However, it
is possible that precursors are only detected in relatively nearby
bursts, which would on average be intrinsically fainter than GRBs in
general.  This could further reduce the amount of energy needed to
produce the observed precursor emission, in the case that precursors
have less energy available than our simulations show or are less
efficient at producing gamma rays.

\subsubsection{Precursor Lorentz Factors}

Table~\ref{precursorenergetics} also contains the ratios of precursor
energies for material with $\gamma_\infty \ge 50$ to $\gamma_\infty
\ge 10$.  As highly relativistic material ($\gamma \sim 100$) is
needed to produce high-energy gamma rays, this ratio should reflect
the spectral hardness of gamma ray emission from the precursor
material.  The precursor material is less relativistic than the jet
material because some stellar material has been mixed into it.
Whereas the jet material typically has $\ga 80\%$ of its energy in
material with $\gamma_\infty \ge 50$, the precursor material has $\sim
60\%$ of its energy in material with $\gamma_\infty \ge 50$.  A larger
percentage of energy in mildly relativistic material could produce a
softer spectrum of emitted radiation, as observed in GRB precursors
(Lazzati 2005) and low-energy GRBs (Galama et al. 1998).

\subsubsection{Isotropic Precursors}
\label{secpr}

When the cocoon breaks out of the star, before the cocoon energy of is
released, there is a thin shell of mildly relativistic material formed
at the leading edge of the expanding material.  This shell provides a
second possible source of precursor emission. The shell is nearly
isotropic and is visible in Fig.~\ref{energy_time_angle_16TIg5}.  Note
that this shell is not well resolved in the simulations presented
here.  The isotropic equivalent energy in this shell is $\sim 10^{47}$
to $\sim 10^{49}$ ergs, far lower than the energy of a typical GRB.
It is therefore unlikely that radiation from this material would be
detected along with a classical GRB.  The shell material also has a
low terminal Lorentz factor, typically around 5, and therefore may not
be able to produce gamma ray photons.  However, this material could
still produce X-ray photons and may be observed as an X-ray flash by
observers far off-axis, who see emission from neither the jet nor the
vortex precursor material.

Between the jet material, vortex material and isotropic shell
material, observers at different angles could see the same event as a
classical GRB without a precursor, a classical GRB with a precursor, a
low-energy GRB, or an X-ray flash.  Figure~\ref{pre_angle_16TIg5}
shows plots of energy flux vs. time for 4 different angles to
illustrate different types of observed events.

At a viewing angle of $1.125\degr$, a large energy flux begins to
arrive about $15$ seconds after the start of the simulation and last for
$18$ seconds.  After this there is a small amount of energy still
arriving (from the unshocked jet) until the end of the simulation.  At
$5\degr$, the precursor is visible as a spike of emission beginning
at $\sim14$s and lasting for about $1$s.  This is followed by $\sim15$
seconds with low energy flux.  Emission then begins again as the edge 
jet comes into view.  After $44$s the emission reaches a lower 
constant level as the unshocked portion of the jet comes into view.  
At $7\degr$, the precursor is again visible starting at $14$s, but 
the jet does not come into view until $\sim20$ seconds later.
At $12\degr$, the precursor is still seen at $14$s, but the jet is 
never visible.

\subsection{Effects of Resolution}

In order to test the effects of resolution on our simulations, we have
carried out a version of 16TIg5 at half the resolution of our other
simulations.  In the low-resolution version of model 16TIg5, there is
not a distinct precursor phase preceding the shocked jet, as can be
seen in Fig.~\ref{energy_time_angle_16TIg5lowres}.  This appears to be
because the resolution of this model is not sufficient for vorticity
to develop at the head of the jet. The thin isotropic shell is,
however, seen in the low resolution model. In the first three seconds
after reaching $2.4 \times 10^{11}$~cm, the jet is wider at lower
energies than at later times and is narrower at high energies.
Although this structure is not a separate precursor as seen in the
high resolution simulations, it could still give rise to precursor
emissions seen off-axis before the jet comes into view.

Despite a lack of distinct vortex structures, the evolution of the
shocked and unshocked phases are very similar to the high resolution
model.  Because material at the head of the jet has not been
deflected, the shocked portion of the jet is larger in the low
resolution model.  In other words, the reverse shock has propagated
farther into the jet at $2.4\times10^{11}$~cm in the low resolution
model.  This would be expected as material is not being shed from the
jet as in the high resolution model.

The breakout times of the cocoon and shocked jet are about 3 
seconds earlier in the low resolution simulation, and unshocked 
jets have nearly identical breakout times (Tab.~\ref{breakout}).  
Figure~\ref{total_energy_angle_16TIg5lowres}
compares the total energy vs. angle at the two resolution and shows
that they are very similar.  Figure~\ref{opeining_angle_16TIg5lowres}
compares the opening angle vs. time of the two resolutions.  The low
resolution plot shows a wider opening angle, particularly at the lower
Lorentz factor cutoffs, but this is expected due to the lower
resolution.  Other than this, the time evolution of the opening angle
is nearly identical in the two simulations.

\section{Analytic Modeling and Interpretation} 
\label{analytic}

More insight into the results of the numerical work can be achieved if
the dominant processes that shape the jet evolution and its
interaction with the star can be singled out. To this aim, we
developed an analytical description of the jet-cocoon-star
interaction. We find that the phase during which the jet is confined
inside the star (before the breakout) and the unshocked jet phase can
be reasonably well approximated with a semi-analytic treatment. This
allows us to compute breakout times, precursor energetics, the amount
of energy given to the star and the late time evolution of the jet
properties.

In LB05 we explored the dynamics of the jet-cocoon interaction under
the \emph{monolithic jet} assumption, in which the jet is assumed to
be uniform across its section. In addition, the jet is assumed to
satisfy Bernoulli conditions, i.e., no significant dissipation by
shocks. As a consequence, the pressure exerted by the jet on the
cocoon material is only the internal pressure (which is
relativistically invariant for a perpendicular Lorentz boost),
\begin{equation}
p_j=\frac{L_j}{4\,\Sigma_j\,c\,\gamma_j^2},
\label{eq:pjet}
\end{equation}
where $L_j$ is the jet luminosity, $\Sigma_j$ is the jet cross section
and $\gamma_j$ is the jet Lorentz factor. Under this approximation,
LB05 found that the jet reaches the stellar surface very narrow and
spreads exponentially afterwards.

\begin{figure}
\epsscale{1.0}
\vspace*{-22mm}
\includegraphics[width={.9\columnwidth}]{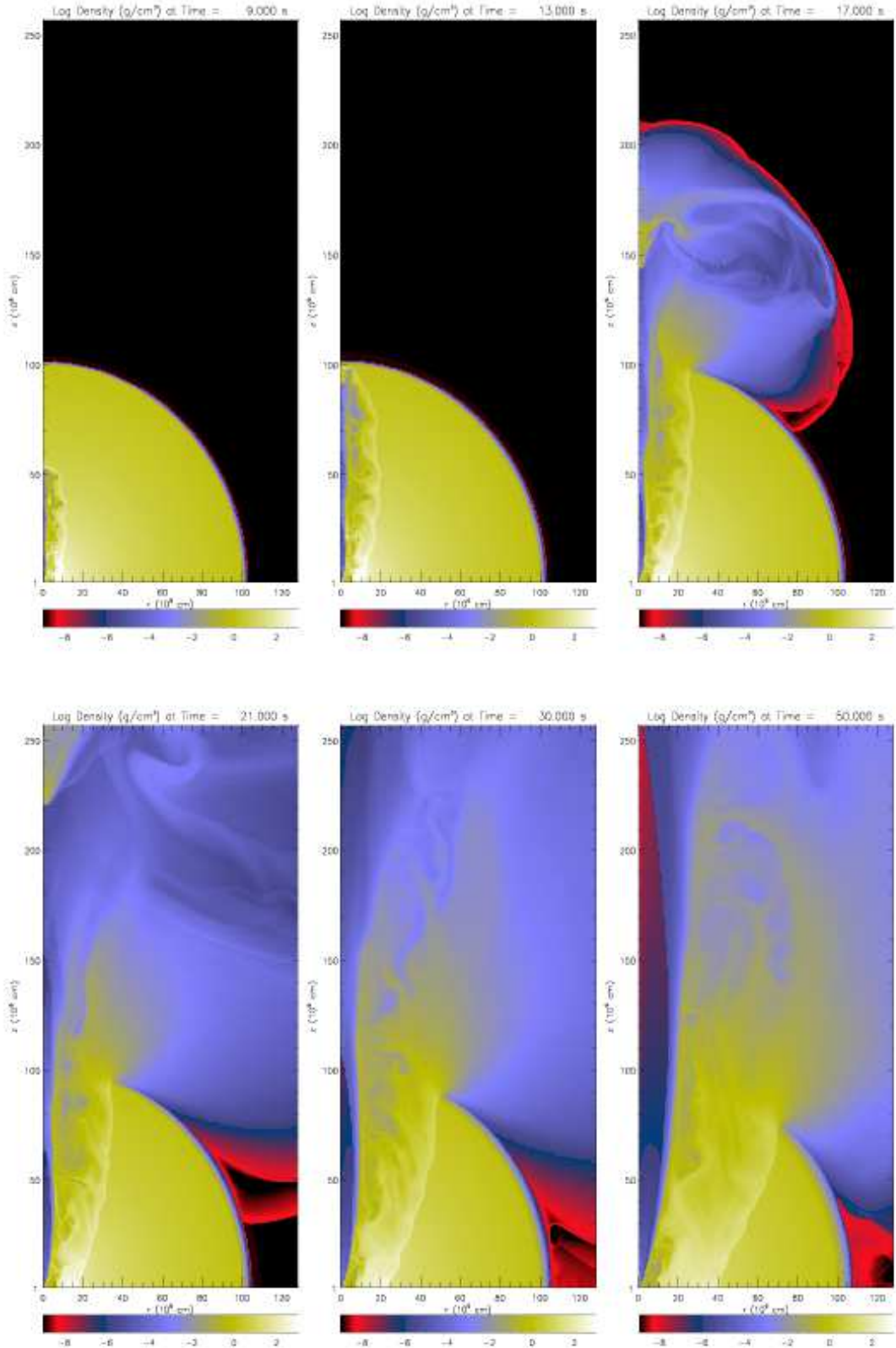}
\caption{\label{timesequence} Time sequence of logarithmic density for
model t10g5 from 9 to 50 seconds.  Initially the jet is confined, and
hot turbulent material is stored in a cocoon (first two panels). When
the jet head reaches the surface, the cocoon is released as a wide
angle outflow (third panel). Immediately afterwards, a heavily shocked
jet flows outside the star (third and fourth panels). Eventually, a
more stable configuration emerges (fifth and sixth panels) in which
the jet is internally free-flowing, and is bounded by a shear layer at
the contact discontinuity with the star.}
\end{figure}

\begin{figure}
\epsscale{1.0}
\includegraphics[width={.9\columnwidth}]{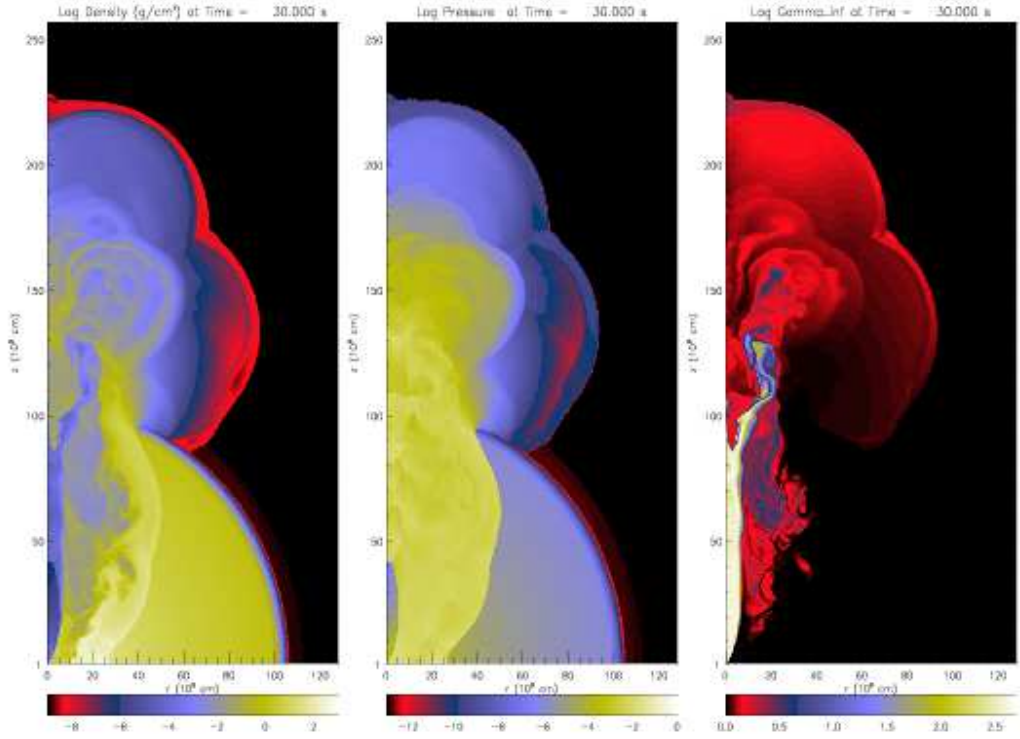}
\caption{\label{rhopg} Logarithmic density, pressure and
$\gamma_\infty$ (left to right) for model t10g2 at 30 seconds. Visible
are the cocoon (high density, high pressure, low $\gamma_\infty$),
precursor region (high $\gamma_\infty$, off axis), shocked jet (low
density, high pressure, high $\gamma_\infty$), and unshocked jet (low
density, low pressure, high $\gamma_\infty$).}
\end{figure}

\begin{figure}
\epsscale{1.0}
%\plotone{phase_definition_16TIg5.eps}
\plotone{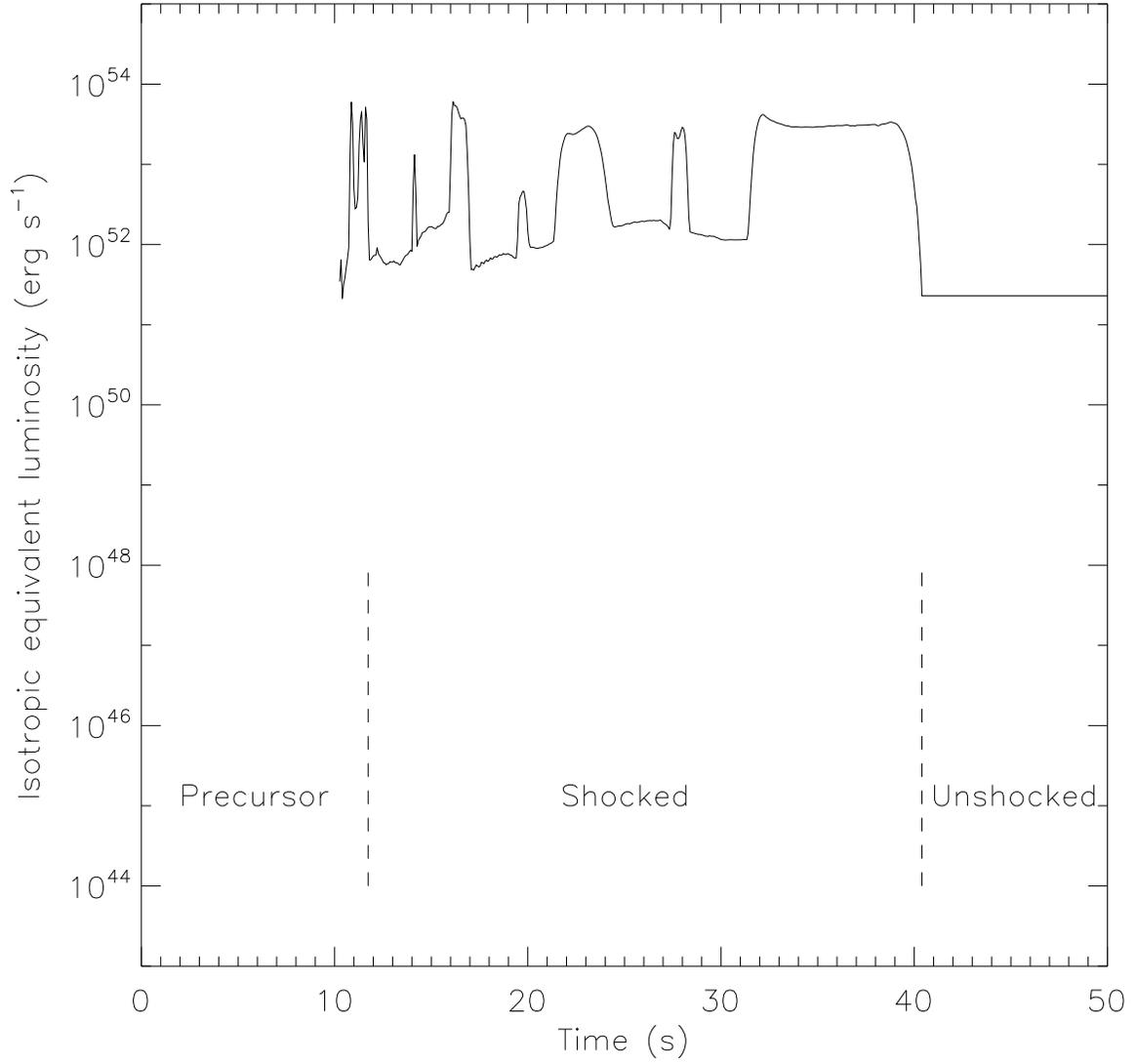}
\caption{\label{fig:phasedef} The definition of the transition times
between the three phases of jet evolution described in the text. The
figure shows the energy flux along the jet axis versus time for model 16TIg5. The
different phases of the jet are shown.}
\end{figure}

\begin{figure}
\epsscale{1.0}
%\plottwo{rhd_jet2.5_9m15t10e250g5low9_hdf5_plt_cnt_Gamma_inf0210.eps}{rhd_jet2.5_9m15t20e250g5low9_hdf5_plt_cnt_Gamma_inf0630.eps}
\plotone{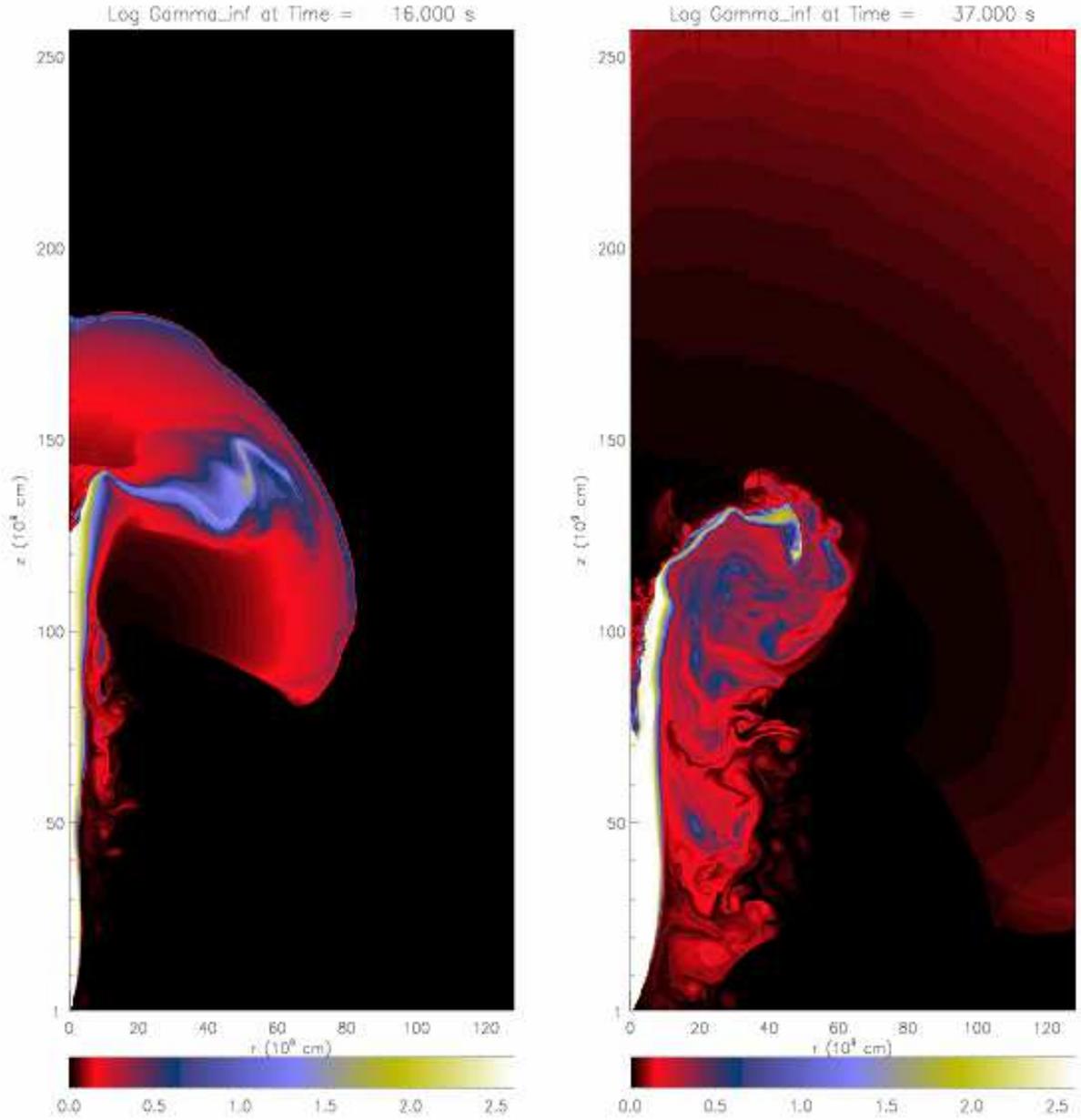}
\caption{\label{t20g5compare} Comparison of $\gamma_\infty$ for models
t10g5 at 16 seconds (left) and t20g5 at 37 seconds (right).  Model
t20g5 is much less collimated and the relativistic material is being
highly deflected even at this late time.}
\end{figure}

\begin{figure}
\epsscale{1.0}
%\plottwo{total_energy_angle_16TIg5.eps}{total_energy_angle_16TIg2.eps}
%\plottwo{total_energy_angle_t10g5.eps}{total_energy_angle_t10g2.eps}
%\plottwo{total_energy_angle_t5g2.eps}{}

\plottwo{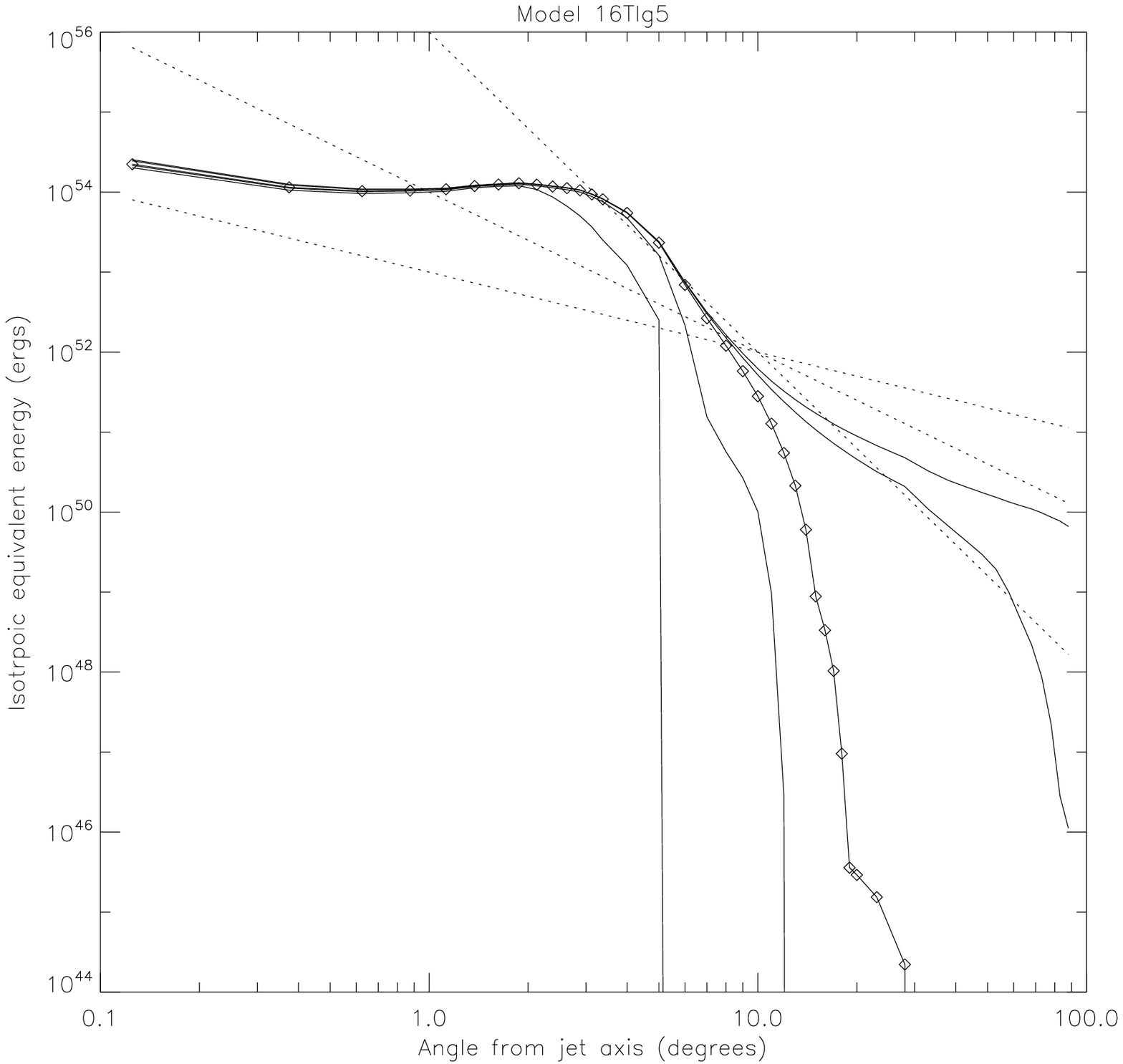}{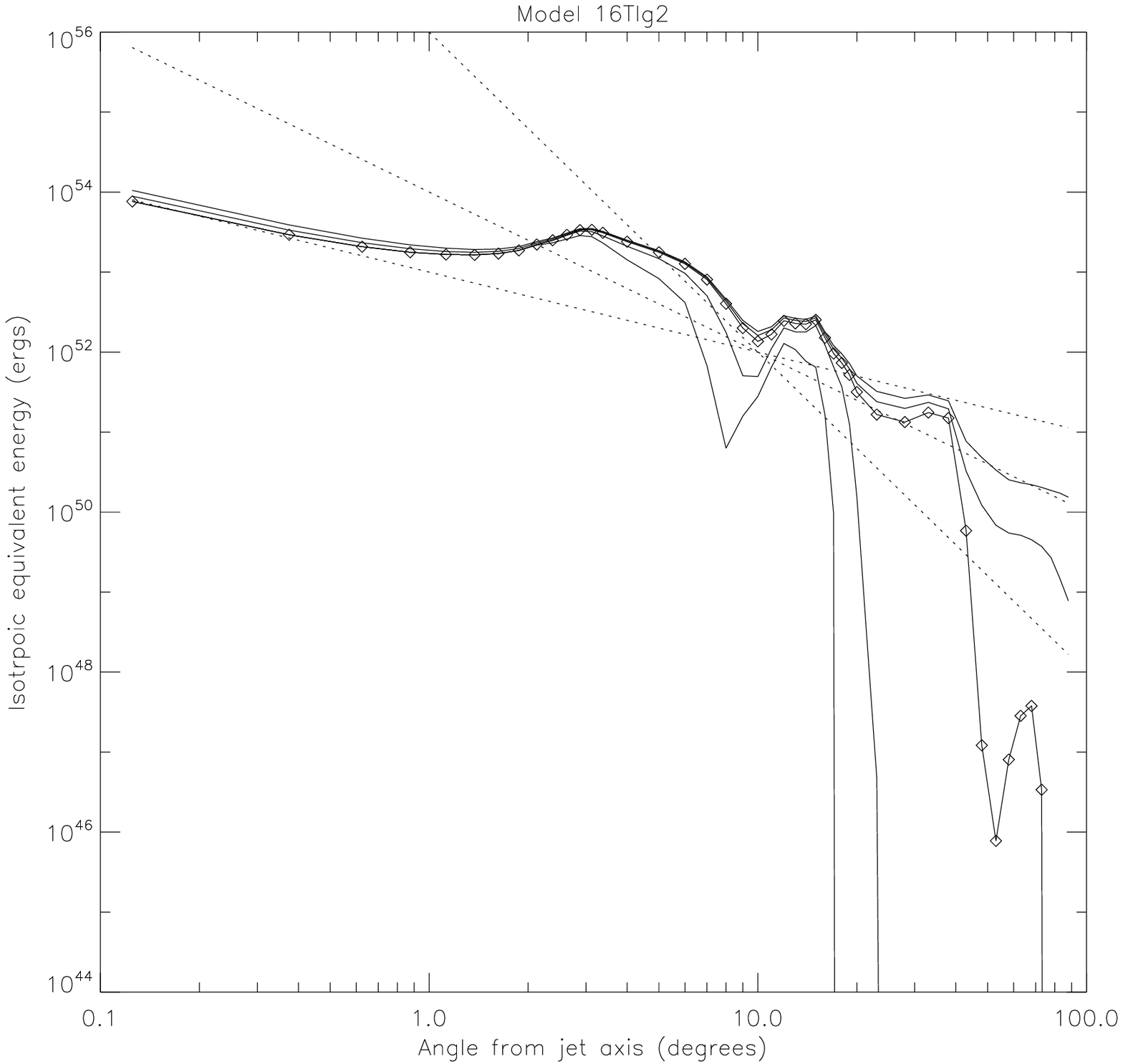}
\plottwo{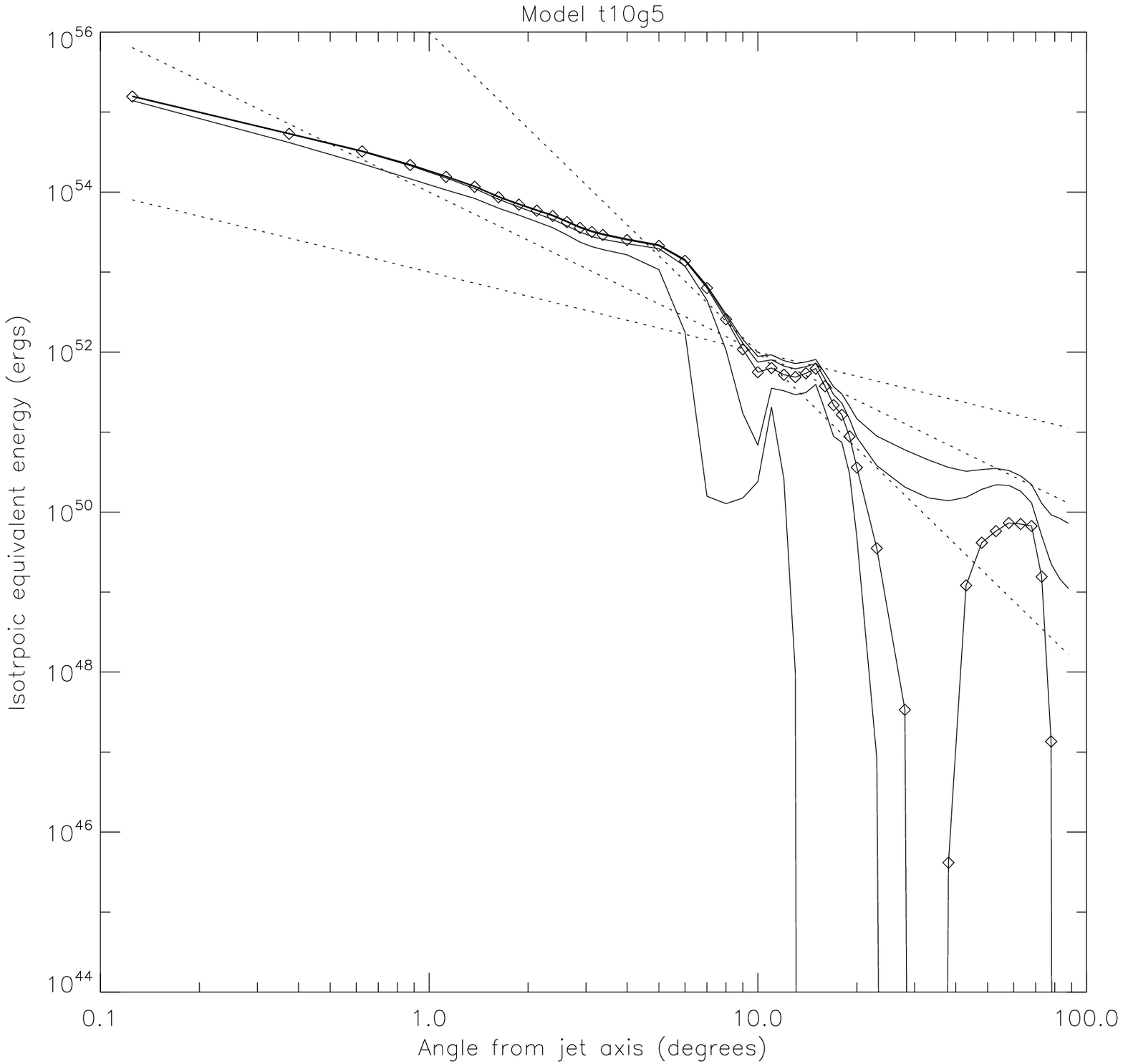}{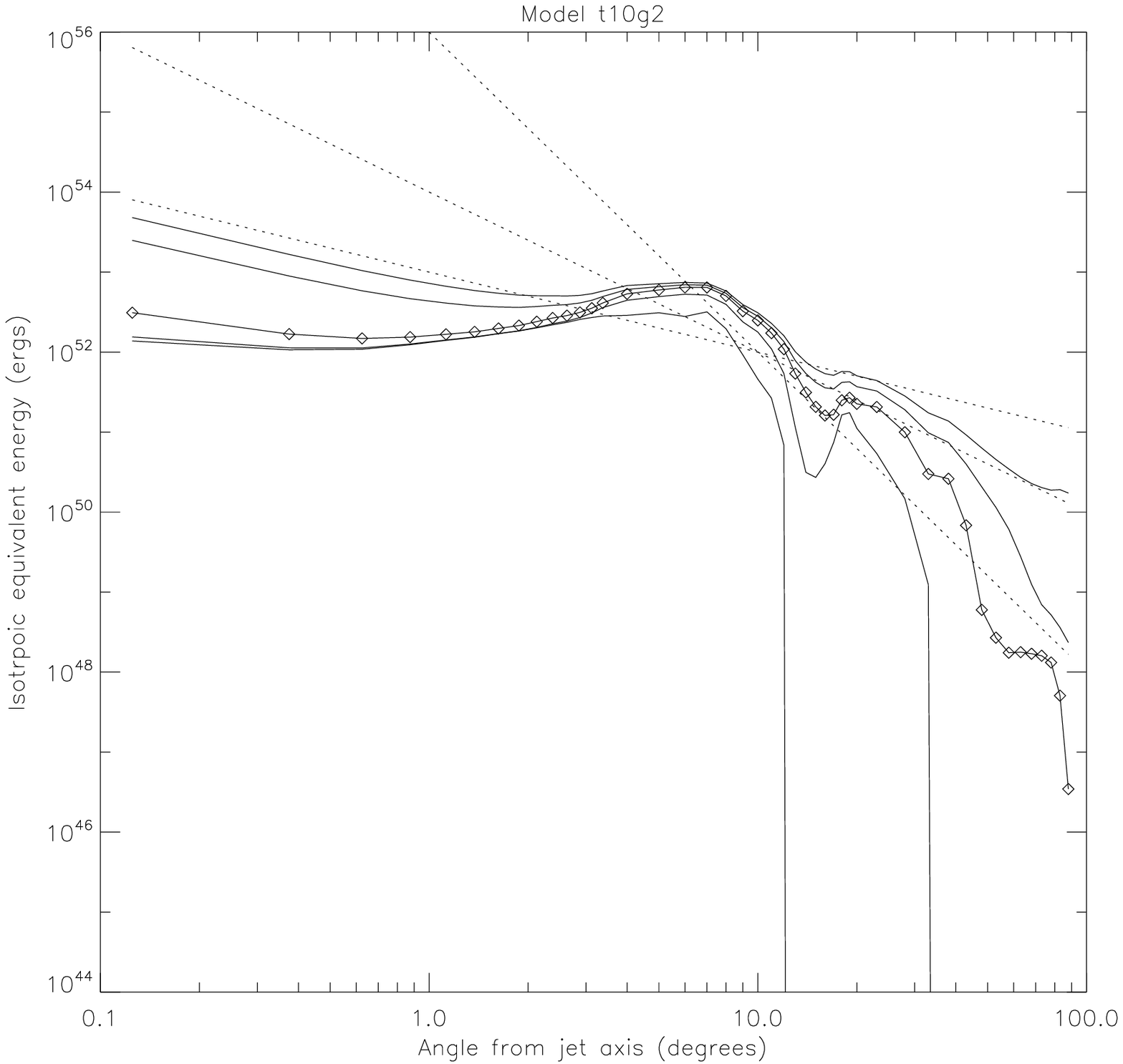}
%\plottwo{total_energy_angle_t5g2.eps}{}
\caption{\label{total_energy_angle_16TIg5} Total isotropic equivalent
energy as a function of angle for models 16TIg5 (upper left), 16TIg2
(upper right), t10g5 (lower left), and t10g2 (lower right).  Different
lines correspond to the amount of energy above a minimum Lorentz
factor of $\gamma_{\infty} = 1.01$, $2$, $10$, $50$, and $200$, from
top to bottom.  Highlighted line corresponds to $\gamma_{\infty} =
10$.  Dotted lines are slopes of $E_{iso} \propto \theta^{-1}$,
$\theta^{-2}$, and $\theta^{-4}$.}
\end{figure}

\begin{figure}
\epsscale{1.0}
%\plottwo{no_pre_energy_angle_16TIg5.eps}{no_pre_energy_angle_16TIg2.eps}
%\plottwo{no_pre_energy_angle_t10g5.eps}{no_pre_energy_angle_t10g2.eps}
%\plottwo{no_pre_energy_angle_t5g2.eps}{}

\plottwo{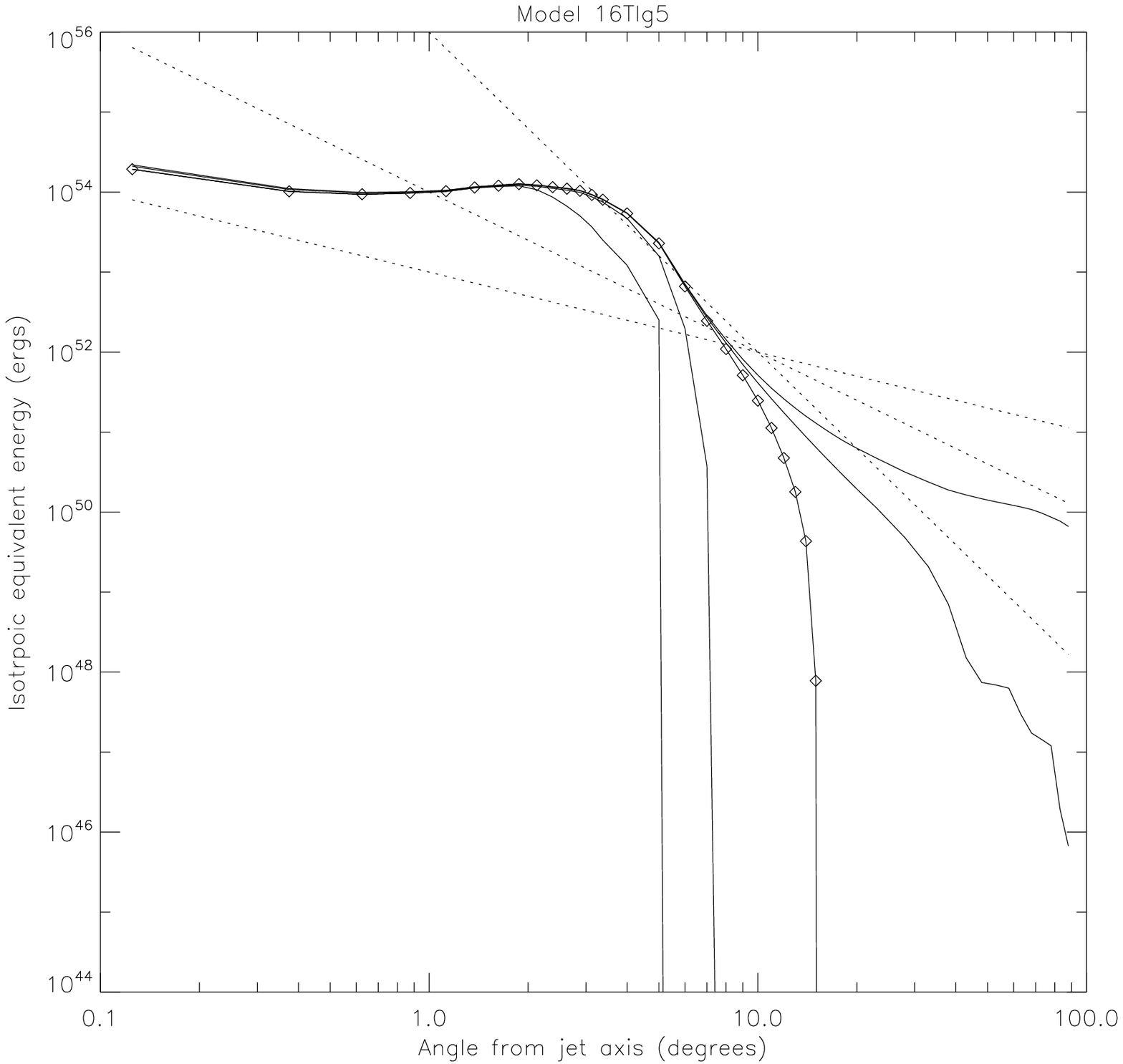}{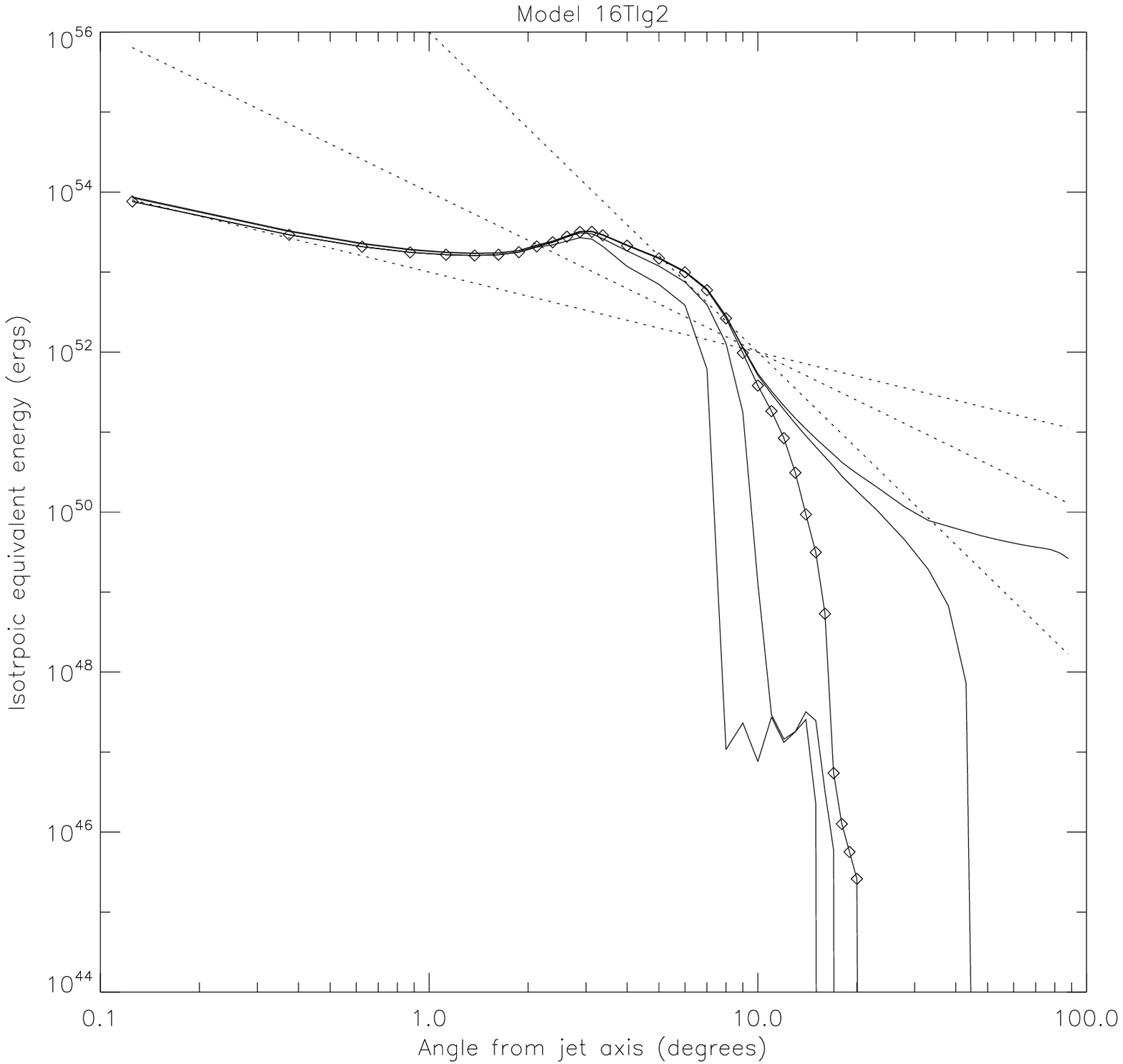}
\plottwo{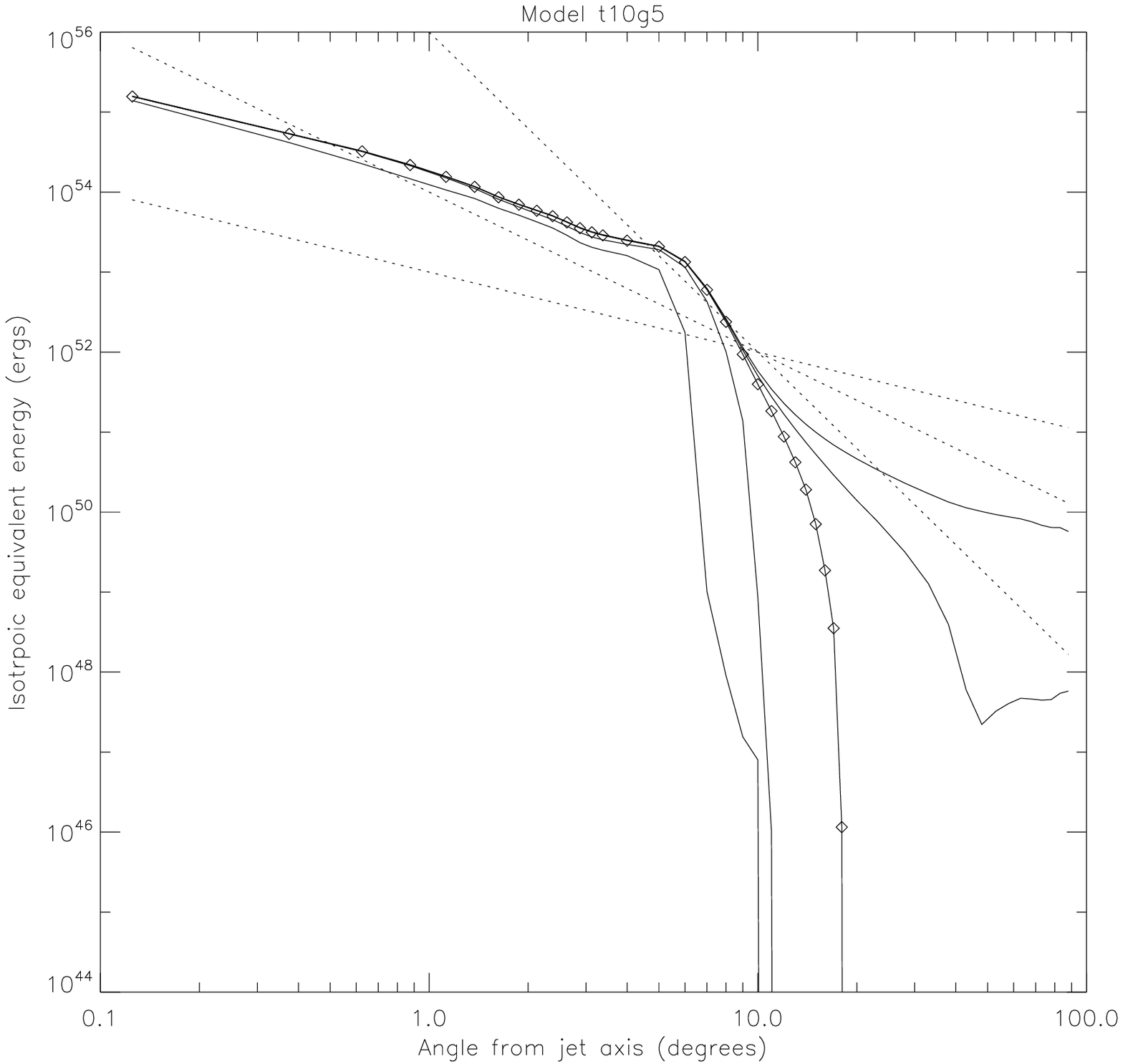}{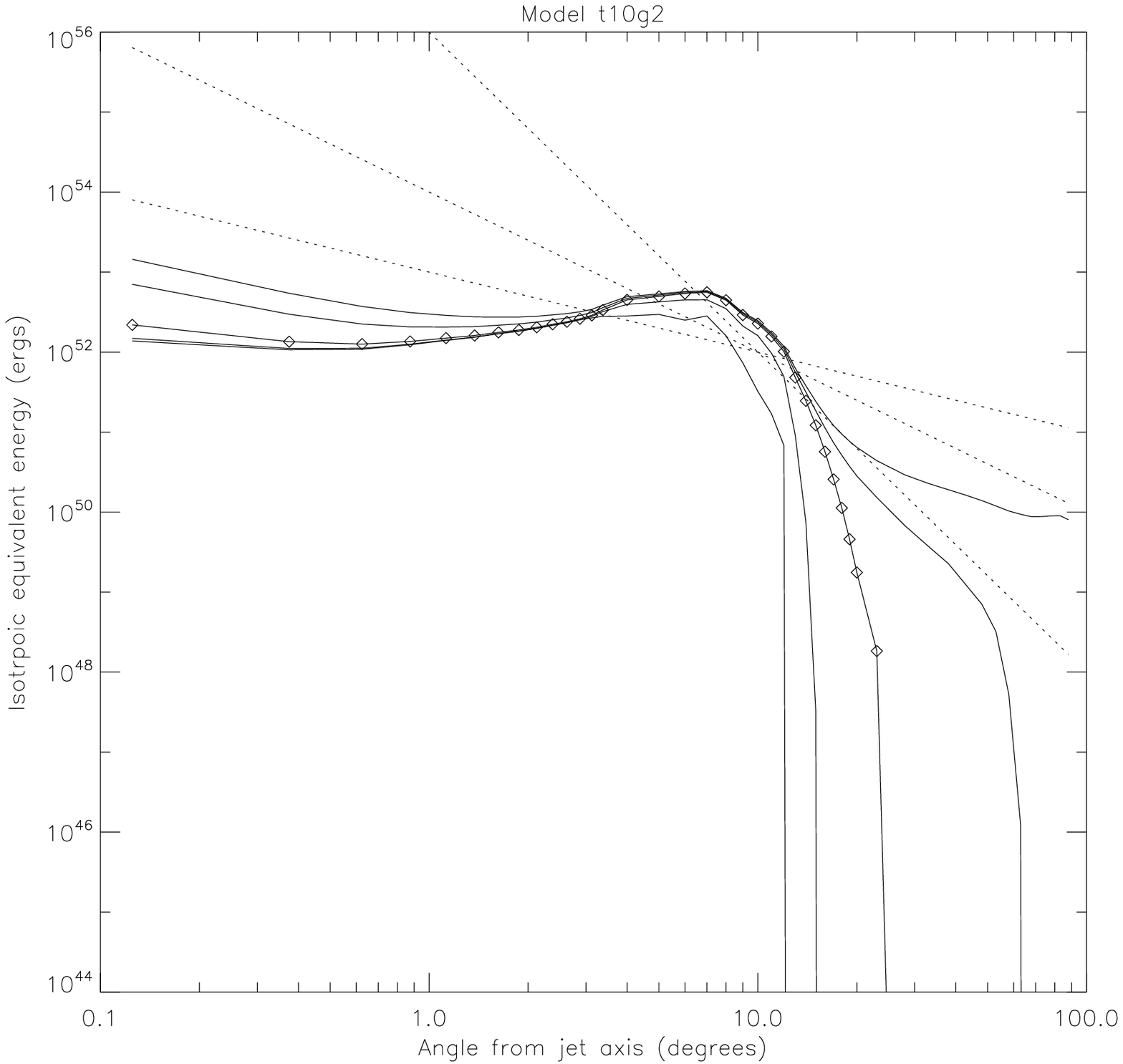}
%\plottwo{no_pre_energy_angle_t5g2.eps}{}
\caption{\label{no_pre_energy_angle_16TIg5} Isotropic equivalent
energy, excluding precursor phase energy, as a function of angle for
models 16TIg5 (upper left), 16TIg2 (upper right), t10g5 (lower left),
and t10g2 (lower right).  Different lines correspond to the amount of
energy above a minimum Lorentz factor of $\gamma_{\infty} = 1.01$,
$2$, $10$, $50$, and $200$, from top to bottom.  Highlighted line
corresponds to $\gamma_{\infty} = 10$.  Dotted lines are slopes of
$E_{iso} \propto \theta^{-1}$, $\theta^{-2}$, and $\theta^{-4}$.}
\end{figure}

\begin{figure}
\epsscale{1.0}
%\plotone{shock_unshock_energy_angle_16TIg5.eps}
\plotone{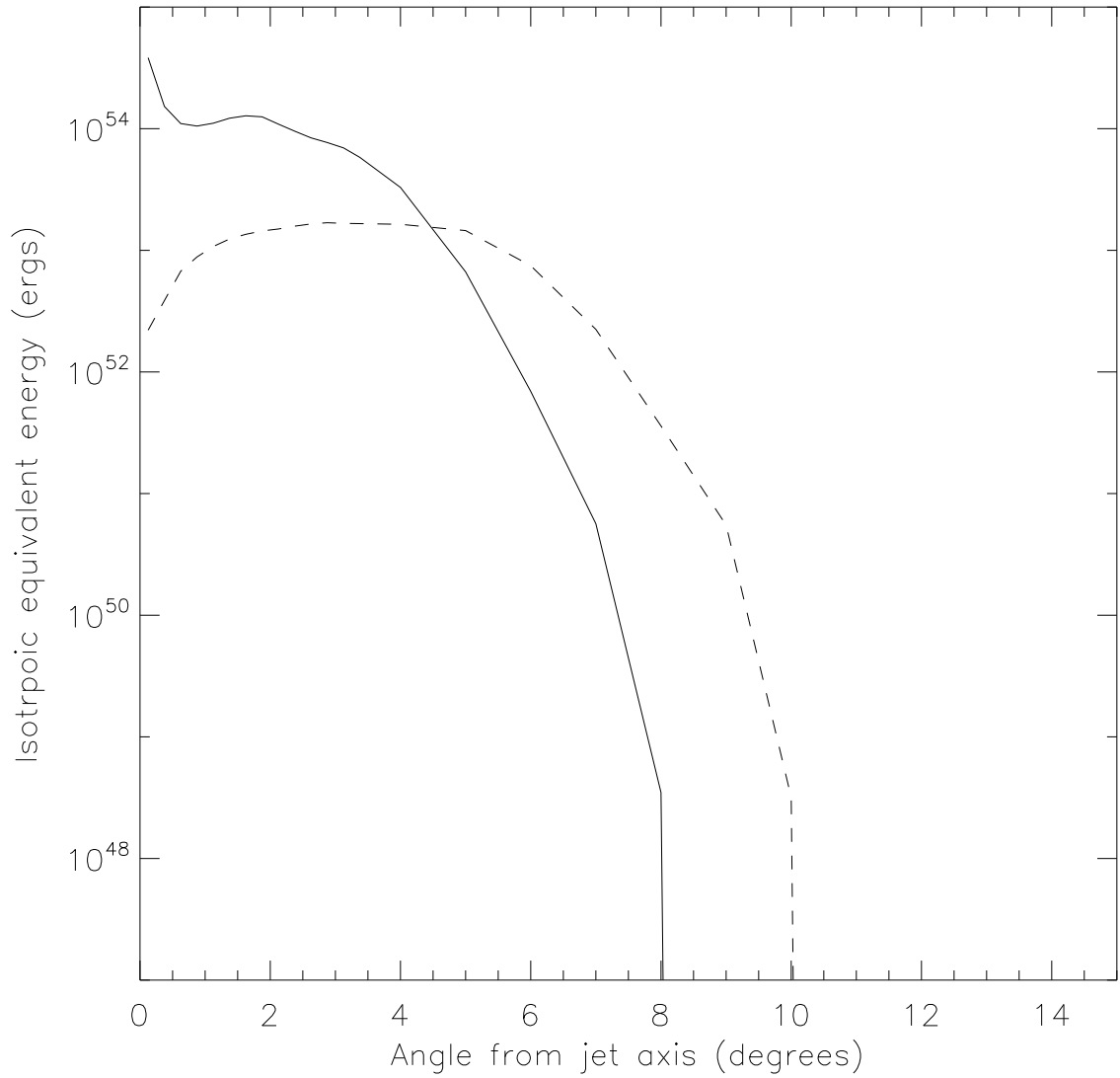}
\caption{\label{shock_unshock_energy_angle_16TIg5} Isotropic
equivalent energy as a function of angle for model 16TIg5 for the
shocked (solid line) and unshocked (dashed line) phases.  Both lines
have a minimum Lorentz factor of $\gamma_\infty = 50$.  }
\end{figure}

\begin{figure}
\epsscale{1.0}
%\plotone{energy_time_angle_16TIg5.eps}
\plotone{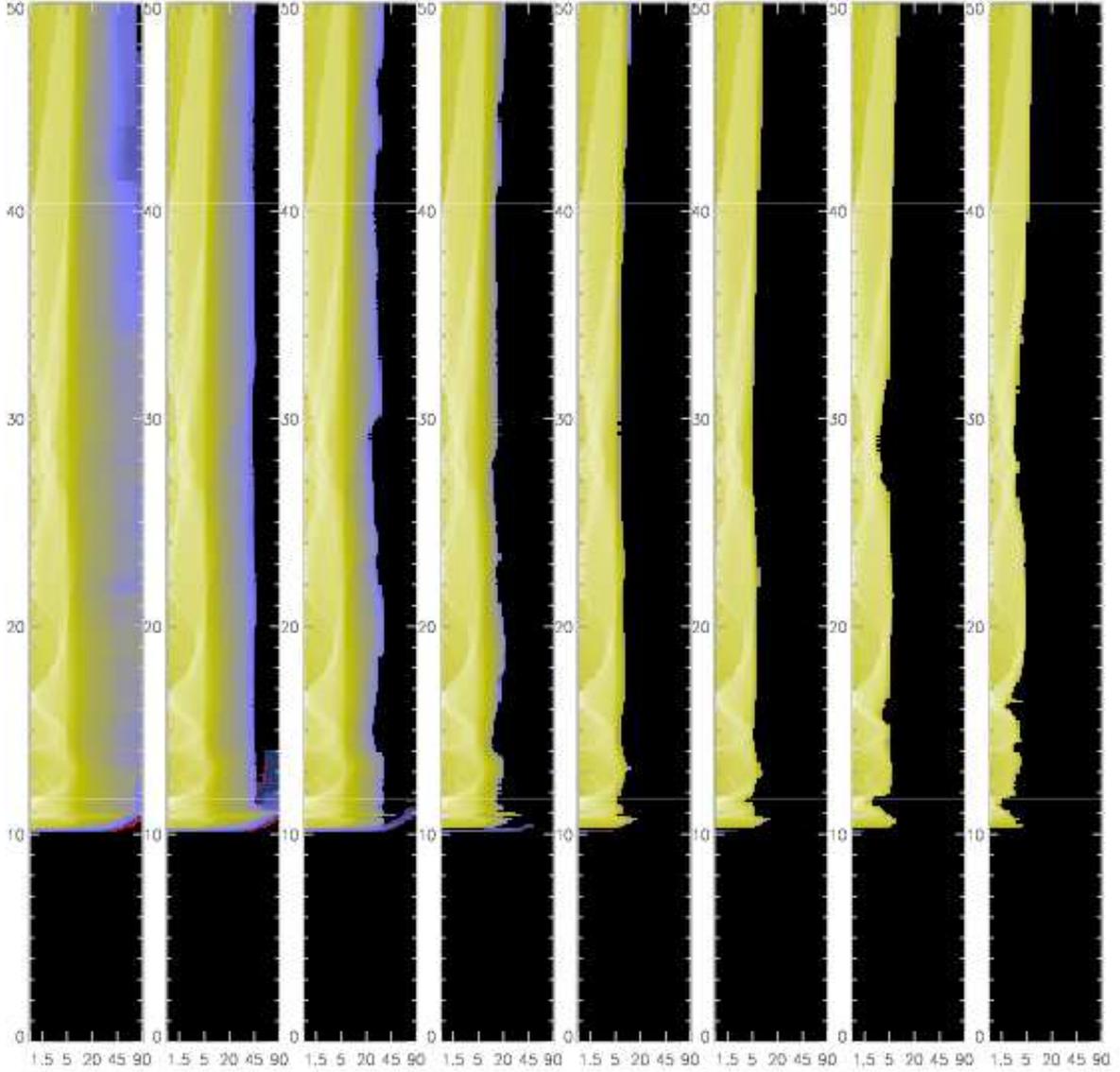}
\caption{\label{energy_time_angle_16TIg5} Time and angular
distribution of (logarithmic) energy from model 16TIg5 measured at
$1.2 \times 10^{11}$~cm.  Y-axis is time in seconds and x-axis is
angle from jet axis in degrees.  The x-axis is not uniform but instead
corresponds to the angular bins described in
\S~\ref{angulardistributionofenergy}.  The panels correspond to the
amount of energy above a minimum Lorentz factor of $\gamma_{\infty} =
1.01$, $2$, $5$, $10$, $30$, $50$, $100$, and $200$, from left to
right. White horizontal lines show the times at which the transitions
between phases take place.}
\end{figure}

\begin{figure}
%\plottwo{breakout_gamma_log_16TIg5.eps}{breakout_pres_log_16TIg5.eps}
\plottwo{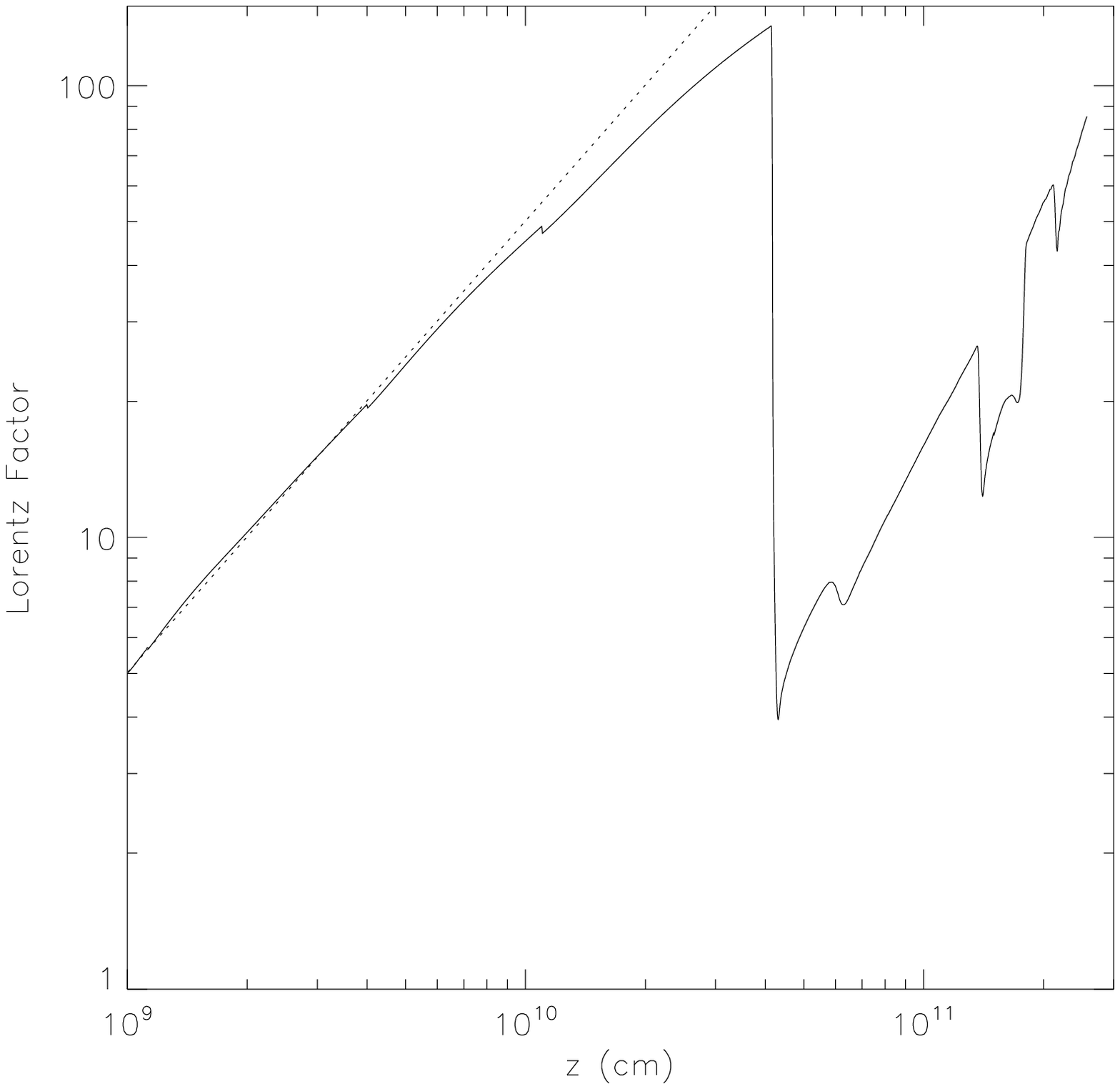}{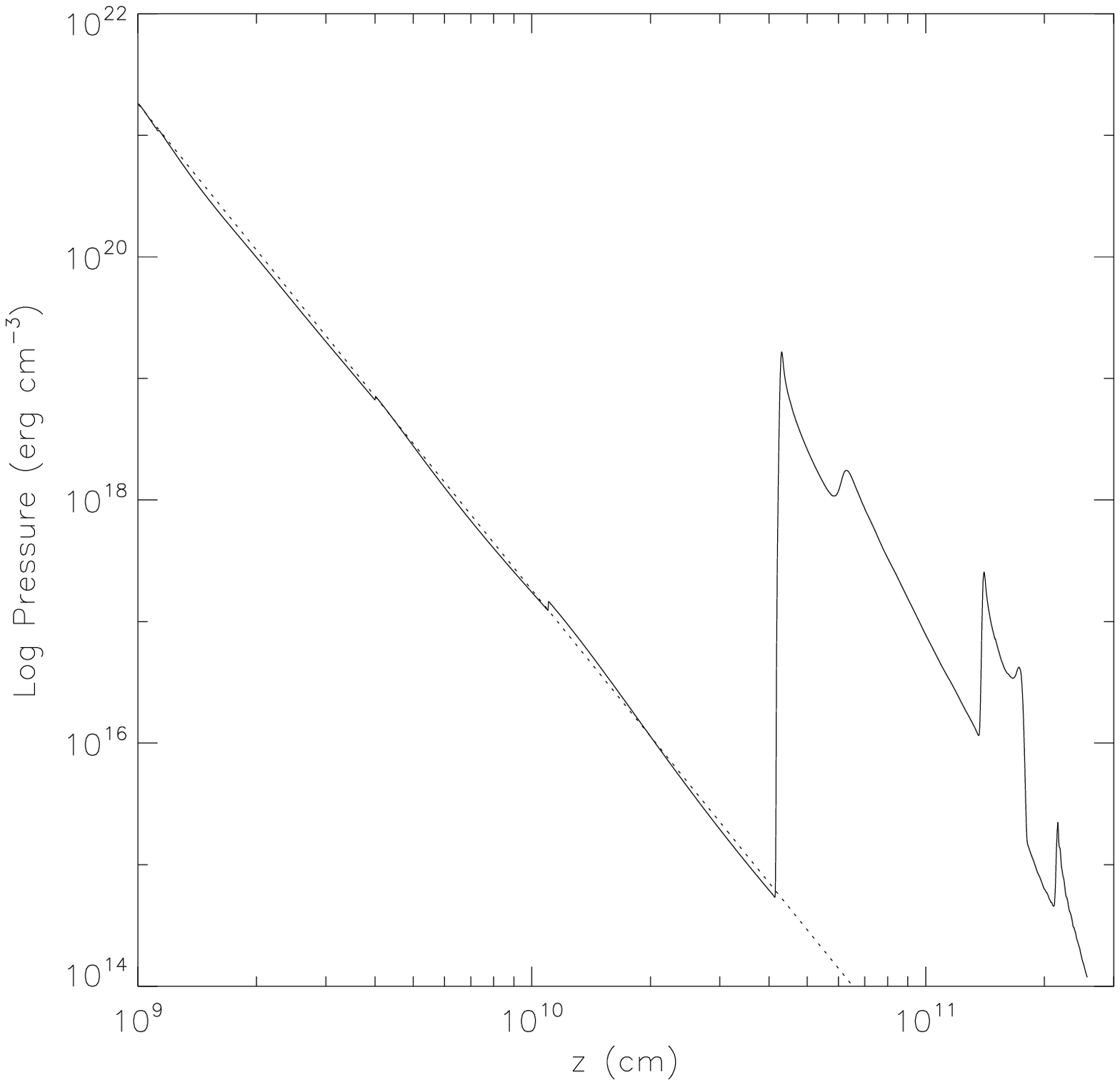}
\caption{\label{fig:frexp} Lorentz factor and pressure along the jet
axis as a function of the distance from the black hole at the time of
the emergence of the unshocked jet phase on the stellar surface for
model 16TIg5. The dotted lines show the theoretical prediction for an
free streaming accelerating jet: $\gamma\propto{}z$ and
$p\propto{z}^{-4}$. This shows that the core of the unshocked jet is
free streaming. Small deviations from the theoretical behavior are due
to the fact that when $\gamma\gtrsim100$ the flow is not
pressure-dominated any more.}
\end{figure}

\begin{figure}
\epsscale{1.0}
%\plottwo{opening_angle_16TIg5.eps}{opening_angle_16TIg2.eps}
%\plottwo{opening_angle_t10g5.eps}{opening_angle_t10g2.eps}
%\plottwo{opening_angle_t5g2.eps}{}

\plottwo{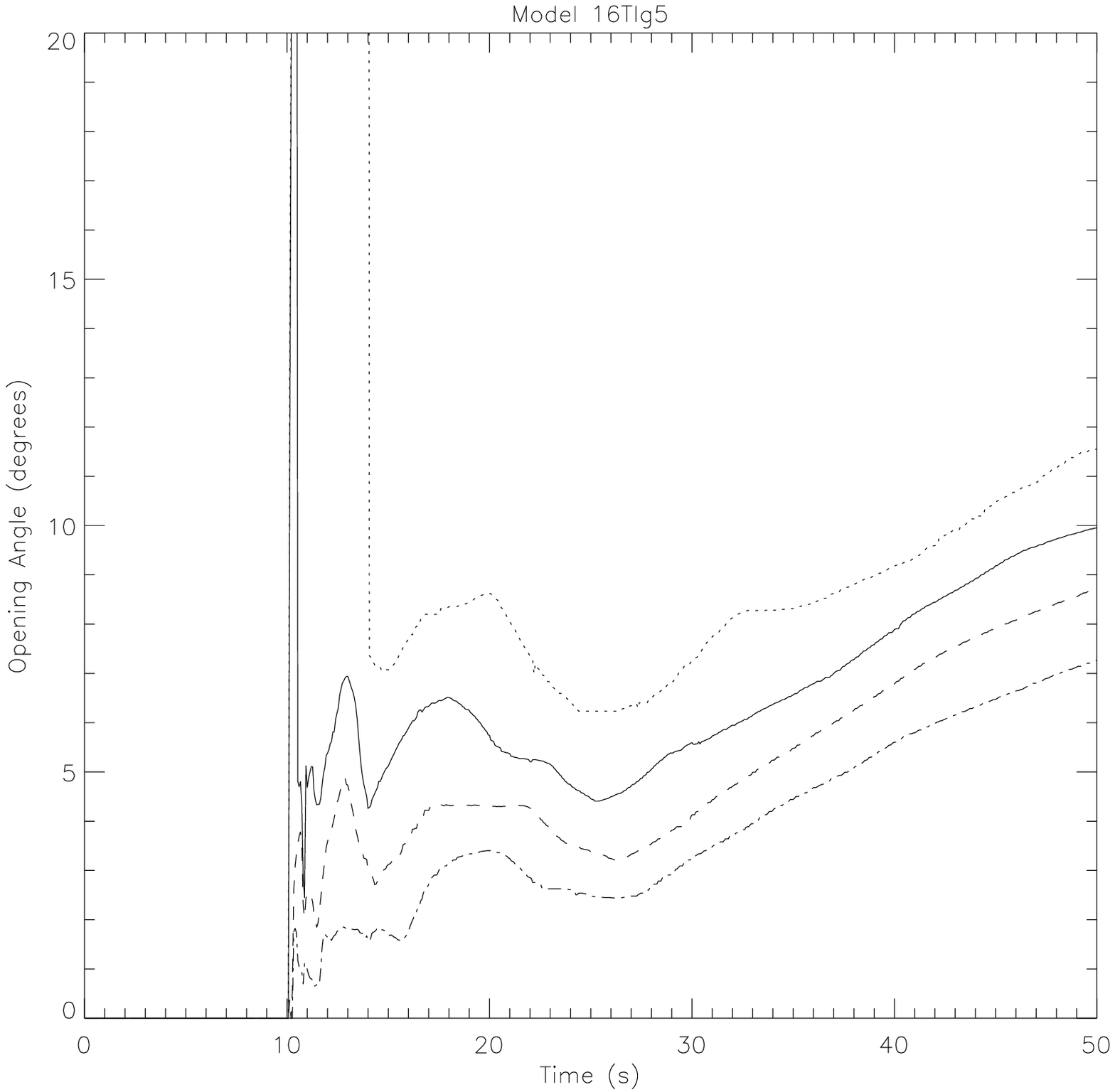}{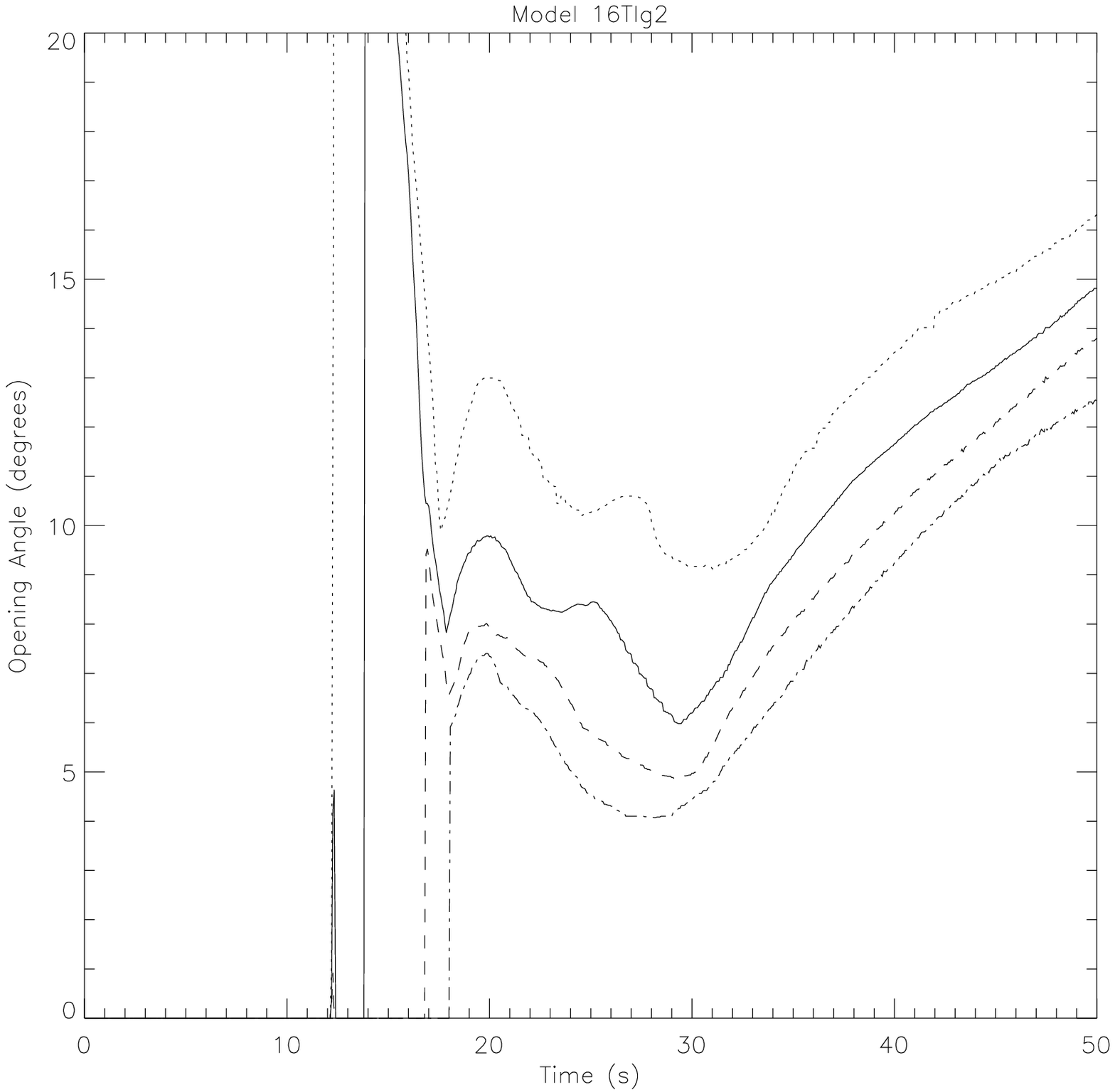}
\plottwo{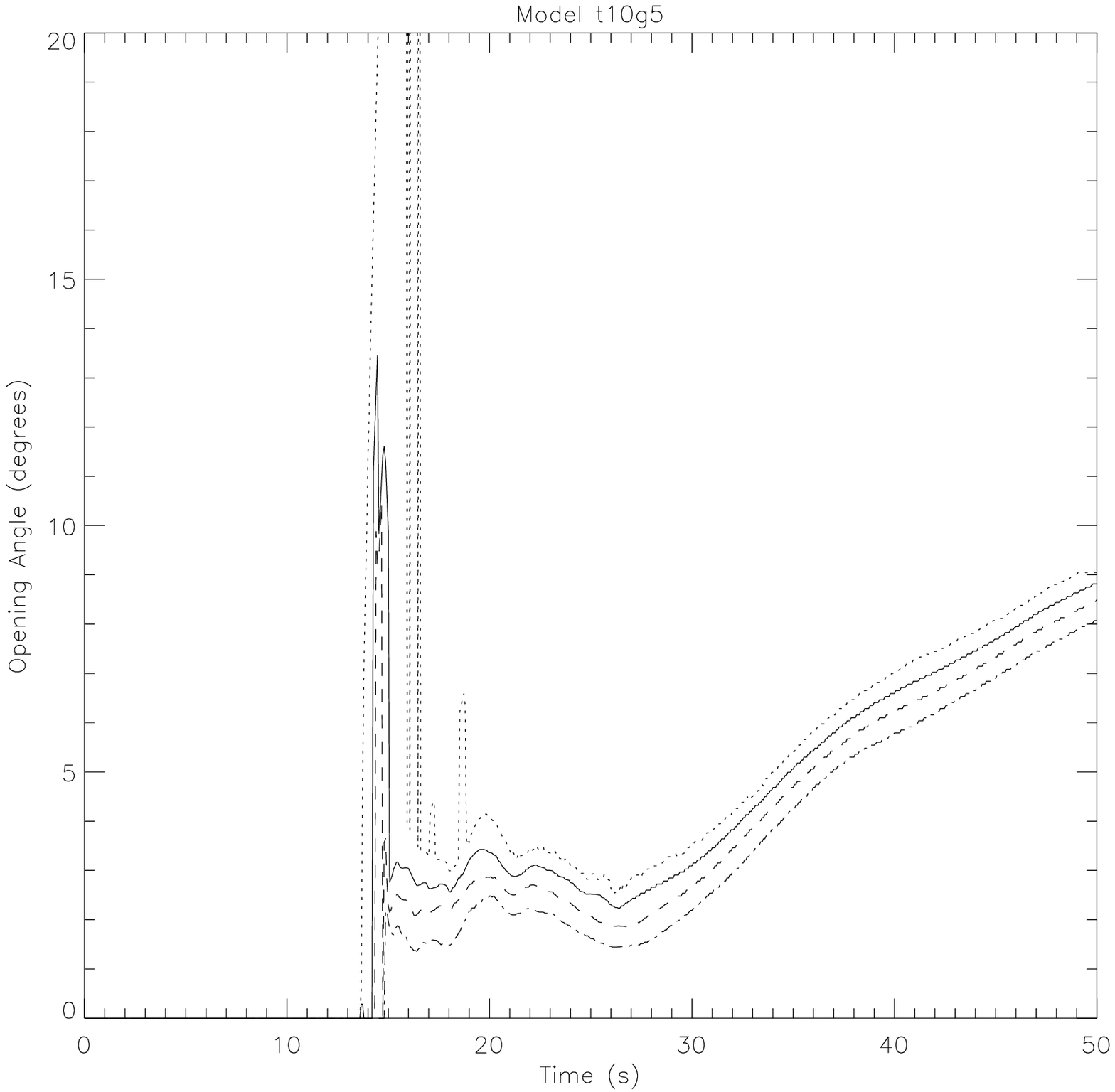}{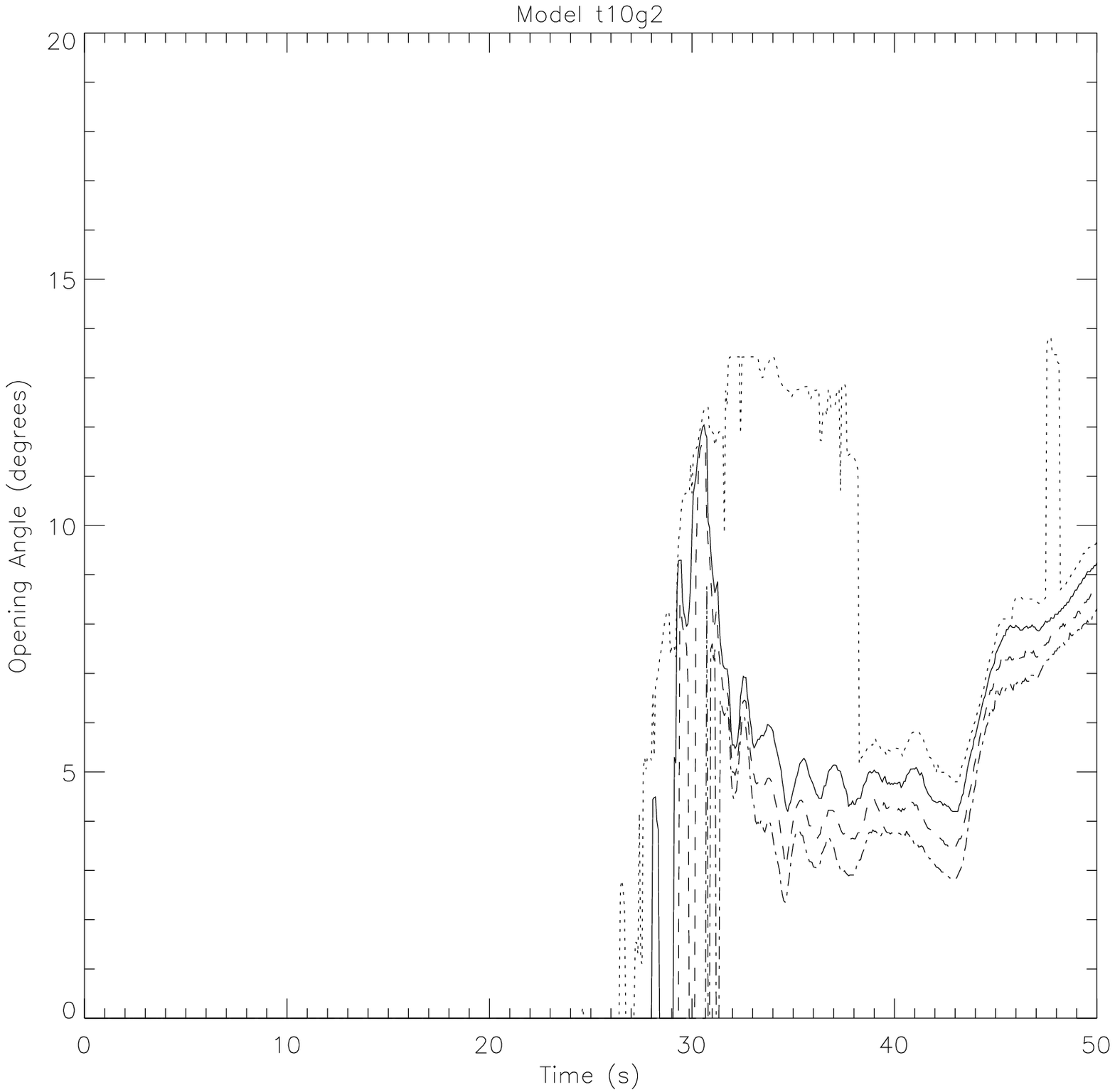}
%\plottwo{opening_angle_t5g2.eps}{}
\caption{\label{opening_angle_16TIg5} Opening angle vs. time at a
radius of $1.2 \times 10^{11}$~cm for models 16TIg5 (upper left),
16TIg2 (upper right), t10g5 (lower left), and t10g2 (lower right).
Different lines correspond to opening angle for material with a
minimum terminal Lorentz factor of $\gamma_{\infty} = 2$ (dotted
line), $10$ (solid line), $50$ (dashed line), and $200$ (dot-dash
line).  The initially large opening angle is due to the passage of the
precursor material.}
\end{figure}

The FLASH numerical simulations we present show that the jet does not
have a uniform distribution of pressure and density but rather
develops a boundary layer structure (see Fig.~\ref{timesequence}). The
core of the jet is freely streaming out to the point at which it
collides with the boundary layer, which in turn flows parallel to the
jet-cocoon boundary. In addition to the jet internal pressure, we have
therefore the ram pressure due to the deflection of the free streaming
jet by the boundary layer. Figure~\ref{fig:sketch} shows a sketch of
the jet geometry that we consider.

Simple geometry allows us to derive the pressure balance equation
\begin{equation}
p_{\rm{cocoon}}=p_j+4\,p_j\,\gamma_j^2\,\sin\left[
{\rm{atan}}\left(\frac{dz}{dr_\perp}\right)-
{\rm{atan}}\left(\frac{z}{r_\perp}\right)
\right],
\label{eq:bal}
\end{equation}
where $z$ is the coordinate along the jet and $r_\perp$ is the
perpendicular size of the jet. Eq.~\ref{eq:bal} can be simplified in
the approximation of a narrow ($z\gg{}r_\perp$) relativistic jet (for
which the ram pressure is much larger than the internal pressure)
yielding a differential equation of the form
\begin{equation}
\frac{dr_\perp}{dz}=\frac{r_\perp}{z}-r_\perp^2\,K,
\label{eq:approx}
\end{equation}
where $K=\pi\,c\,p_{\rm{cocoon}}/L_j$ is a constant related to the
ratio of the ram to cocoon pressures. Eq.~\ref{eq:approx} has an
analytic solution of the form
\begin{equation}
r_\perp=\frac{2\,z}{K\,z^2+C},
\label{eq:solu}
\end{equation}
where $C$ is a constant of integration. Eq.~\ref{eq:solu} can be
rewritten in a clearer form as
\begin{equation}
\theta_j=\frac{2\,\theta_0}{2+K\,\theta_0\,(z^2-z_0^2)}.
\label{eq:theta}
\end{equation}
In this form is easy to see how the opening angle of the jet is
initially constant, but decreases as the jet propagates. For very
large values of $z$ the jet tends to close on itself. This equation
will be used in the following to study the time dependence of the jet
opening angle. We now concentrate on the parameter $K$. The missing
piece of information to derive it is the cocoon pressure.

To compute the cocoon pressure we develop the approximations of
Begelman \& Cioffi (1989) and the jet propagation description of
Matzner (2003). The cocoon evolution is governed by the first
principle of thermodynamics which, for a relativistic temperature
cocoon, reads
\begin{equation}
d\rho_{\rm{cocoon}}=\frac{dQ-4/3\rho_{\rm{cocoon}}dV_{\rm{cocoon}}}
{V_{\rm{cocoon}}},
\label{eq:thermo}
\end{equation}
where the evolution of the cocoon volume is computed as
\begin{equation}
\frac{dV_{\rm{cocoon}}}{dt} = 
2\pi\,\int_{z_0}^{z_h}r_{\rm{cocoon},\perp}\,v_{\rm{sh}}\,dz.
\label{eq:vol}
\end{equation}
Here $v_{\rm{sh}}=\sqrt{\rho_{\rm{cocoon}}/3\rho_\star}$ is the cocoon
expansion velocity (Begelman \& Cioffi 1989; $\rho_\star$ is the
matter density of the star), $r_{\rm{cocoon},\perp}$ is the transverse
size of the cocoon, $z_0$ is the location of the base of the jet and
$z_h$ is the position of the head of the jet.

The energy input into the cocoon, $dQ$, is calculated differently in
the two phases: when the whole jet is inside the star (the cocoon is
bounded) and when the jet has broken out (the cocoon is unbounded). In
the first phase, the energy input is from the dissipation of the jet
energy. In the second phase, energy is lost through a channel at the
stellar surface, that we assume to have a surface area equal to half
the jet cross section\footnote{The factor of two is calibrated through
simulations. Note that this approximation is different from that of
LB05, who assumed a constant aperture of the cocoon.}. We obtain
\begin{equation}
dQ=\left\{
\begin{array}{cc}
L_j\,(1-\beta_h) & {\rm jet~in~star}\\
-\rho_{\rm{cocoon}}\,\Sigma_j/2\,c_s\,dt & {\rm jet~outside~the~star}
\end{array}\right.,
\label{eq:dq}
\end{equation}
where $\beta_h$ is the speed of the head of the jet in units of $c$.

Equations~\ref{eq:pjet}, \ref{eq:bal}, \ref{eq:thermo}, \ref{eq:vol},
and~\ref{eq:dq} constitute a solvable system of equations.  To check
the validity of the assumptions we have computed the solution of the
equations and run simulations for the same progenitor and engine
parameters.

Figure~\ref{fig:jet} shows a comparison of the simulation results
(model 10g5) with the semi-analytic predictions. During the confined
jet phase, before the breakout, the motion of the head of the jet is
reproduced with reasonable accuracy. The energy stored in the cocoon
at the time of breakout and the energy given to the stellar material
(potentially powering a supernova explosion) are accurate within
$10\%$, while the breakout time is reproduced within 20\%. Comparison
with other simulations (16TIg5, t10g2 and t5g2) show that the
semi-analytic results are accurate to within $20\%$ with respect to
the numerical treatment. The analytic treatment tends to be more
accurate for jets injected with higher values of $\gamma\theta_0$,
i.e., jets that have lost causal contact. This is not surprising since
our analytic treatment does not account for the spreading of the jet
due to internal motions, which is relevant in the
$\gamma<\theta_0^{-1}$ case.
 
\section{Afterglows}

\begin{figure}
\epsscale{1.0}
%\plotone{pre_angle_16TIg5.eps}
\plotone{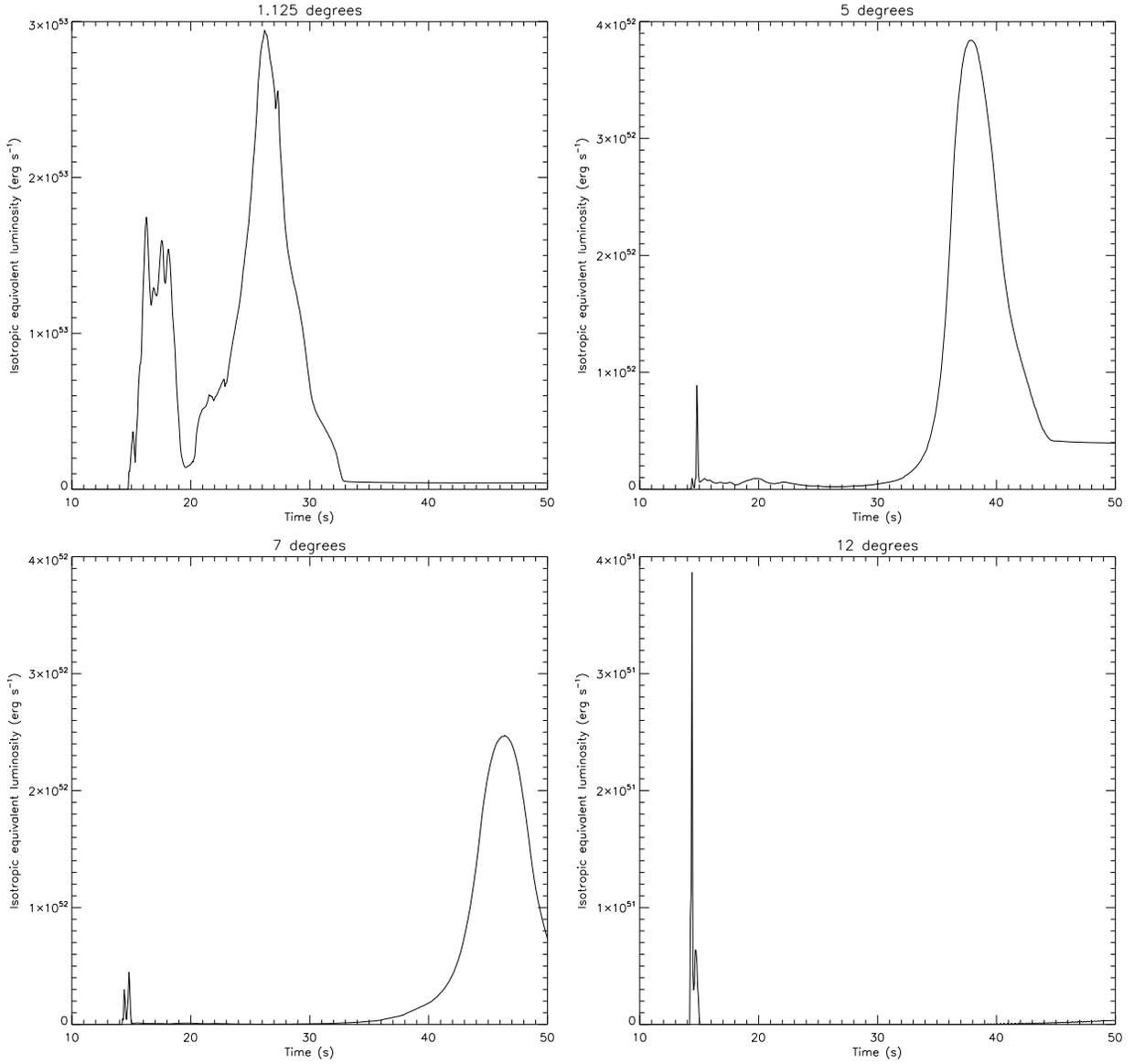}
\caption{\label{pre_angle_16TIg5} Energy flux vs. time for model
 t10g5 at 4 different angles.  The energy flux is for material with a
 minimum terminal Lorentz factor of $\gamma_{\infty} = 10$.  Angles
 represented are $1.125\degr$ (upper left), $5\degr$ (upper right),
 $7\degr$ (lower left), and $12\degr$ (lower right).}
\end{figure}

\begin{figure}
\epsscale{1.0}
%\plottwo{pre_energy_angle_16TIg5.eps}{pre_energy_angle_16TIg2.eps}
%\plottwo{pre_energy_angle_t10g5.eps}{pre_energy_angle_t10g2.eps}
%\plottwo{pre_energy_angle_t5g2.eps}{}

\plottwo{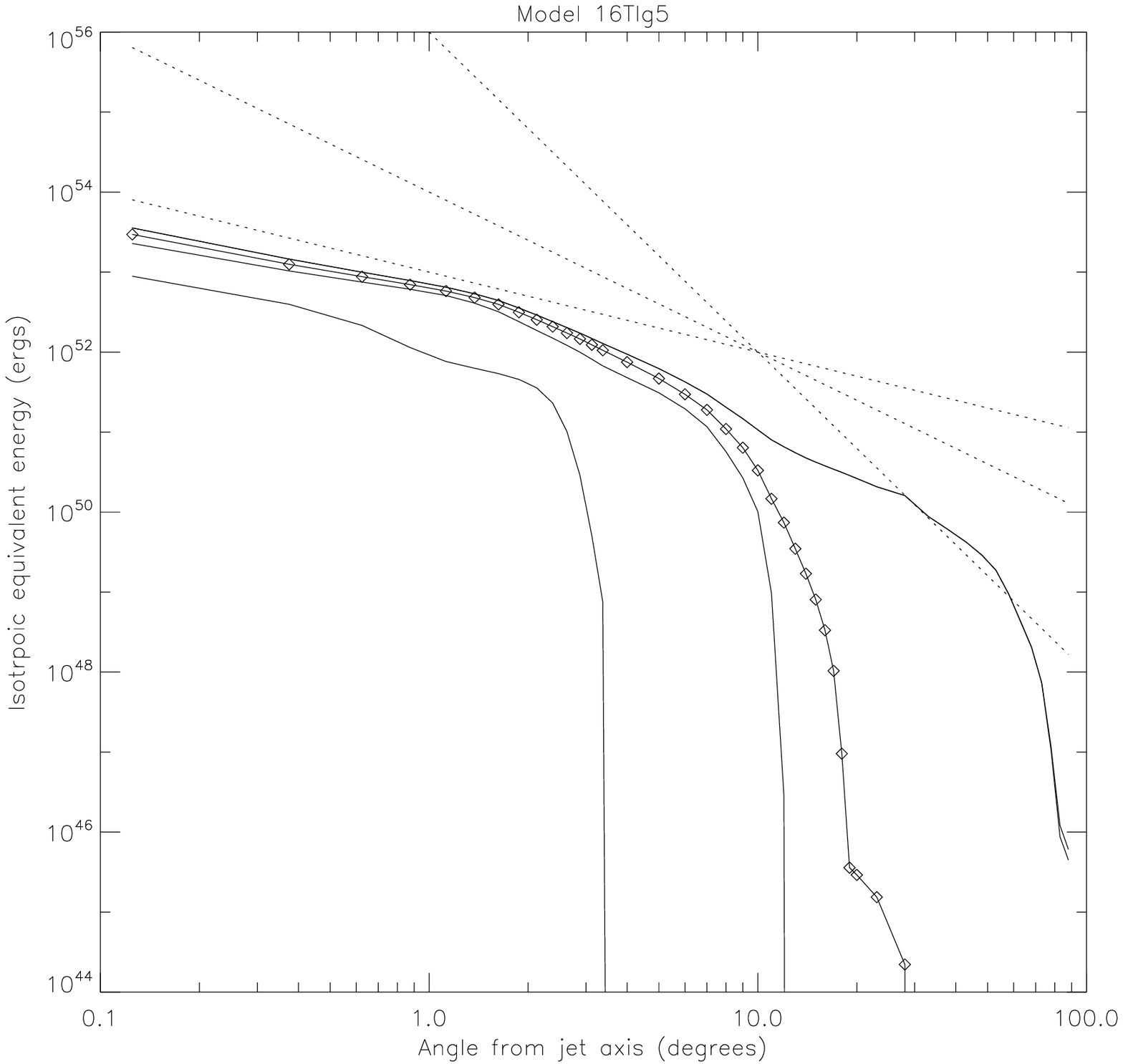}{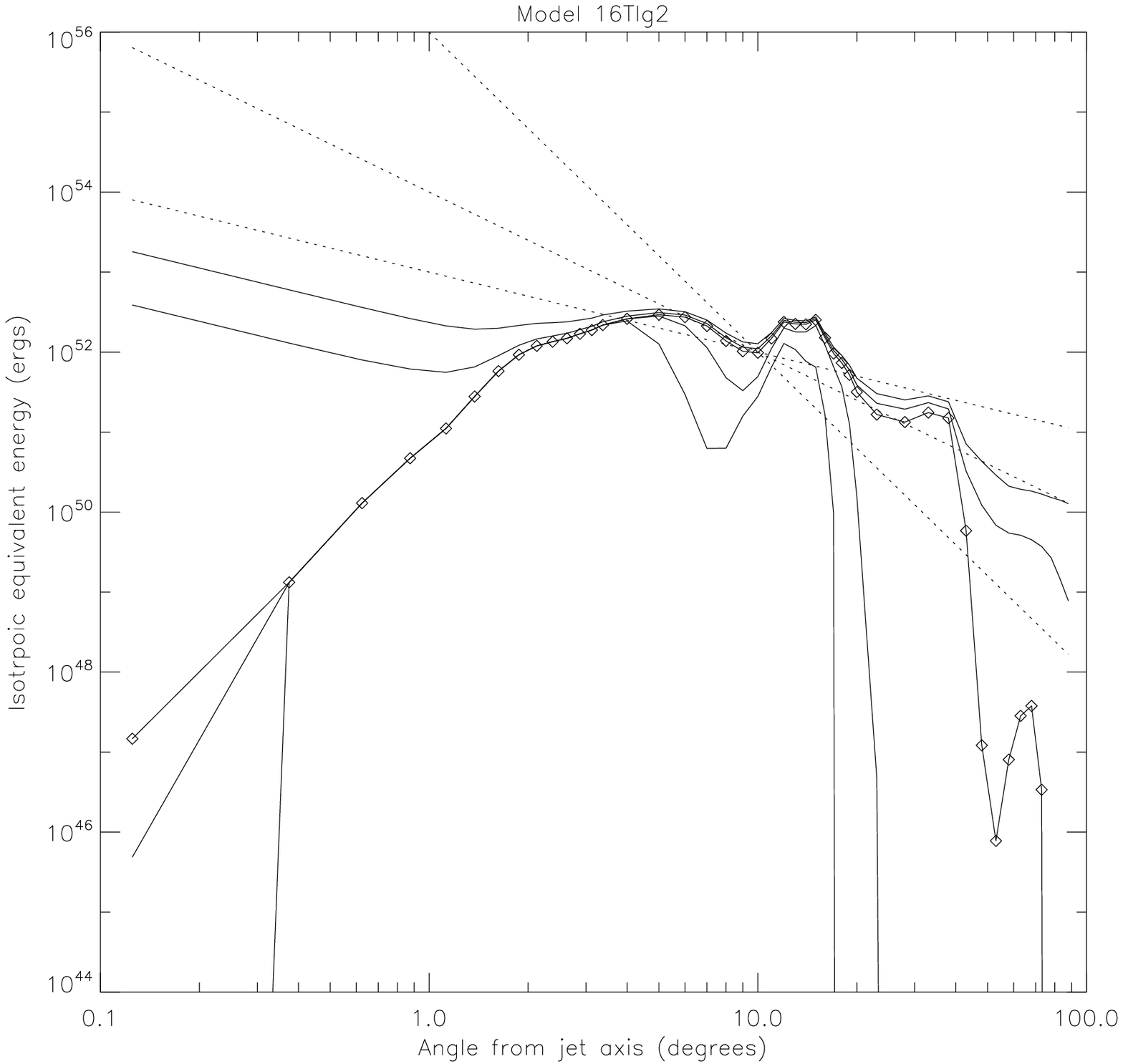}
\plottwo{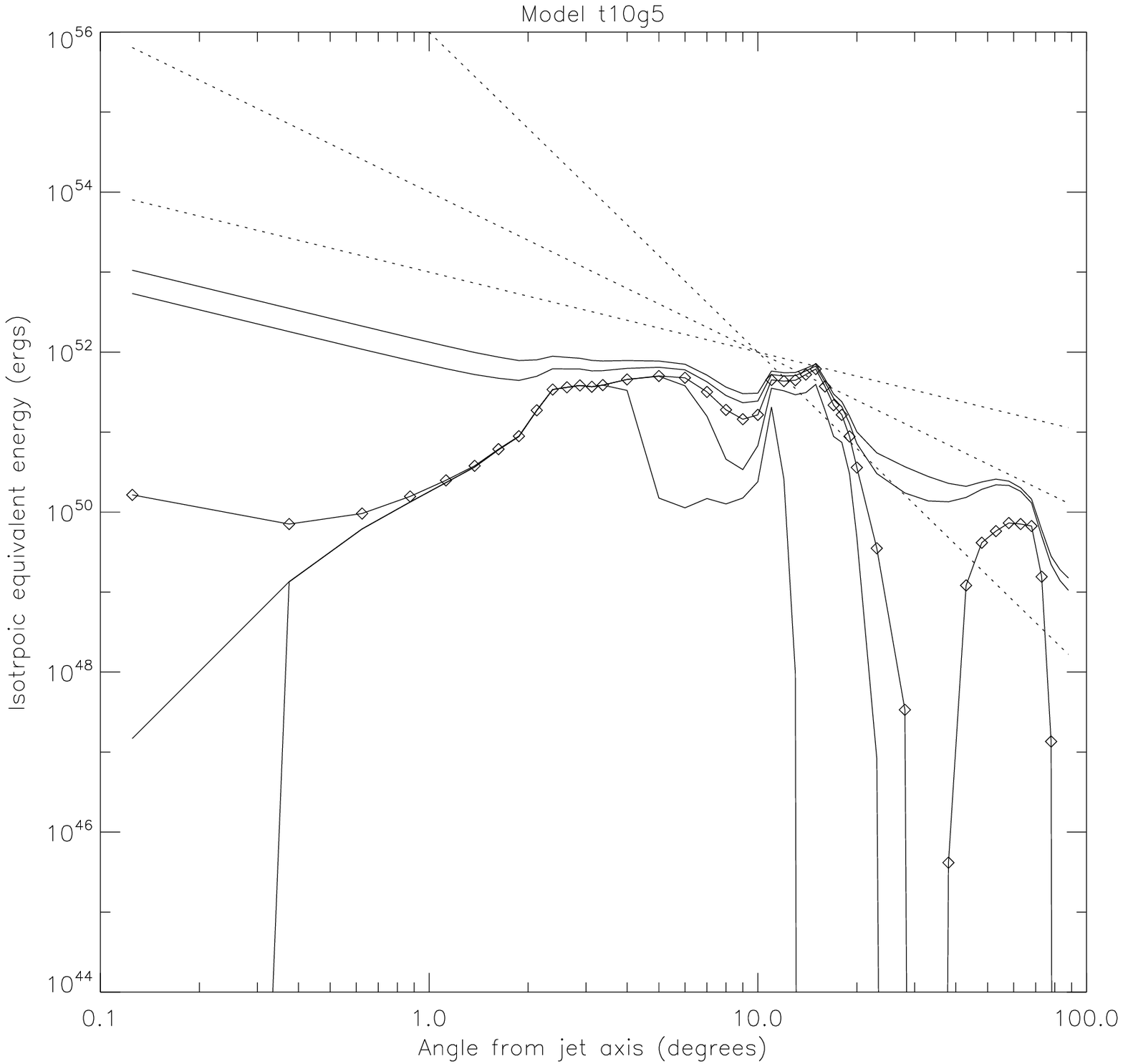}{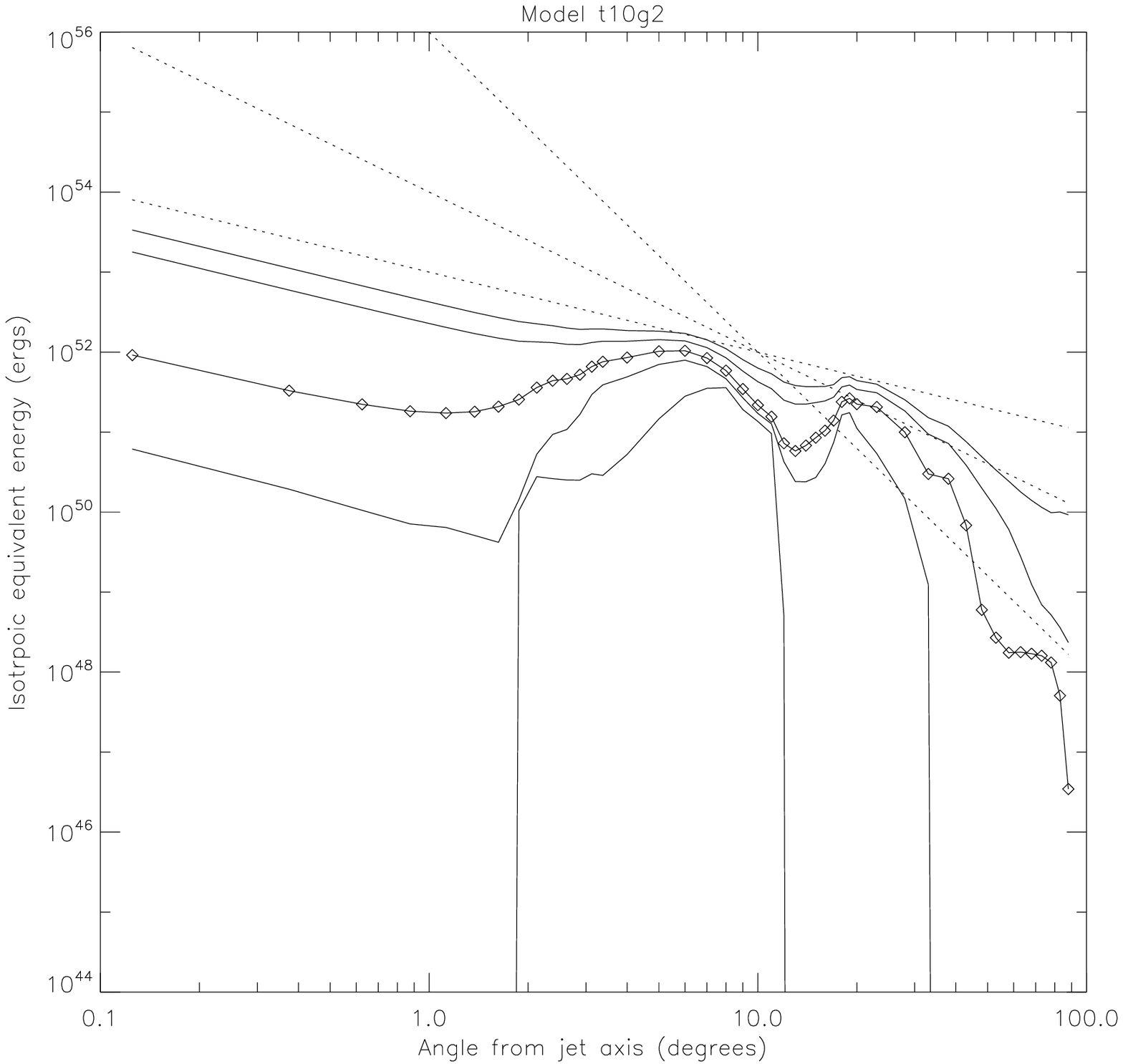}
%\plottwo{pre_energy_angle_t5g2.eps}{}
\caption{\label{pre_energy_angle_16TIg5} Isotropic equivalent energy
 as a function of angle from precursor phase for models 16TIg5 (upper
 left), 16TIg2 (upper right), t10g5 (lower left), and t10g2 (lower
 right).  Different lines correspond to the amount of energy above a
 minimum Lorentz factor of $\gamma_{\infty} = 1.01$, $2$, $10$, $50$,
 and $200$, from top to bottom.  Highlighted line corresponds to
 $\gamma_{\infty} = 10$.  Dotted lines are slopes of $E_{iso} \propto
 \theta^{-1}$, $\theta^{-2}$, and $\theta^{-4}$.}
\end{figure}

\begin{figure}
\epsscale{1.0}
%\plotone{energy_time_angle_16TIg5lowres.eps}
\plotone{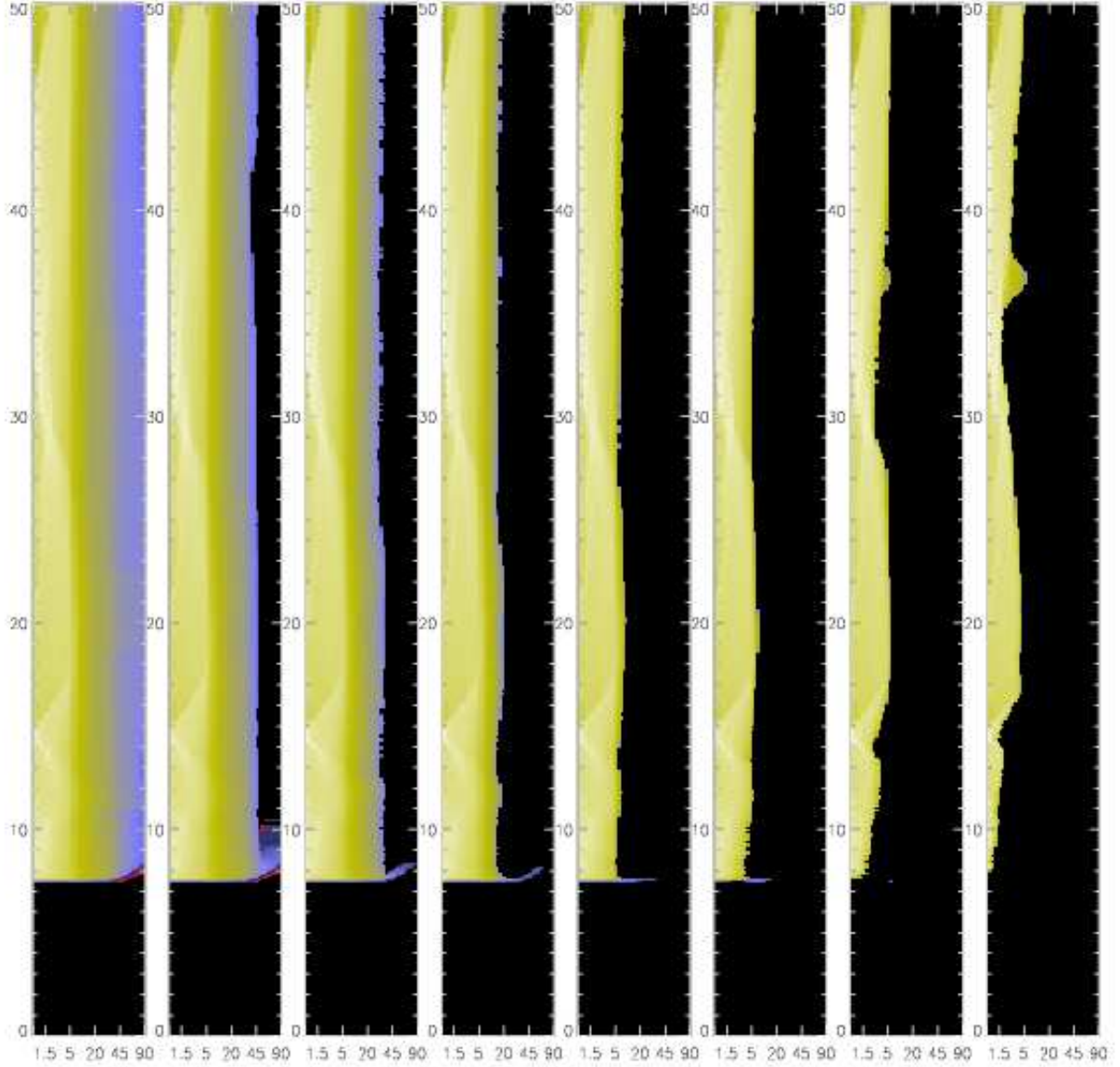}
\caption{\label{energy_time_angle_16TIg5lowres} Time and angular
distribution of (logarithmic) energy from low resolution version of
model 16TIg5 measured at
$1.2 \times 10^{11}$~cm.  Y-axis is time in seconds and x-axis is
angle from jet axis in degrees.  The x-axis is not uniform but instead
corresponds to the angular bins described in
\S~\ref{angulardistributionofenergy}.  The panels correspond to the
amount of energy above a minimum Lorentz factor of $\gamma_{\infty} =
1.01$, $2$, $5$, $10$, $30$, $50$, $100$, and $200$, from left to
right.}
%Same as
%figure~\ref{energy_time_angle_16TIg5} but for the low resolution run
%of model TI16g5.}
% Time and angular
%distribution of (logarithmic) energy from the low resolution version
%of model 16TIg5 measured at $2.4 \times 10^{11}$~cm.  Y-axis is time in
%seconds and x-axis is angle from jet axis in degrees.  The x-axis is
%not uniform but instead corresponds to the angular bins described in
%section \ref{angulardistributionofenergy}.  The panels correspond to
%the amount of energy above a minimum Lorentz factor of
%$\gamma_{\infty} = 1.01$, $2$, $5$, $10$, $30$, $50$, $100$, and
%$200$, from left to right.}
\end{figure}

\begin{figure}
\epsscale{1.0}
%\plottwo{total_energy_angle_16TIg5.eps}{total_energy_angle_16TIg5lowres.eps}
\plottwo{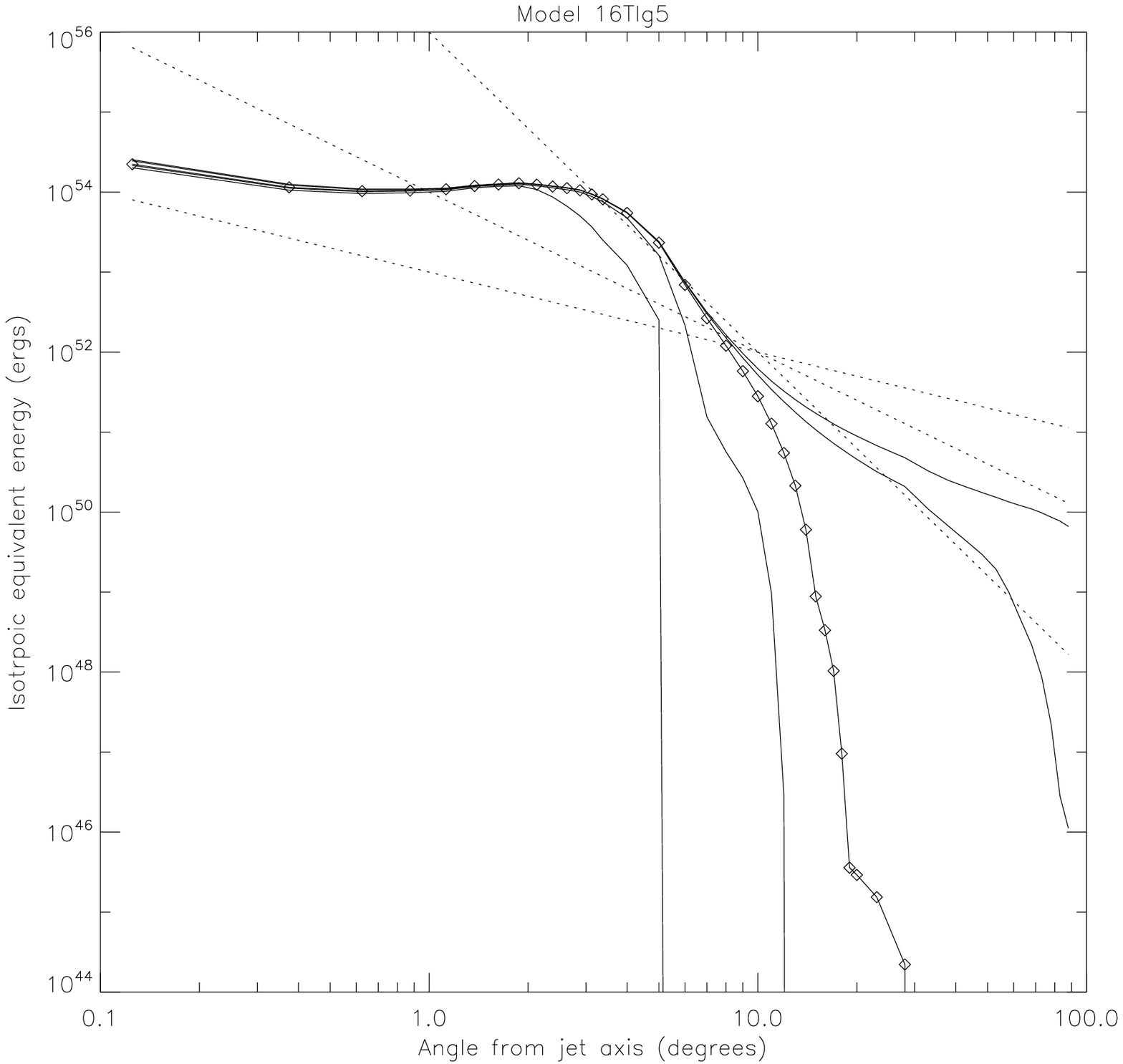}{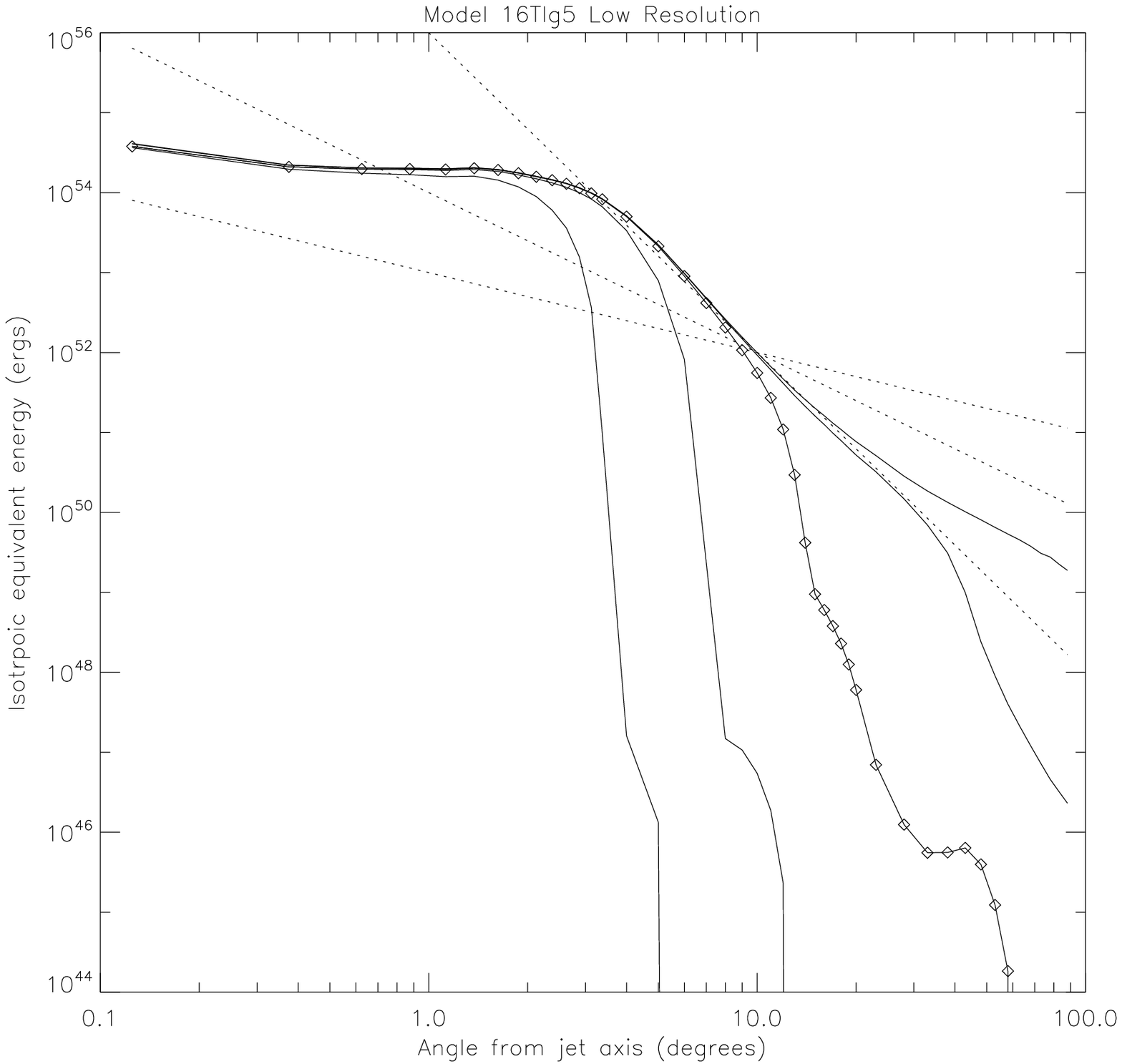}
\caption{\label{total_energy_angle_16TIg5lowres} Total isotropic
equivalent energy as a function of angle for high resolution (left)
and low resolution (right) versions of model 16TIg5.  Different lines
correspond to the amount of energy above a minimum Lorentz factor of
$\gamma_{\infty} = 1.01$, $2$, $10$, $50$, and $200$, from top to
bottom.  Highlighted line corresponds to $\gamma_{\infty} = 10$.
Dotted lines are slopes of $E_{iso} \propto \theta^{-1}$,
$\theta^{-2}$, and $\theta^{-4}$.}
\end{figure}

\begin{figure}
\epsscale{1.0}
%\plottwo{opening_angle_16TIg5.eps}{opening_angle_16TIg5lowres.eps}
\plottwo{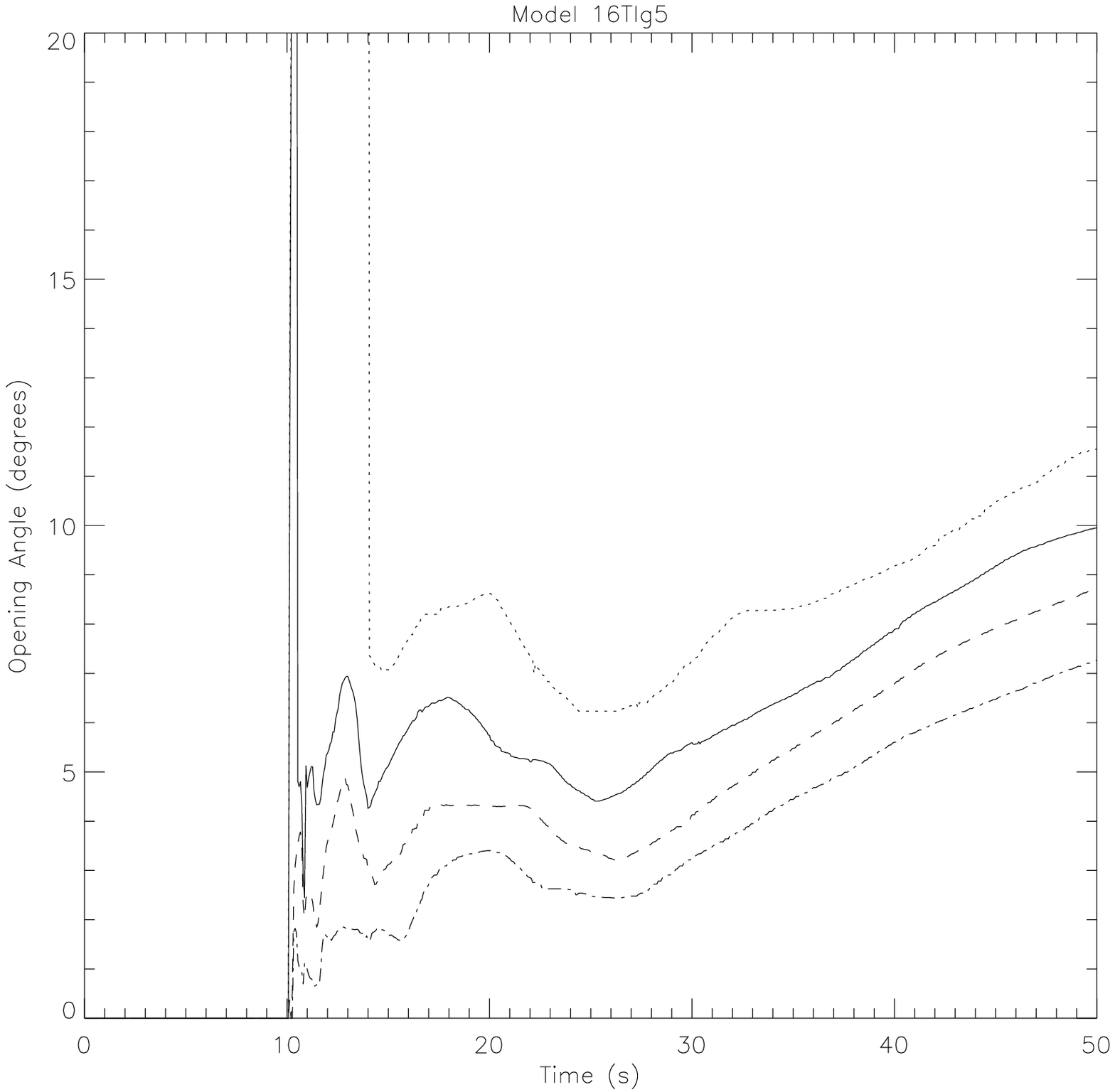}{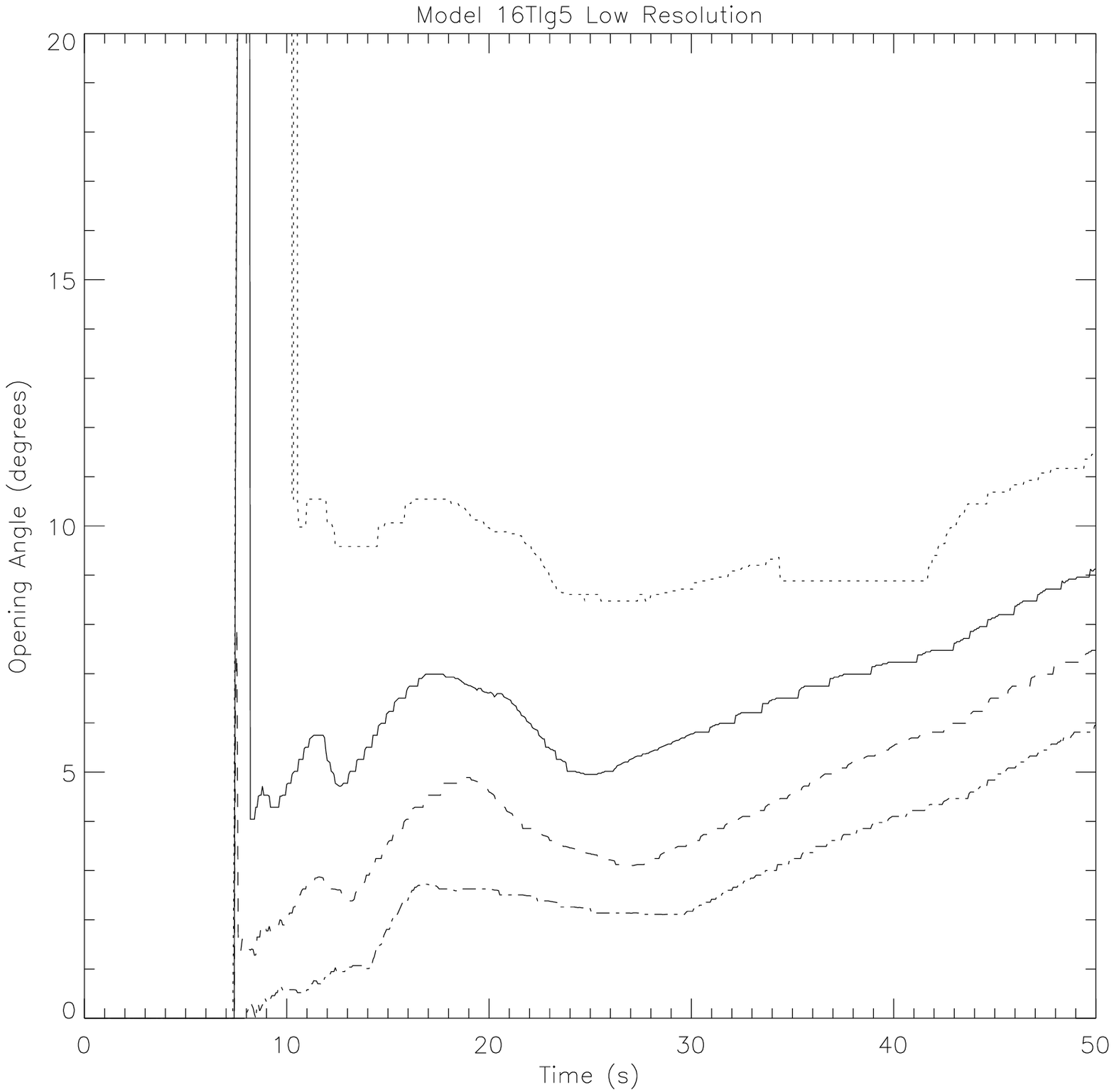}
\caption{\label{opeining_angle_16TIg5lowres} Opening angle vs. time
for high resolution (left) and low resolution (right) versions of
model 16TIg5.  Different lines correspond to opening angle for
material with a minimum terminal Lorentz factor of $\gamma_{\infty} =
2$ (dotted line), $10$ (solid line), $50$ (dashed line), and $200$
(dot-dash line).  The initially large opening angle is due to the
passage of the precursor material.}
\end{figure}

\begin{figure}
\epsscale{1.0}
%\plotone{sketch.ps}
\plotone{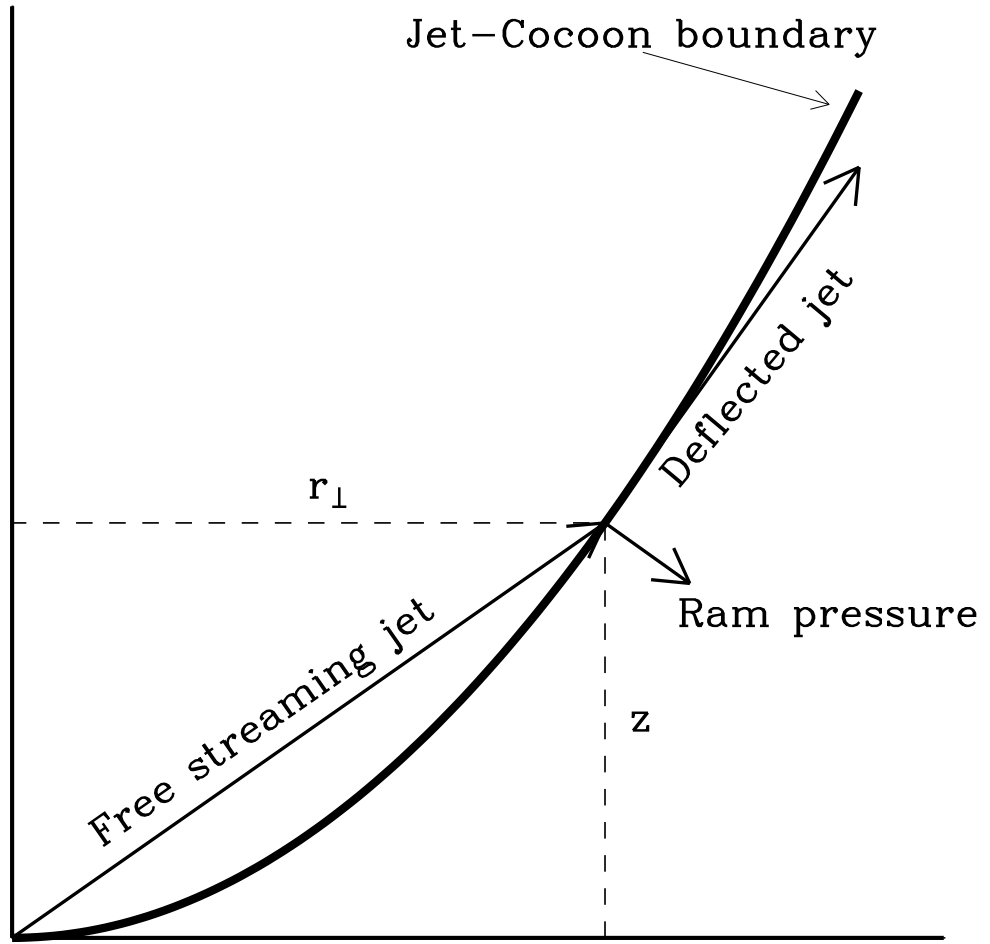}
\caption{Cartoon of the geometry used in the analytic approximations.
\label{fig:sketch}}
\end{figure}

\begin{figure}
\epsscale{1.0}
%\plotone{jet.ps}
\plotone{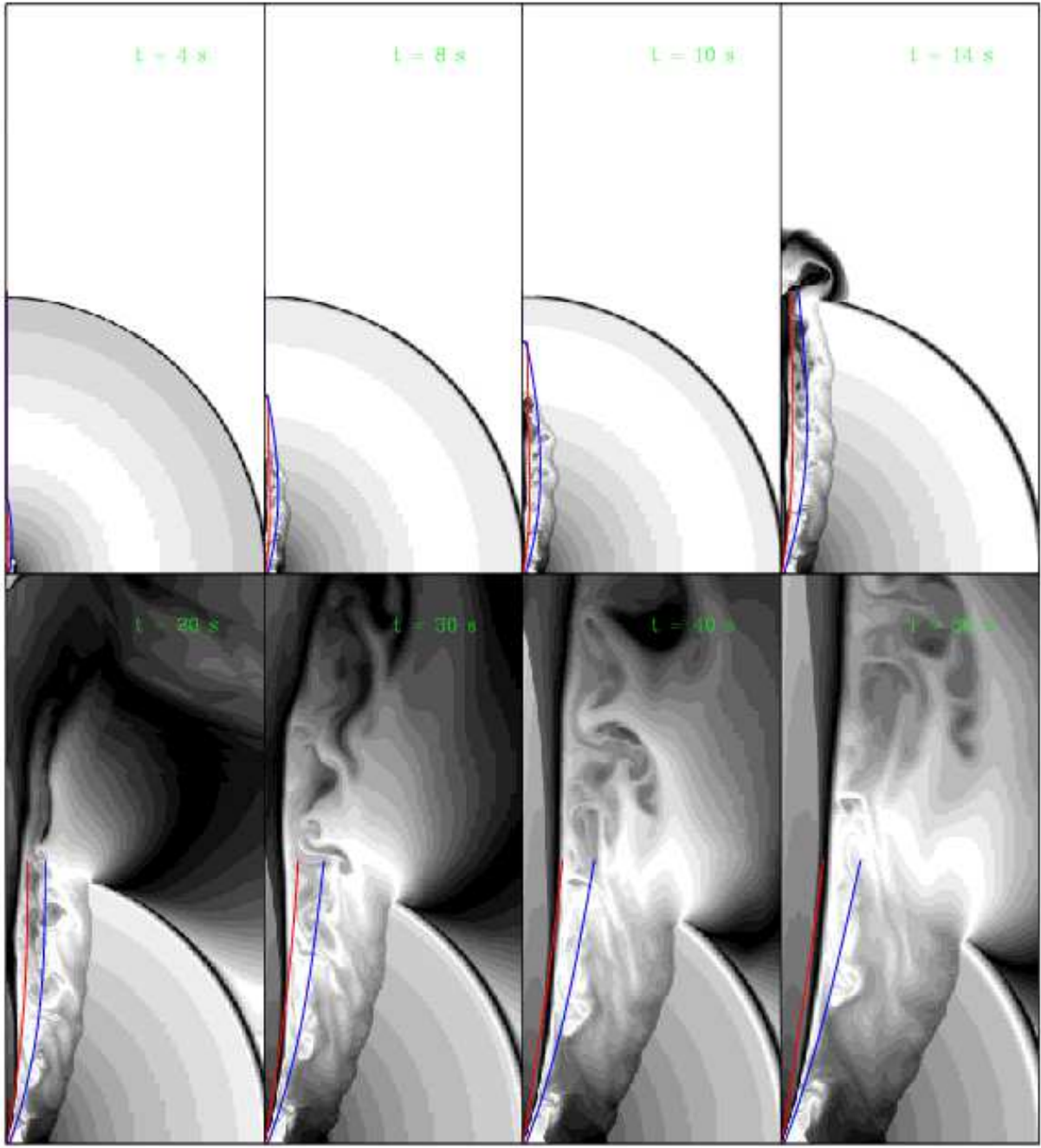}
\caption{Comparison between the analytic and numerical results. The
red (inner) line shows the jet boundary while the blue (outer) line
shows the cocoon boundary. The simulations stills are from model t10g5.
\label{fig:jet}}
\end{figure}

Besides light curves of the prompt emission
(Fig.~\ref{pre_angle_16TIg5}), we can compute afterglow light curves
based on the energy distributions obtained from these simulations. To
this aim, we adopt the afterglow code of Rossi et al. (2004). We input
into the code the energies from model 16TIg5 and t10g5 with a
lower-limit on the Lorentz factor $\gamma_\infty=5$ and we assume a
constant density of the ISM with $n=10$~cm$^{-3}$. The flow is
supposed to propagate from the low radii of our outer boundary to the
external shock radius without re-adjusting the energy distribution and
no sideways expansion of the external shock is assumed. We also
simplify the computation by assuming a Lorentz factor $\gamma=400$ for
the whole fireball. This simplification is due to the fact that it is
not possible to input, for a given angle, material with different
Lorentz factors, as obtained in the simulations. The effect of this
approximation should be to slightly modify the shape of the light
curve around the peak of the afterglow emission.

\begin{figure}
\epsscale{1.0}
%\plottwo{afte1.ps}{afte2.ps}
\plottwo{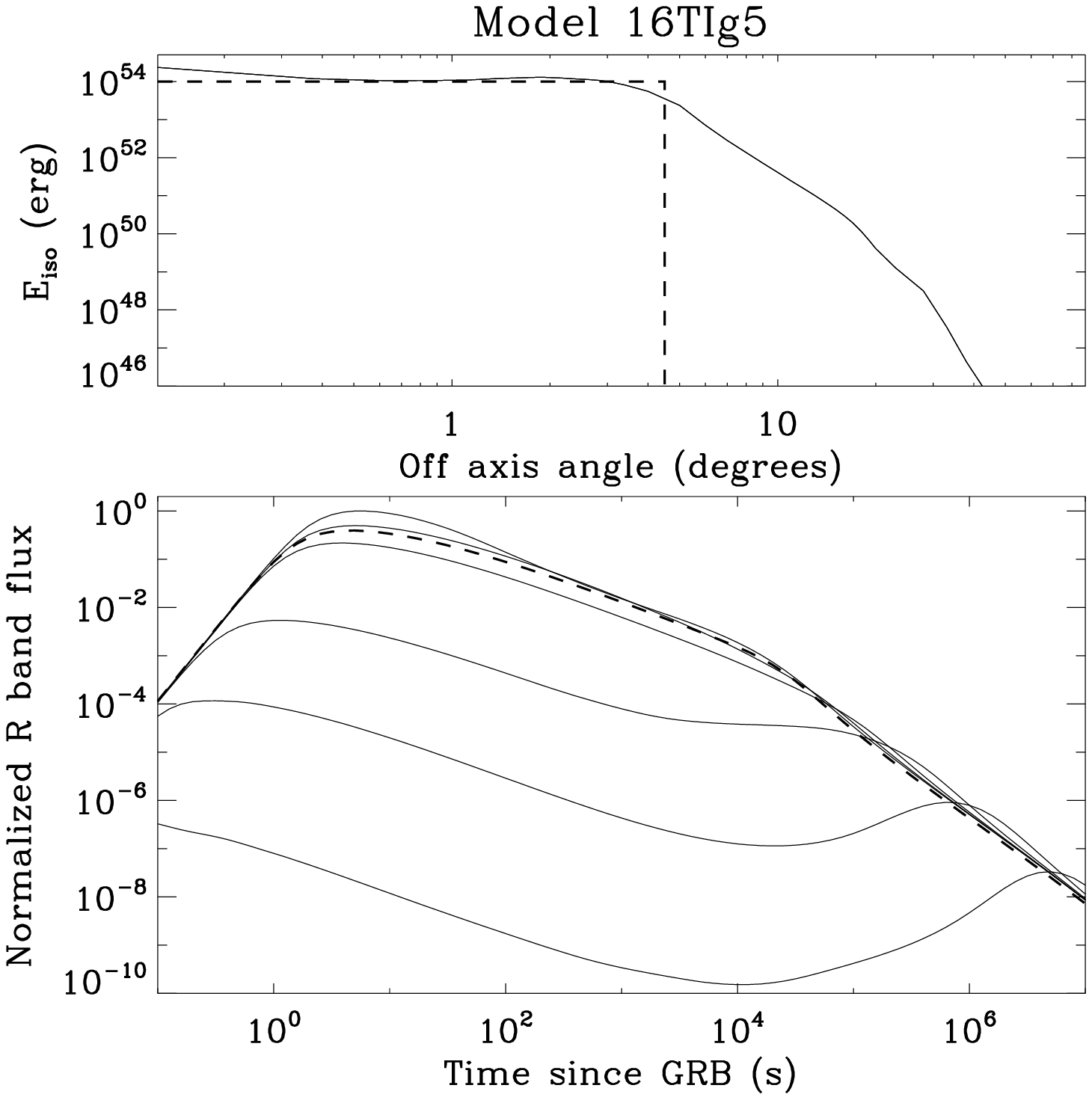}{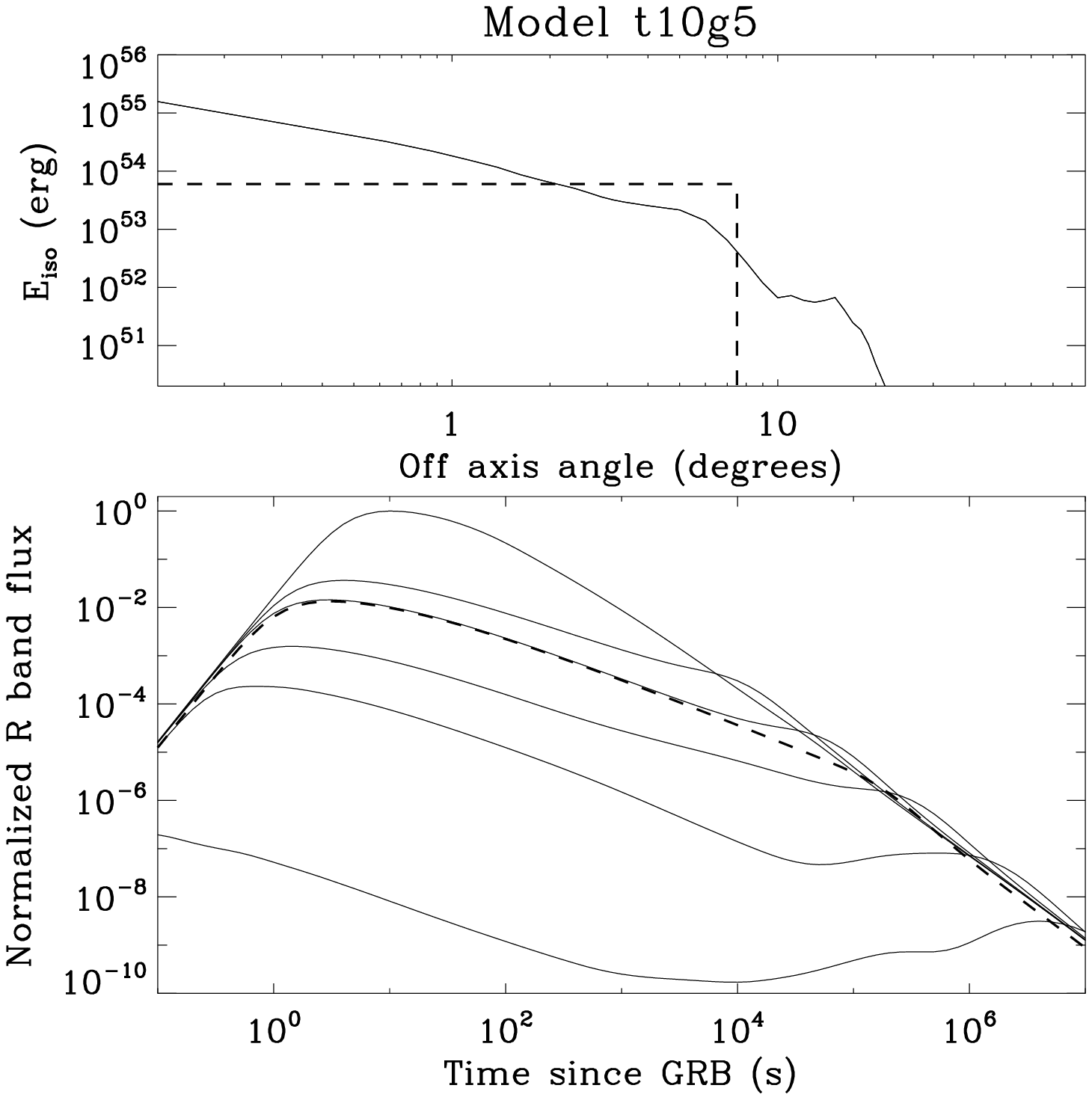}
\caption{\label{fig:after} Afterglow light curves in the R band for the
energy distribution obtained in model 16TIg5 (left panel) and t10g5
(right panel). Each panel shows the energy distribution in the top and
the afterglow light curves on the bottom. Viewing angles $\theta_0=0$,
2, 4, 8, 16, and 32 degrees off-axis are shown, from top to bottom. A
dashed curve shows, for comparison purposes, the afterglow of a flat
energy distribution with an opening angle of $4\degr$ and $7\degr$ for
model 16TIg5 and t10g5, respectively. All light curves have been
normalized to the maximum of the on-axis light curve.}
\end{figure}

Results of the afterglow calculation are shown in
Fig.~\ref{fig:after}. Light curves have been computed for different
observing off-axis angles $\theta_0=0$, 2, 4, 8, 16, and 32 degrees
and for two simulations: model 16TIg5 and model t10g5. The jet
properties are the same in the two models, but the progenitor stars
are different. In each panel we plot also the afterglow from a
standard top-hat jet for comparison. The behavior of the two
simulations is quite different. Model 16TI has a very flat energy
distribution in its center. For this reason, the inner light curves
resemble very well those of a top hat jet. Only the on-axis light
curve deviates due to its high energy. While the slopes are different
from those of the top-hat afterglow, the spectra are the same. This
implies that the use of the so-called afterglow closure
relations\footnote{Closure relations are simple equations that
associate a temporal decay of the afterglow to a given spectral slope
and to a given distribution of the ambient medium into which the
external shock runs.}  (Price et al. 2002) cannot be blindly applied
to afterglow with the angular energy distributions derived from these
simulations.

Owing to its more centrally condensed distribution, model t10g5 has
afterglows for small off-axis angles that differ from each other and
from the top hat example. Qualitatively, the curves are however
similar to one another, with an early shallow decay followed by a
sharper decay when the jet reaches causal contact. In both models 16TI
and t10g5, curves at large off-axis angles show a plateau or even a
bump several hours to weeks after the explosion. This is due to the
fact that the radiation from the brighter jet core enters into the
line of sight at late times. Comparison of the two lower panels of
Fig.~\ref{fig:after} teaches us how the progenitor structure is
important not only for the prompt GRB emission, but also for the
ensuing afterglow radiation.

\section{Summary and Conclusions}

We have presented high resolution 2D simulations of the propagation of
light relativistic jets inside the cores of massive stars. We use an
adaptive mesh code (FLASH) that allows us to study the behavior of the
jet-star interaction over a long timescale and a wide spatial
range. Different stellar progenitors as well as initial conditions of
the jet are explored. Thanks to the high resolution and large spatial
and temporal domain of our simulations we can confirm and study in
more detail jet features discussed in previous work (MacFadyen \&
Woosley 1999; Aloy 2000; MacFadyen et al. 2001; Zhang et al. 2003,
2004; Mizuta et al. 2006), as well as identify new phenomenologies.

The main conclusion of this paper is that, even if the central engine
is stationary, the jet that propagates out of the progenitor star is
characterized by three phases, all of which display significant
variability. The first phase is a wide angle release of mildly to
moderately relativistic material, which we call the precursor. This
phase is due to the release of the turbulent shocked material that
accumulates around the jet due to vortex shedding before it breaks out
of the star (Ramirez-Ruiz et al. 2002; Zhang et al. 2003). This
initial phase is followed by a shocked jet phase. In this second phase
the jet material that flows out of the star has been heavily
shocked. The energy flow is highly variable in this phase, and no
temporal trend in the properties can be identified. During this phase
the jet is most highly collimated. Finally, as the pressure of the
cocoon decreases, the jet settles into a stable configuration with a
freely expanding core surrounded by a shocked shear layer at the
boundary with the cocoon material. In this phase the energy flow is
almost constant, and the jet opening angle increases logarithmically
with time.

This temporal evolution of the jet is also associated with the angular
distribution of energy, and therefore determines what different
observers see from different directions. The precursor is
characterized by a wide opening angle and can be seen from most
directions. The shocked jet phase is the most concentrated and can be
seen only by observers within several degrees of the axis. As a
consequence, such observers will see a very bright event. Observers
who lie a few degrees outside the shocked jet cone will see the
precursor, followed by a dead time of several tens of
seconds. Eventually, when the jet opens to contain their line of
sight, they will see a second phase of emission. These observers will
therefore measure dead-times much longer than any timescale of the
inner engine, as found in several BATSE light curves (Lazzati 2005).

The overall angular distribution of energy is complex and does not
seem to follow any simple correlation with the jet or progenitor
properties. In some cases, the jet is characterized by a flat core
with a sudden cutoff, very similar to a top-hat jet (16TIg5), while in
other cases the distribution is more centrally peaked. It seems,
however, that our high resolution simulations were not able to
reproduce previous analytic or numerical results like the
$\theta^{-2}$ universal jet of LB05 or the $\theta^{-3}$ distribution
found by Zhang et al. (2003). Such behaviors are observed only in a
limited range of angles from the jet axis. We computed afterglow light
curves from our energy distributions showing that differences in the
afterglow phase can result from the different properties of the
progenitor and/or of the jet in the core of the star. We also
developed further the analytic model of LB05, refining some
approximations to obtain a model that can reproduce such basic
features of the simulations as the propagation of the bow shock inside
the star and the precursor energy, as well as the qualitative
evolution of the jet opening angle.

It is easy to find paths along which this work can be developed
further. Higher resolution and dimensionality will certainly be worth
exploring (Zhang et al. 2004), as well as the consequences of a more
accurate equation of state capable of describing both relativistic and
non-relativistic material. Including a non-relativistic wind from the
accretion disk is also worthwhile, as this seems to be a ubiquitous
outcome of jet-launching simulations (Proga et al. 2003; de Villiers
et al. 2006). This wind component may alter the cocoon properties,
which are so important in defining the behavior of the precursor and
shocked jet phases. Simulations also suggest that the jet luminosity
should be highly variable if not intermittent. Preliminary studies
(Aloy et al. 2000) suggest that this may enhance the propagation of
the jet through the star, but higher resolution simulations are
required. An interesting hypothesis is that the three phases of jet
propagation will respond differently to an intermittent engine, with
the precursor virtually unaffected while the unshocked jet should
retain most of the engine variability.

\bigskip

The software used in this work was in part developed by the
DOE-supported ASC/Alliance Center for Astrophysical Thermonuclear
Flashes at the University of Chicago.  We thank Alex Heger for
providing us with the tabulated properties of his stellar models,
Andrew MacFadyen for useful discussions and advice on the testing of
the relativistic FLASH code and Miguel Aloy for useful discussions.
This work was supported by NSF grant AST-0307502, NASA Astrophysical
Theory Grant NNG06GI06G, and Swift Guest Investigator Program
NNX06AB69G.

\clearpage
%Tables

\begin{deluxetable}{lllll}
\tablecaption{Model Parameters \label{modelparameters}}
\tablehead{
\colhead{Model} & \colhead{Stellar Model} & \colhead{Input Power (erg s$^{-1}$)} & \colhead{Injection Opening Angle} & \colhead{Injection Lorentz Factor} }

\startdata
16TIg5 & Realistic & $5.32 \times 10^{50}$ & $10\degr$  & $5$  \\
16TIg2 & Realistic & $5.32 \times 10^{50}$ & $10\degr$  & $2$  \\
t10g5  & Power Law & $5.32 \times 10^{50}$ & $10\degr$  & $5$  \\
t10g2  & Power Law & $5.32 \times 10^{50}$ & $10\degr$  & $2$  \\
t5g2   & Power Law & $5.32 \times 10^{50}$ & $5\degr$   & $2$  \\
t20g5  & Power Law & $5.32 \times 10^{50}$ & $20\degr$  & $5$  \\

\enddata

\end{deluxetable}

\begin{deluxetable}{llll}
\tablecaption{Breakout Times of Cocoon, Shocked Jet and Unshocked Jet
at Surface of Star \label{breakout}} 
\tablehead{ \colhead{Model} &
\colhead{Cocoon Breakout (s)} & \colhead{Shocked Jet Breakout (s)} &
\colhead{Unshocked Jet Breakout (s)}}

%\startdata
%16TIg5 & $5.87$   & $6.33$ & $13.6$ \\
%16TIg2 & $9.66$   & $12.0$ & $25.0$ \\
%t10g5  & $10.2$   & $10.7$ & $20.0$ \\ 
%t10g2  & $24.0$   & $31.0$ & $48.0$ \\ 
%t5g2   & $21.2$   & $31.8$ & $46.0$ \\
%16TIg5lowres & $5.66$   & $5.73$ & $13.8$ \\
%
%\enddata

\startdata
16TIg5 & $7.53$   & $7.73$ & $23.4$ \\
16TIg2 & $9.66$   & $12.0$ & $25.0$ \\
t10g5  & $13.1$   & $14.0$ & $29.0$ \\ 
t10g2  & $24.0$   & $31.0$ & $48.0$ \\ 
t5g2   & $21.2$   & $31.8$ & $46.0$ \\
16TIg5lowres & $4.93$   & $5.00$ & $23.9$ \\

\enddata

\end{deluxetable}

%\begin{deluxetable}{lll}
%\tablecaption{Start Times of Shocked and Unshocked Phases at
%$2.4\times10^{11}$~cm \label{shocktimes}}
%\tablehead{
%\colhead{Model} & \colhead{Shocked Phase Start (s)} & \colhead{Unshocked Phase Start (s)}}
%
%\startdata
%16TIg5  & $14.66$s  & $25.2$s \\
%16TIg2 & $33.87$s & $44.47$s \\
%t10g5  & $16.6$s   & $39.47$s \\ 
%t10g2  & $40.33$s  & \nodata \\ 
%t5g2  & $47.2$s  & \nodata  \\
%16TIg5  & $12.0$s  & $29.93$s \\
%\enddata
%
%\end{deluxetable}

\begin{deluxetable}{llllllll}
\tablecaption{Precursor Energetics. Columns report the model
identification, the energy emitted with $\gamma_\infty>10$, the angle
at which the energy is maximized and the width of $dE/d\Omega$ at one
tenth of its maximum, the energy emitted with $\gamma_\infty>50$, the
angle at which the energy is maximized and the width of $dE/d\Omega$
at one tenth of its maximum. The last column reports the ratio of the
energies for the two limiting Lorentz factors.
\label{precursorenergetics}}
\tablehead{
\colhead{Model} & \colhead{$E_{\gamma_\infty \ge 10}$ (erg)} & \colhead{$\theta_{\max}$} & \colhead{$\theta_\frac{1}{10}$} & \colhead{$E_{\gamma_\infty \ge 50}$ (erg)} & \colhead{$\theta_{\max}$} & \colhead{$\theta_\frac{1}{10}$} & \colhead{$E_{\gamma_\infty \ge 50}/E_{\gamma_\infty \ge 10}$}}

%\startdata
%16TIg5 &  $1.079\times10^{50}$ & $6\degr$ & $19\degr$ &  $3.424\times10^{49}$ & $6\degr$ & $3\degr$ & $0.317$ \\
%16TIg2 &  $5.213\times10^{50}$ & $15\degr$ & $34\degr$ &  $2.941\times10^{50}$ & $15\degr$ & $14\degr$ & $0.564$ \\
%t10g5  &  $9.318\times10^{49}$ & $9\degr$ & $31\degr$ &  $4.669\times10^{49}$ & $9\degr$ & $12\degr$ & $0.501$ \\
%t10g2  &  $2.384\times10^{50}$ & $23\degr$ & $34\degr$ &  $9.392\times10^{49}$ & $23\degr$ & $24\degr$ & $0.394$ \\
%t5g2   &  $5.758\times10^{50}$ & $38\degr$ & $43\degr$ &  $4.09\times10^{50}$ & $18\degr$ & $44\degr$ & $0.711$ \\
%
%\enddata

\startdata
16TIg5 &  $3.467\times10^{50}$ & $4\degr$  & $10\degr$ &  $2.470\times10^{50}$ & $4\degr$  & $9\degr$  & $0.712$ \\
16TIg2 &  $4.563\times10^{51}$ & $15\degr$ & $34\degr$ &  $2.613\times10^{51}$ & $15\degr$ & $14\degr$ & $0.573$ \\
t10g5  &  $7.948\times10^{50}$ & $15\degr$ & $54\degr$ &  $4.602\times10^{50}$ & $15\degr$ & $14\degr$ & $0.579$ \\
t10g2  &  $1.181\times10^{51}$ & $23\degr$ & $34\degr$ &  $5.281\times10^{50}$ & $23\degr$ & $24\degr$ & $0.447$ \\
t5g2   &  $2.560\times10^{51}$ & $7\degr$  & $39\degr$ &  $1.898\times10^{51}$ & $7\degr$  & $39\degr$ & $0.741$ \\

\enddata

\end{deluxetable}

\clearpage

%FIGURES

\clearpage
\thispagestyle{empty}

%\begin{figure}
%\epsscale{1.0}
%\plotone{opening_angle_fit_16TIg5.eps}
%\caption{\label{opening_angle_fit_16TIg5} Opening angle (vertical
%axis) vs. time (horizontal axis) for model 16TIg5.  Solid lines
%correspond to opening angle for material with a minimum terminal
%Lorentz factor of $\gamma_{\infty} = 10$.  Dotted line is a fit to the
%data in the unshocked phase of a function $\theta = a_0 + a_1 \ln
%(t)$.  In this case, $a_0 = -18.8$ and $a_1 = 7.90$.}
%\end{figure}

\clearpage

%\begin{figure}
%\epsscale{1.0}
%\plottwo{pre_region_contour_mod_16TIg5.eps}{pre_region_contour_mod_t5g2.eps}
%\caption{\label{energy_time_angle_16TIg5lowres} Time and angular
%distribution of (logarithmic) energy from the low resolution version
%of model 16TIg5 measured at $2.4 \times 10^{11}$~cm.  Y-axis is time
%in seconds and x-axis is angle from jet axis in degrees.  The x-axis
%is not uniform but instead corresponds to the angular bins described
%in section \ref{angulardistributionofenergy}.  The panels correspond
%to the amount of energy above a minimum Lorentz factor of
%$\gamma_{\infty} = 1.01$, $2$, $5$, $10$, $30$, $50$, $100$, and
%$200$, from left to right.}
%\end{figure}

\clearpage

%%%%%%%%%%%%%%%%%%%%
%Appendix

\appendix
\section{Code Testing \label{codetesting}}

In order to evaluate the accuracy of the FLASH relativistic
hydrodynamics module, we have carried out a series of tests and
compared the results to analytic solutions where possible. These
results have also been compared to the weighted essentially
non-oscillatory (WENO) scheme used by Zhang \& MacFadyen (2006) in
their Relativistic Adaptive Mesh code (RAM hereafter). All test
problems were run using FLASH's Riemann-type solver along with a
piecewise-parabolic reconstruction scheme, identical to that used for
the simulations presented in this paper. Although this scheme is
formally less accurate and more diffusive than that used in RAM, our
testing shows that the errors produced by FLASH are similar to those
produced by RAM.

\subsection{1D Riemann Problem 1}

\begin{figure}
\epsscale{1.0}
%\plotone{riemann1_400.eps}
\plotone{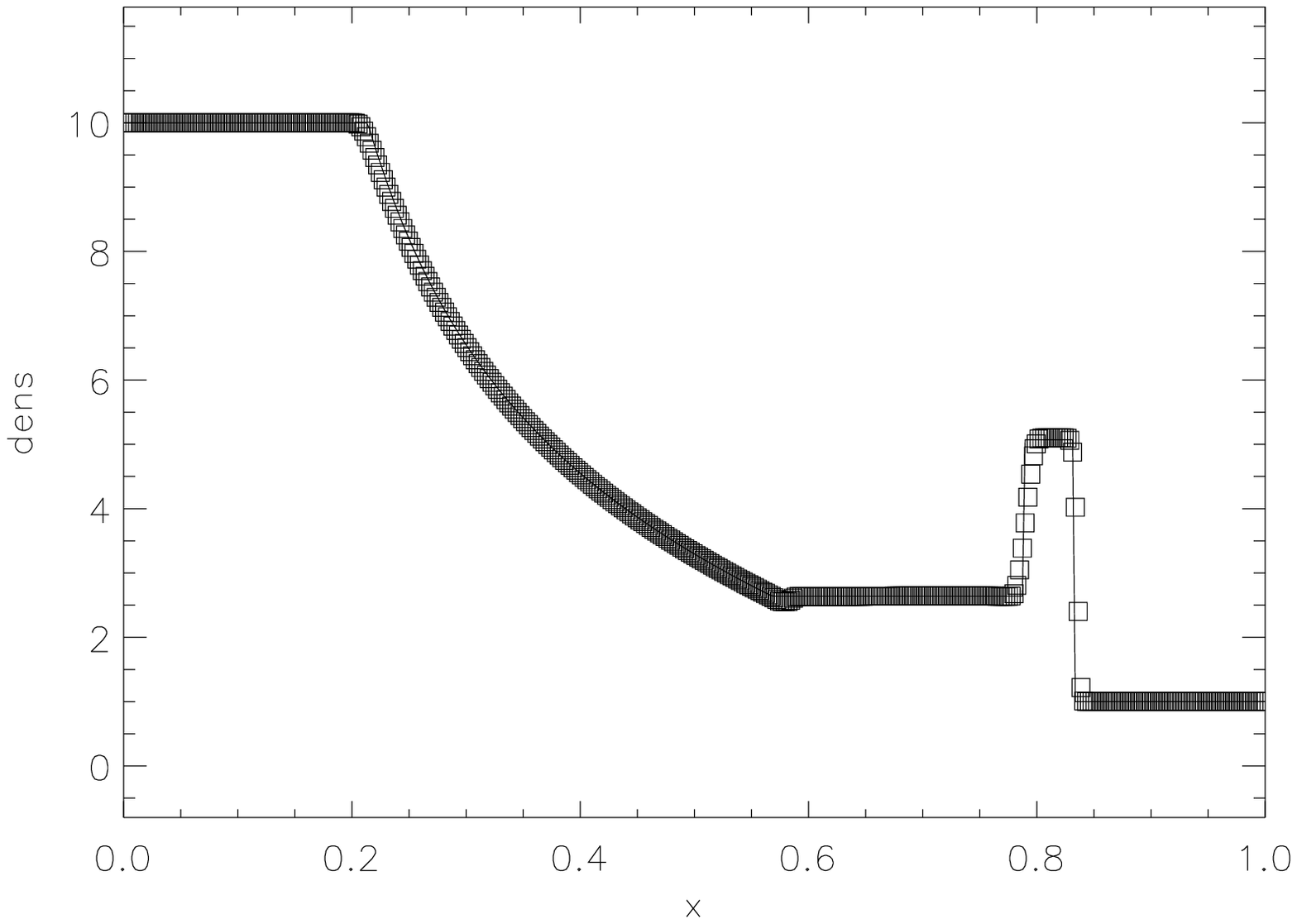}
\caption{\label{riemann1pic} 1D Riemann Problem 1.  Density for the
analytic solution (solid line) and numerical solution (squares) with a
resolution of 400 grid points is plotted at $t=0.4$. }
\end{figure}

For the 1D Riemann problem, there is a discontinuity between two
states at $x=0.5$ on a grid ranging from $x=0$ to $1$. The values of
pressure, density, perpendicular ($v_x$) and transverse ($v_y$)
velocity can be different for the left and right states. For the first
Riemann problem considered, the left state has a pressure of $13.33$
and density of $10$ while the right state has a pressure of $10^{-8}$
and density of $1$.  The $x$ and $y$ velocity in both states is
initially $0$.  For this problem an adiabatic index of $\Gamma=5/3$ 
and CFL number of $0.5$ are used. 
As time progresses, a mildly relativistic shock wave forms
traveling to the right and a rarefaction wave travels to the left.
The result at $t=0.4$ is then compared to the analytic solution to the
identical problem (Pons et al. 2000), and the L1 error calculated. The
L1 error is defined as $L1 = \Sigma _j \Delta x_j |u_j - u(x_j)|$ where
$x_j$ is the coordinate of grid point $j$ and $u(x_j)$ and $u_j$ are the
analytic and numerical values of proper density at grid point $j$ and $\Delta x_j$ is
the grid spacing. This is identical to the definition of L1 error in
RAM. Figure~\ref{riemann1pic} shows a comparison between the numerical
and analytic solutions at a resolution of 400 grid
points. Table~\ref{riemann1table} lists the L1 errors and convergence
rates for resolutions from 100 to 3200 grid points for this model, and
compares it with the error and convergence rate from RAM.  Convergence
rate is defines as $[\ln(L1_{i/2}/L1_i)]/[\ln{2}]$ where $L1_i$ is the
$L1$ error at resolution $i$.

For Riemann type problems, the L1 error is dominated by the few points
near the shock front.  This limits the convergence rate of any code to
about 1.  It also makes the L1 error sensitive to the particular slope
limiter being used.  Therefore, small differences in the L1 error
should not be taken to indicate a difference in the fundamental
accuracy achievable by the two codes.

\subsection{1D Riemann Problem 2}

\begin{figure}
\epsscale{1.0}
%\plotone{riemann2_400.eps}
\plotone{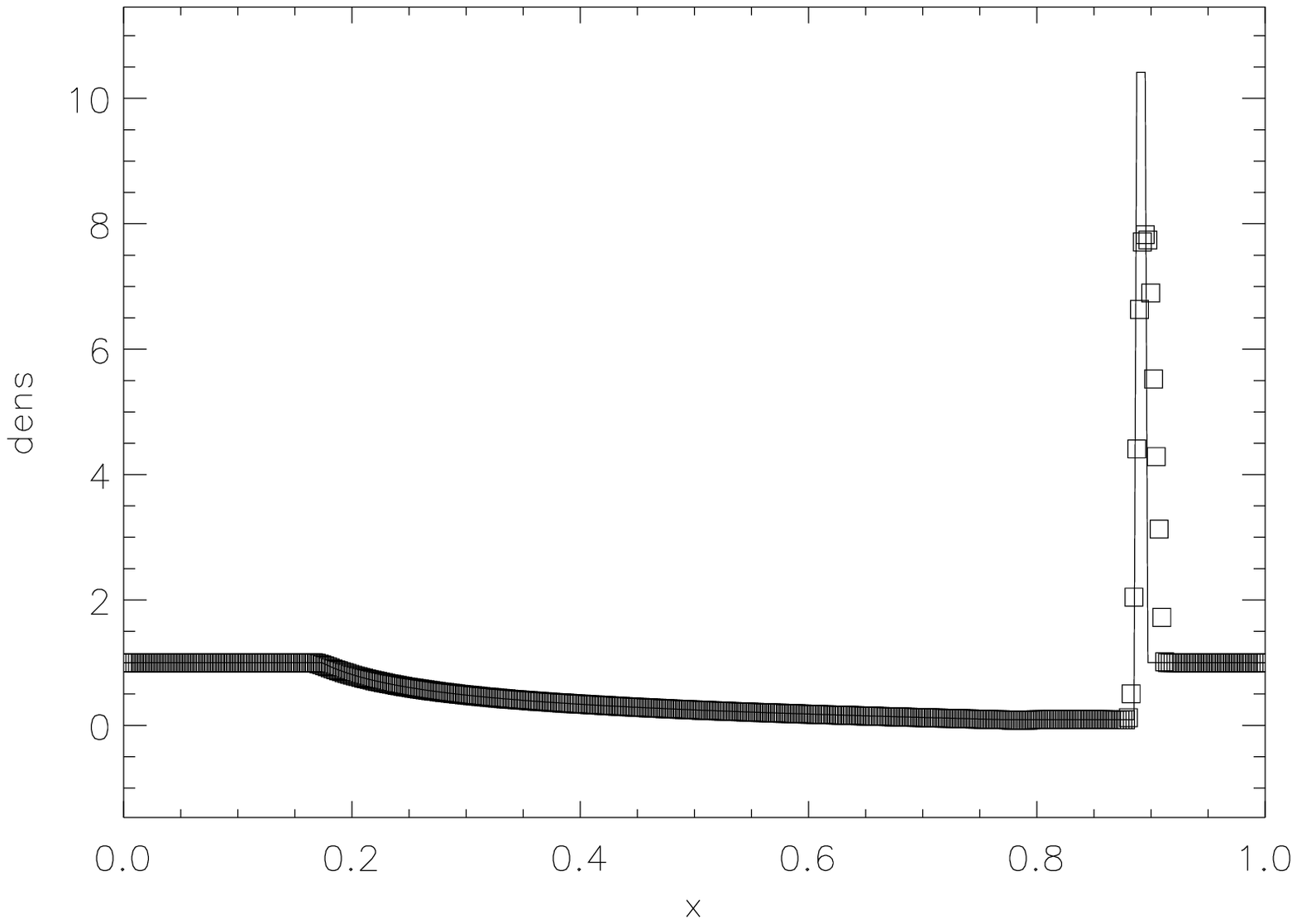}
\caption{\label{riemann2pic} 1D Riemann Problem 2.  Density for the
analytic solution (solid line) and numerical solution (squares) with a
resolution of 400 grid points is plotted at $t=0.4$. }
\end{figure}

For the second Riemann test, pressure is set to $1000$ in the left
state and $10^{-2}$ in the right state.  Density in both states in
initially $1$. The velocities in both states are again set to $0$ with
an adiabatic index of $5/3$ and CFL number of $0.5$.  
These conditions give rise to higher
velocities than Riemann problem 1 and a narrower shock.
Figure~\ref{riemann2pic} shows a comparison between the numerical and
analytic solutions at a resolution of 400 grid points at
$t=0.4$. Table~\ref{riemann2table} lists the L1 errors and convergence
rates for resolutions from 100 to 3200 grid points for this model, and
compares it with the error and convergence rate from RAM.

\subsection{1D Riemann Problem 3}

\begin{figure}
\epsscale{1.0}
%\plotone{riemann3_400.eps}
\plotone{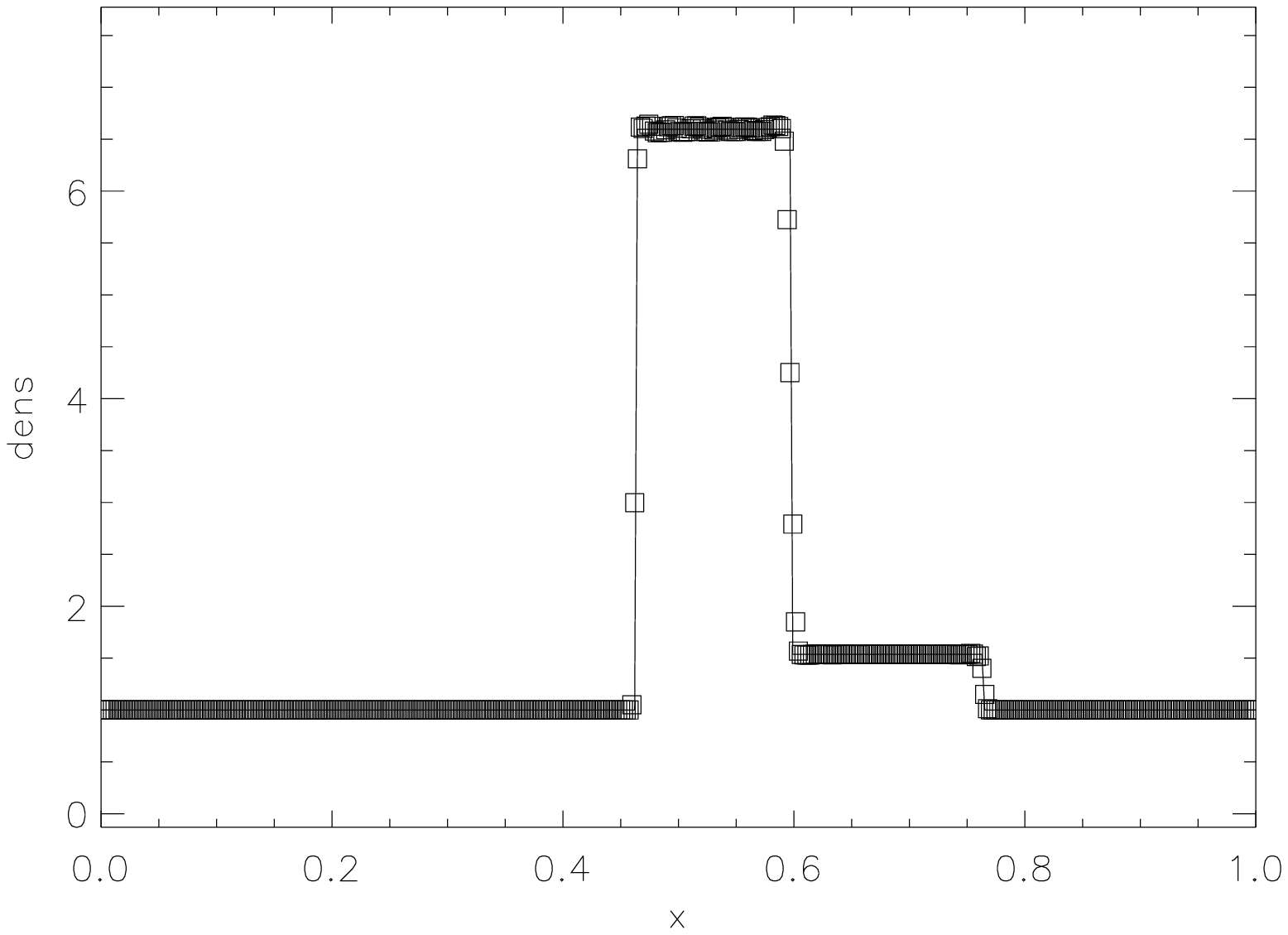}
\caption{\label{riemann3pic} 1D Riemann Problem 3.  Density for the
analytic solution (solid line) and numerical solution (squares) with a
resolution of 400 grid points is plotted at $t=0.4$. }
\end{figure}

For the third standard Riemann test problem, a situation is set up
in which a strong reverse shock develops.  Pressure is set to $1$ in
the left state and $10$ in the right state.  Density in both states in
initially $1$. Velocity in the $x$ direction is set to $0.9$ in the
left state and $0$ in the right state. Transverse velocity ($v_y$) is
$0$ in both states.  An ultra-relativistic adiabatic index
($\Gamma=4/3$) is used in this test along with a CFL number of $0.5$. 
Figure~\ref{riemann3pic} shows a
comparison between the numerical and analytic solutions at a
resolution of 400 grid points at $t=0.4$. Oscillations in the
post-shock density of $\sim1\%$ are visible. Table~\ref{riemann3table}
lists the L1 errors and convergence rates for resolutions from 100 to
3200 grid points for this model, and compares it with the error and
convergence rate from RAM.

\subsection{1D Riemann Problem 4: "Easy" Shear Velocity Test}

\begin{figure}
\epsscale{1.0}
%\plotone{riemann4_400.eps}
\plotone{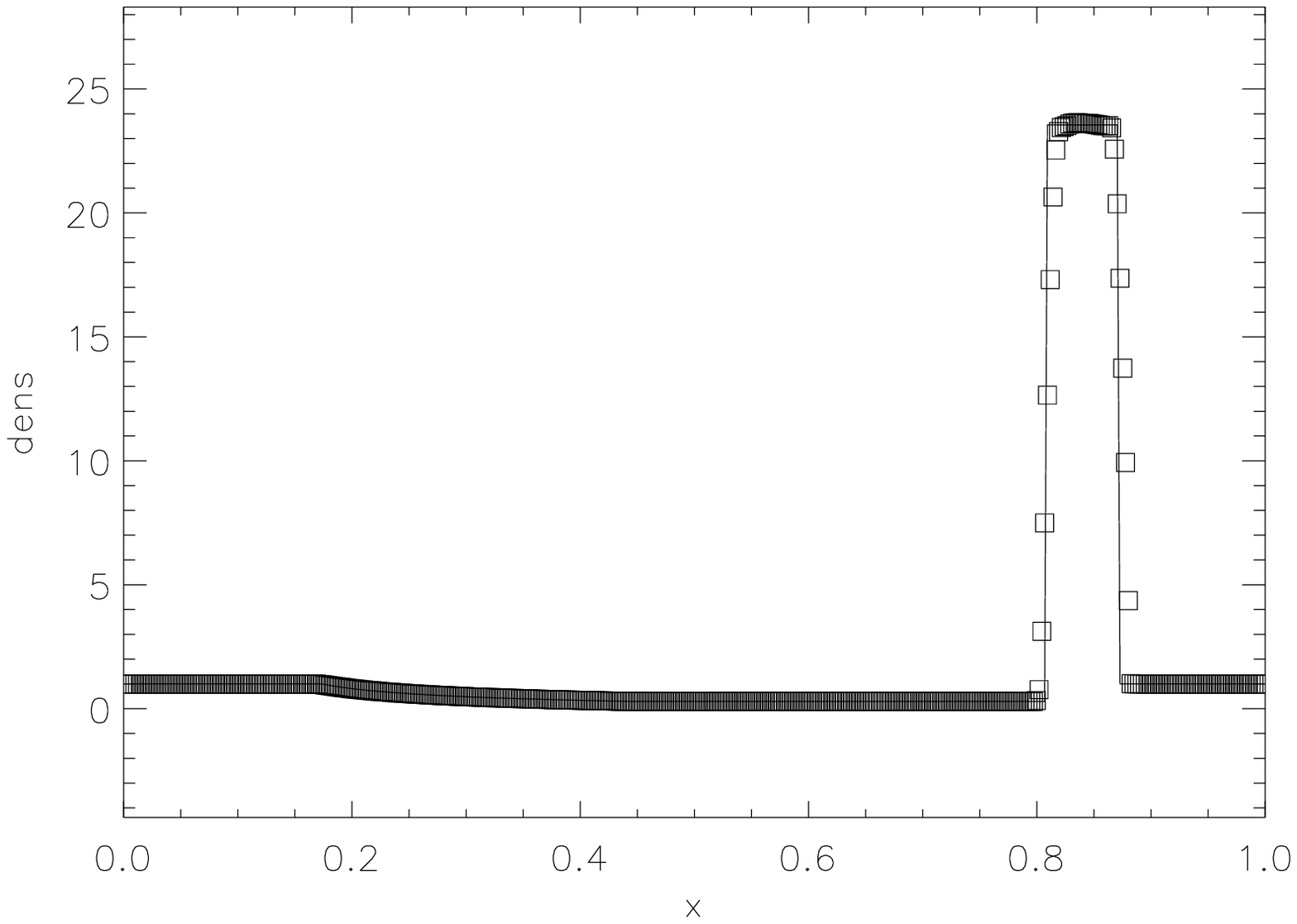}
\caption{\label{riemann4pic} 1D Riemann Problem 4: "Easy" Shear
Velocity Test.  Density for the analytic solution (solid line) and
numerical solution (squares) with a resolution of 400 grid points is
plotted at $t=0.4$. }
\end{figure}

Analytical solutions can also be obtained for a 1D Riemann problem
with a transverse (shear) velocity (Pons et al. 2000). For the "easy"
version of this problem, the setup is identical to that of Riemann
problem 2 above, but with a shear velocity of $v_y = 0.99$ in the
right state and $v_y = 0$ in the left state.  Figure~\ref{riemann4pic}
shows a comparison between the numerical and analytic solutions at a
resolution of 400 grid points at $t=0.4$. Table~\ref{riemann4table}
lists the L1 errors and convergence rates for resolutions from 100 to
3200 grid points for this model, and compares it with the error and
convergence rate from RAM.

\subsection{1D Riemann Problem 5: "Hard" Shear Velocity Test}

\begin{figure}
\epsscale{1.0}
%\plotone{riemann5_400.eps}
\plotone{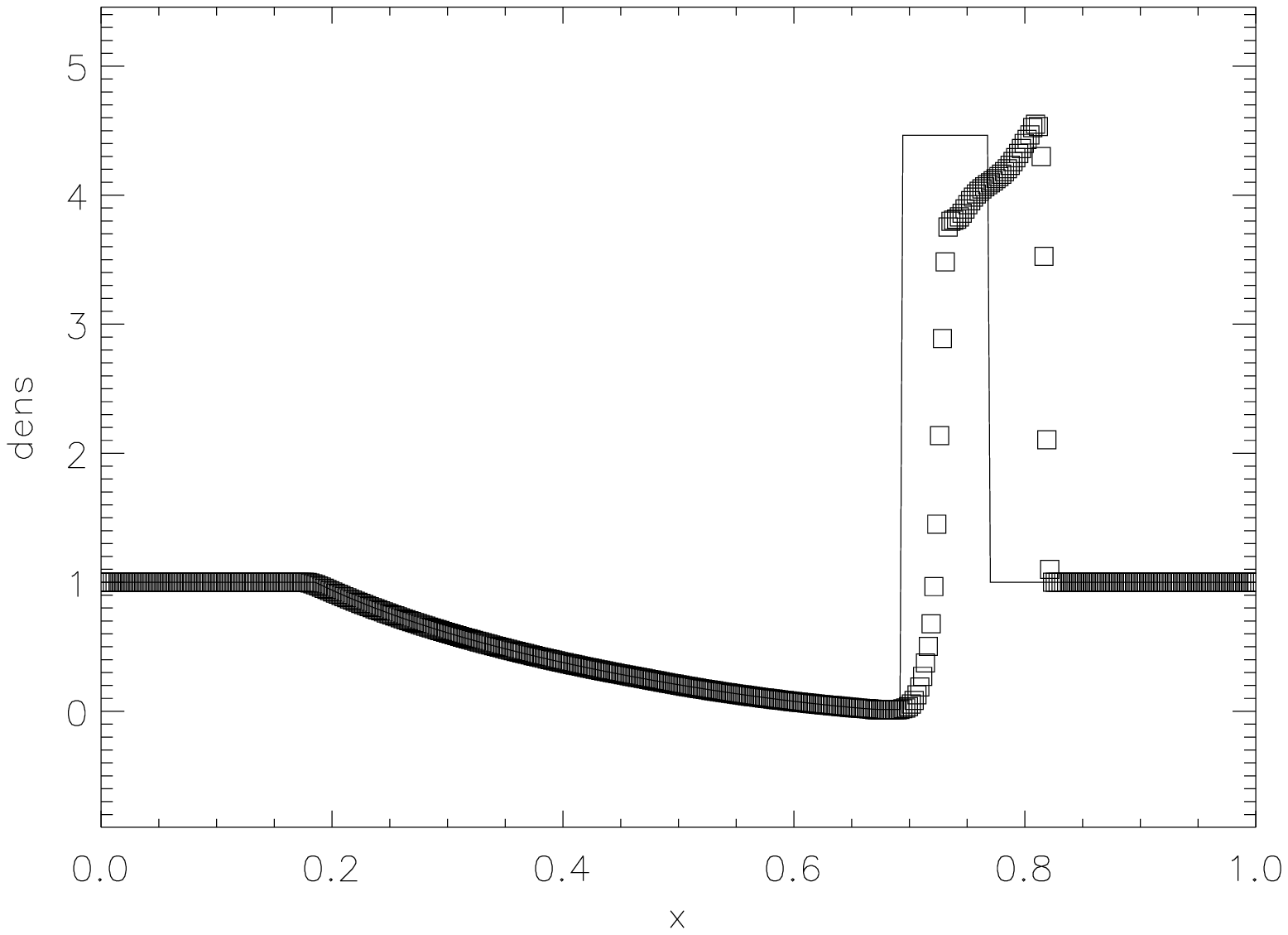}
\caption{\label{riemann5pic} 1D Riemann Problem 5: "Hard" Shear
Velocity Test.  Density for the analytic solution (solid line) and
numerical solution (squares) with a resolution of 400 grid points is
plotted at $t=0.6$. }
\end{figure}

For the "hard" shear velocity test, the setup is again identical to
that of Riemann problem 2 above, but with a shear velocity of $v_y =
0.9$ in both the left and right states.  This test is more difficult
for numerical codes due to the shear velocity in the high-pressure
left state.  As the numerical solution poorly matches the analytic
solution at low resolution, this test is run on higher resolution
grids.  Figure~\ref{riemann5pic} shows a comparison between the
numerical and analytic solutions at a resolution of 400 grid points at
$t=0.6$. Table~\ref{riemann5table} lists the L1 errors and convergence
rates for resolutions from 100 to 6400 grid points for this model, and
compares it with the error and convergence rate from RAM.  It should
be noted that for the last 3 Riemann problems the absolute error using
FLASH is significantly small than for the F-WENO scheme in RAM.  In
particular, comparing figure \ref{riemann5pic} to figure 9 in RAM it
is apparent that FLASH produces a better fit to the analytic solution
at low resolution.  This is likely due to the different slope limiters
being used by the two codes rather than fundamental limitations of the
relativistic solvers they employ.

\subsection{1D Isentropic Flow}

\begin{figure}
\epsscale{1.0}
%\plotone{isenfig.eps}
\plotone{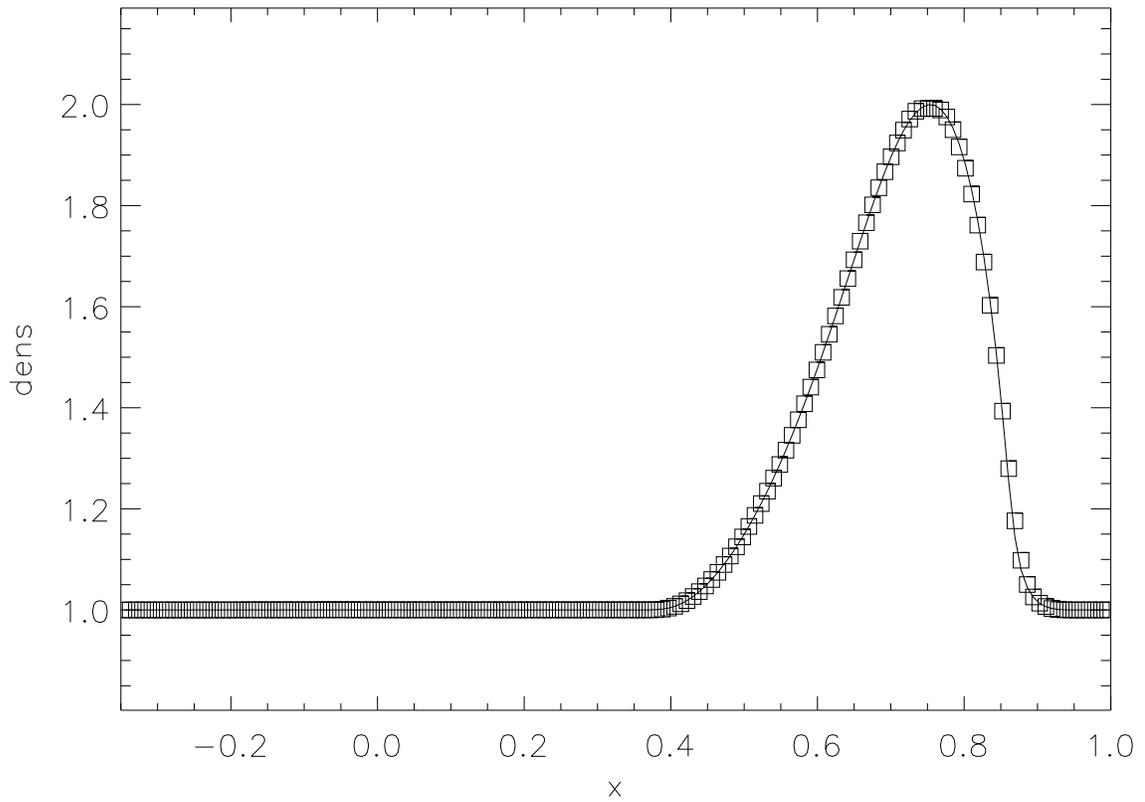}
\caption{\label{isen1Dpic} 1D isentropic flow.  Density for the
analytic solution (solid line) and numerical solution (squares) with a
resolution of 160 grid points is plotted at $t=0.8$. }
\end{figure}

For all the Riemann problems, the error is dominated by the region
near the shock, which gives approximately first order convergence
regardless of the scheme used.  To examine the behavior of the code in
smooth flow regions, we evaluate an isentropic smooth flow set up in a
uniform reference state.  The initial density for this problem is
given by

\begin{equation} \label{eqnrho0}
\rho_0 (x) = \rho_{ref} (1+\alpha f(x)),
\end{equation}
where $\rho_{ref}$ is the density of the reference state and

\begin{equation}
 f(x) = \left \{ \begin{array}{lll}
                   ((x/L)^2 - 1)^4 & : & |x|<L \\
                   0 & : & $otherwise,$ \end{array} \right.
\end{equation}
$\alpha$ and $L$ are the amplitude and width of the pulse.  Pressure
is given by the adiabatic equation of state $p = K\rho^\Gamma$, where
$K$ is a constant.  The initial velocity in the reference state ($|x|
> L$) is $0$.  The initial velocity inside the pulse is found by
assuming that one of the two Riemann invariants is constant,

\begin{equation}
J_- = \frac{1}{2} \ln(\frac{1+v}{1-v}) - \frac{1}{\sqrt{\Gamma-1}} 
\ln(\frac{\sqrt{\Gamma-1}+c_s}{\sqrt{\Gamma-1}-c_s}) = const
\end{equation}
where 

\begin{equation} \label{cs}
c_s = \sqrt{\Gamma\frac{p}{\rho+\frac{\Gamma}{\Gamma-1}p}}
\end{equation}
is the sound speed.  Solving for $v$ gives

\begin{equation}
v=\frac{e^{2(J_- + \frac{1}{\sqrt{\Gamma-1}} 
\ln(\frac{\sqrt{\Gamma-1}+c_s}{\sqrt{\Gamma-1}-c_s})} - 1}
{e^{2(J_- + \frac{1}{\sqrt{\Gamma-1}} 
\ln(\frac{\sqrt{\Gamma-1}+c_s}{\sqrt{\Gamma-1}-c_s})} + 1}
\end{equation}
where $J_-$ is a constant calculated from the reference state.  For
the 1D isentropic test, our computational domain extends from
$x=-0.35$ to $x=1.0$.  The reference state is set to $\rho_{ref}=1$, 
$v_{ref} = 0$, $p_{ref}=100$, $K=100$, with the pulse amplitude 
$\alpha=1.0$ and pulse length $L=0.3$.  An adiabatic index of 
$\Gamma=5/3$ and a CFL number of $0.5$ are used.  
The test is run until $t=0.8$, which is before a shock
develops in the flow.  An analytic solution for the density at this
time can be found by characteristic analysis.  A comparison of
numerical and analytic results for density at $t=0.8$ is shown in
figure \ref{isen1Dpic}.  L1 errors and convergence rates for
resolutions from 80 to 5120 grid points are shown in
Tab.~\ref{isen1Dtable}.  This table shows that FLASH has a convergence
rate of 2 for smooth flows, as expected for the solver being used
(formally 2nd order accurate in time and space).  The F-WENO scheme
used in RAM has a convergence rate of $\sim3$, also as expected
(formally 5th order accurate in space, 3rd order accurate in time).
RAM is clearly more accurate in smooth flow regions, although both
codes converge toward the correct solution.

\subsection{2D Isentropic Flow}

The isentropic flow problem can also be used to test the behavior of
the FLASH code in a two dimensional situation.  For this test, the
computational region goes from $0.0 \le x \le 3.75$ and $0.0 \le y \le
5.0$ in 2D Cartesian coordinates.  The boundary conditions of the grid
are periodic.  Periodic isentropic waves with a spacing of 3.0 are
place in the grid such that ${\bf k}$, the normal vector perpendicular
to the wave front, is ${\bf k}=(4/5,3/5)$. Note that the spatial
periods in the $x$ and $y$ direction are 3.75 and 5.0, so that
periodic boundary conditions are appropriate.  The initial density
profile is given by $\rho_0(d)$ (eqn. \ref{eqnrho0}) where $d
=$mod$({\bf k}\cdot{\bf r}+S/2,S)-S/2$ and ${\bf r}=(x,y)$.  The
reference state is set to $\rho_{ref}=1$, $v_{ref} = 0$,
$p_{ref}=100$, $K=100$, with the pulse amplitude $\alpha=1.0$ and
pulse length $L=0.9$.  An adiabatic index of $\Gamma=5/3$ and a CFL 
number of $0.5$ are used.  
The test is run until $t=2.4$.  Table~\ref{isen2Dtable} shows the total L1
errors and convergence rates for grid resolutions from $48\times64$ to
$768\times1024$ grid points.  The convergence rate for FLASH is again
about 2, as expected.

The above results confirm that the relativistic FLASH module used here
converges to the correct solution for problems with known analytic
results and has a comparable level of accuracy to other available
relativistic hydro codes.

%START OF NONANALYTIC TESTS

\subsection{2D Riemann Problem}

\begin{figure}
\epsscale{1.0}
%\plotone{riemann2D_contour.eps}
\plotone{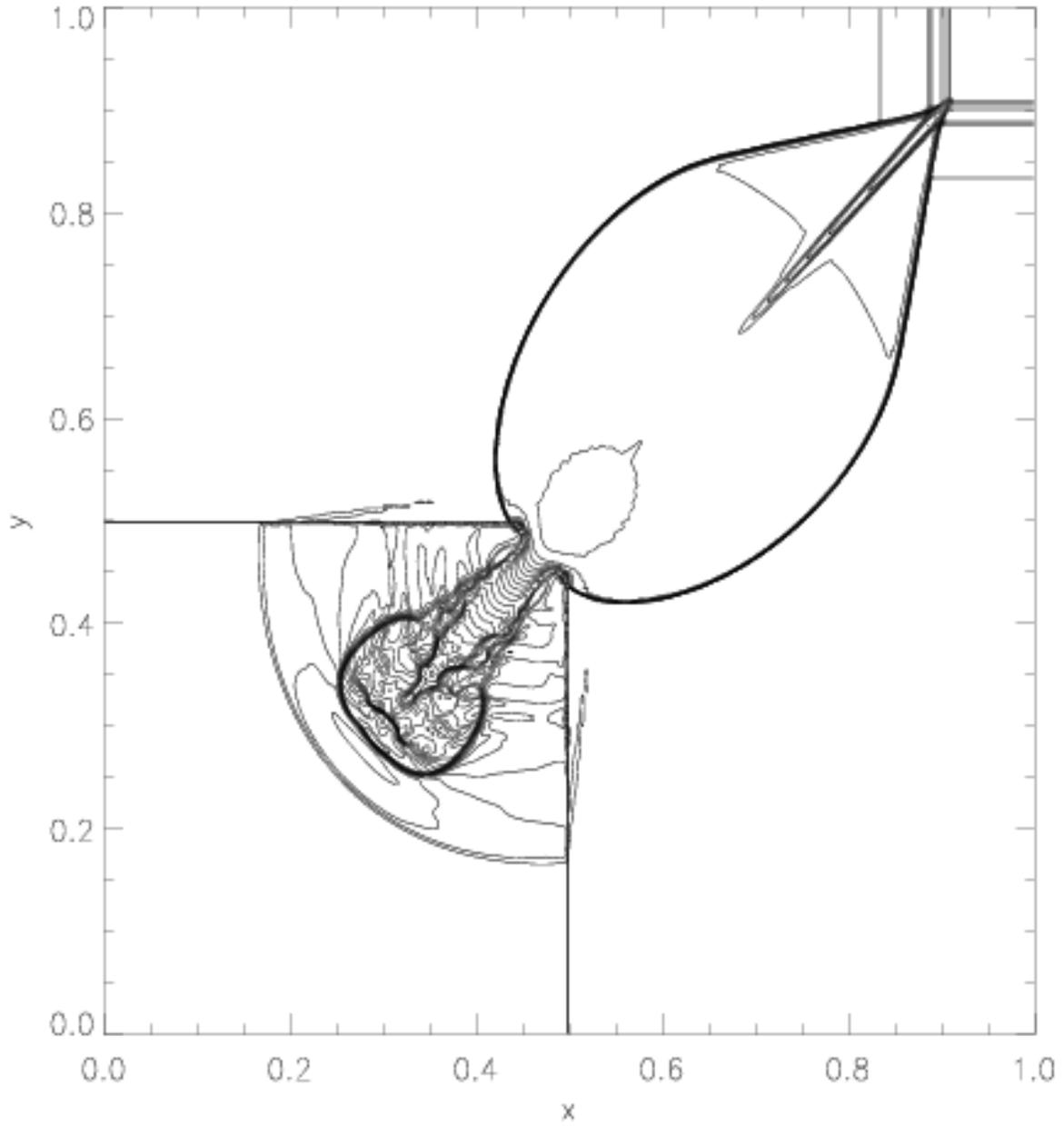}
\caption{\label{riemann2Dpic} 2D Riemann problem at $t=0.4$.  Thirty
logarithmically spaced contours of density are plotted for a uniform
grid of $512\times512$ grid points.}
\end{figure}

The Riemann problem can be extended to two dimensions by setting up a
square grid in which each quarter has different initial conditions.
Although there is no analytic solution to this problem, it has been
well studied as a test of relativistic codes (Del Zanna \& Bucciantini
2002; Lucas-Serrano et al. 2004; Zhang \& MacFadyen 2006).  The
computational domain extends from $0.0 \le x \le 1.0$ and $0.0 \le y
\le 1.0$ in 2D Cartesian coordinates.  The initial conditions for the
four quarters of the grid are

\begin{displaymath}
\begin{array}{llll}
 (\rho,v_x,v_y,p)=(0.1,0,0,0.01)  & $for$ & 0.5 \le x \le 1.0 & 0.5 \le y \le 1.0 \\
 (\rho,v_x,v_y,p)=(0.1,0.99,0,1.) & $for$ & 0.0 \le x \le 0.5 & 0.5 \le y \le 1.0 \\
 (\rho,v_x,v_y,p)=(0.5,0,0,1.)    & $for$ & 0.0 \le x \le 0.5 & 0.0 \le y \le 0.5 \\
 (\rho,v_x,v_y,p)=(0.1,0,0.99,1.) & $for$ & 0.5 \le x \le 1.0 & 0.0 \le y \le 0.5 
\end{array}
\end{displaymath}
An adiabatic index of $\Gamma=5/3$ is used with a CFL number of $0.2$
on a grid of $512\times512$ grid points.  Figure~\ref{riemann2Dpic}
shows logarithmically spaced density contours of the test at $t=0.4$.
Our test correctly reproduces the two curved shock fronts and sharp
density spike in the upper right portion of the grid seen in other
codes.  The low density flow moving toward the lower left is also
seen, although it appears more turbulent than in other codes.
Symmetry is not perfectly preserved across the diagonal from lower
left to upper right.  This is because FLASH uses operator splitting
rather than a Runge-Kutta integration scheme such as that used in RAM.
However, this loss of symmetry is not important as our code is being
used to examine jets in 2D cylindrical coordinates where there is no
symmetry to preserve.

\subsection{Isentropic Pulse}

\begin{figure}
\epsscale{1.0}
%\plottwo{isen_pulse_00.eps}{isen_pulse_07.eps}
%\plottwo{isen_pulse_15.eps}{isen_pulse_20.eps}

\plotone{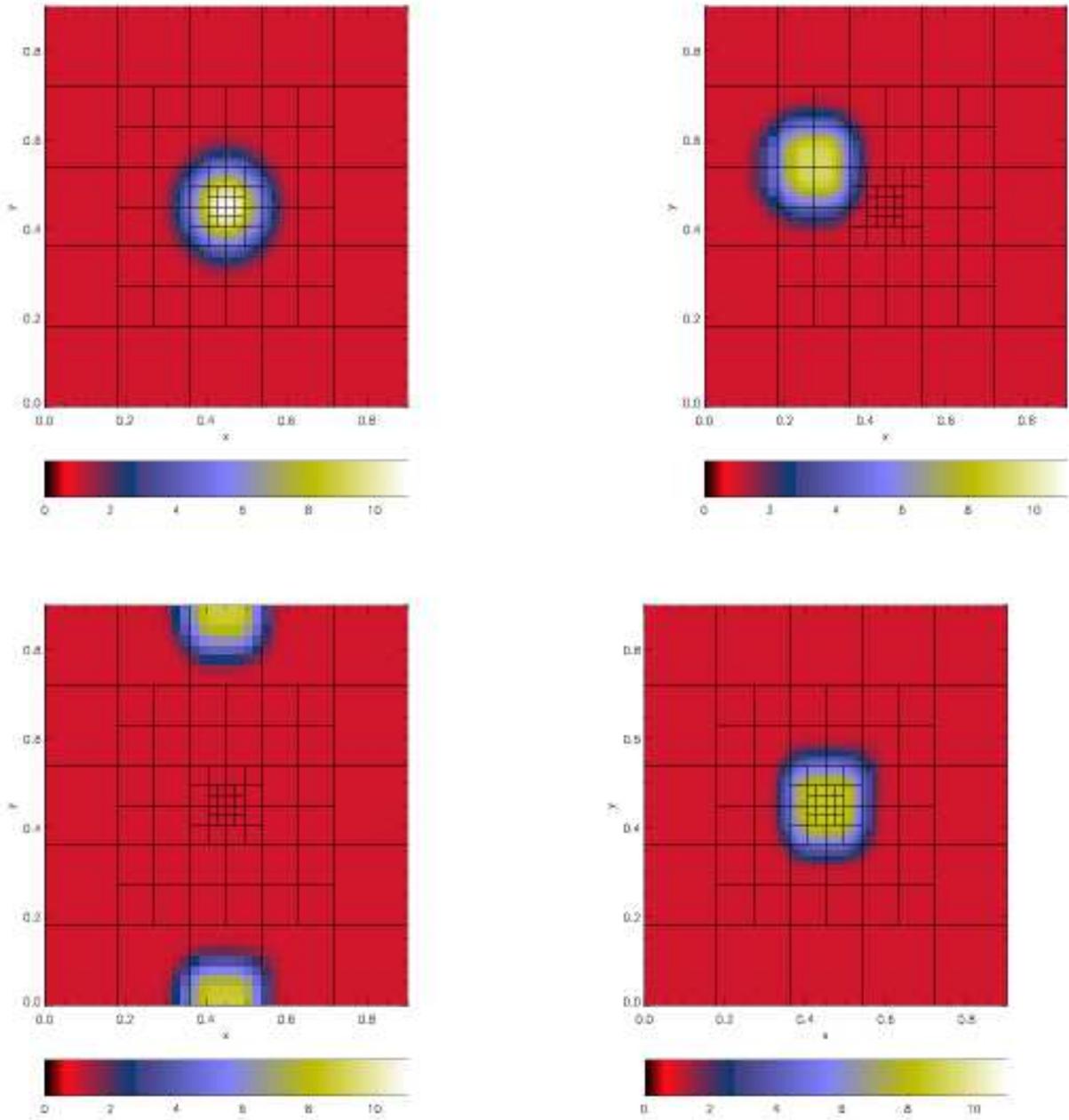}
\caption{\label{isenpulsepic} Isentropic pulse density at $t=0$ (upper
left), $3.5$ (upper right), $7.5$ (lower left), and $10$ (lower left).
Block structure of the mesh is shown by black lines.  Each block
contains $8\times8$ grid points.  The pulse moves up and to the right
at a speed of $v=0.9$ at an angle of $\arctan(4/3)$. }
\end{figure}

To examine what happens when material crosses a change in grid
refinement, we set up a 2D isentropic pulse in density only that is
advected across changes in refinement and eventually returned to its
original location.  For this problem the computational domain ranges
from $0.0 \le x \le 0.9$ and $0.0 \le y \le 0.9$ in 2D Cartesian
coordinates with periodic boundaries.  The resolution of the mesh
varies from $\Delta x=0.0225$ at the edges to $\Delta x=0.0028125$ in
the innermost portion of the grid (see figure \ref{isenpulsepic} for
block structure).  The structure of the grid is fixed and does not
change as the density pulse moves.  The pulse in initially centered at
$x=0.45$, $y=0.45$ and is set by eqn. \ref{eqnrho0} with $\alpha=10$
and $L=0.2$.  Velocity everywhere is set to $v_x=0.72$ and $v_y=0.54$,
for a total velocity of $v=0.9$.  Pressure everywhere is $p=1$, the
adiabatic index is $\Gamma=5/3$ and the CFL number is $0.5$.  
The test is run until $t=10$, at
which time the pulse is again centered at the center of the
computational domain.  As can be seen in figure \ref{isenpulsepic},
the density at the center of the pulse has been flattened and the
shape has become somewhat more square due to changes in refinement as
the pulse has advected.  However, there are no spurious waves and the
size of the pulse has not changed significantly.  At $t=10$, no
fluctuations in pressure are detectable.  A comparison with figure 10
in RAM for an identical test shows that while their pulse has
increased in width by $\sim35\%$, there has been almost no increase in
width in our test, indicating that, in this case, FLASH has less
numerical diffusion.

\subsection{Emery Step}

\begin{figure}
\epsscale{1.0}
%\plotone{windtunnel_dens_contours.eps}
%\plotone{windtunnel_dens_contours_amr.eps}

\plotone{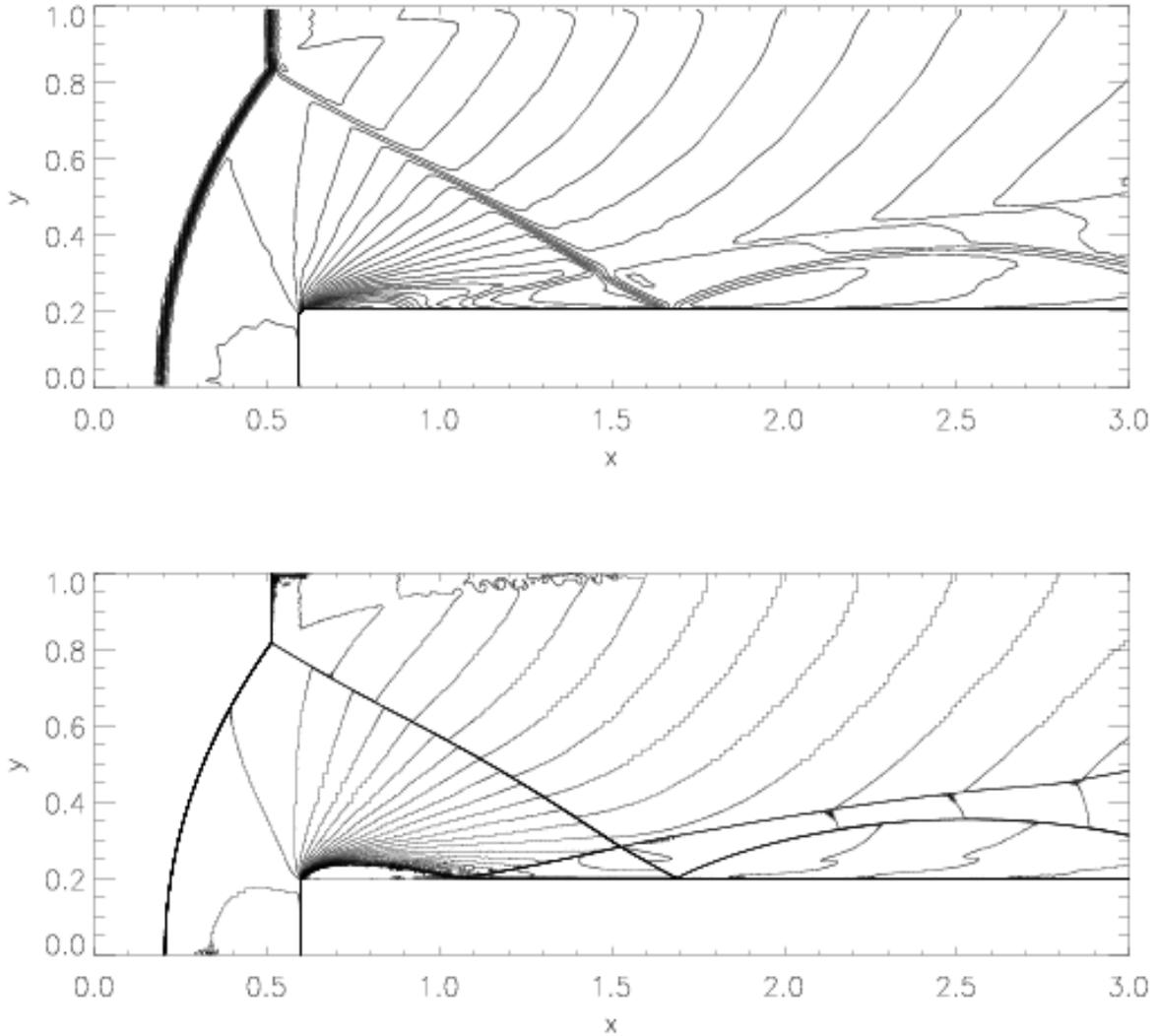}
\caption{\label{wind1pic} Emery step at $t=4$.  Thirty logarithmically
spaced contours of density are plotted for a uniform grid of
$240\times80$ grid points (upper panel) and for an adaptive grid of up
to $3840\times1280$ grid points.  All the same features are apparent
in the low resolution test, and there are no artifacts or
instabilities in the smooth downstream region.  Also note the
development of a Kelvin-Helmholtz instability along the contact
discontinuity originating from the vertical Mach stem in the lower
panel.}
\end{figure}

The Emery step problem (Emery 1968; Woodward \& Colella 1984) consists
of a wind flowing into a sharp vertical step in a wind tunnel with
reflective upper and lower boundaries.  The step causes a reverse
shock to propagate into the wind.  The shock will eventually collide
with the upper boundary and reflect, giving rise to a Mach stem which
is initially nearly vertical.  By the end of the test a portion of the
reflected shock has again reflected, this time off the lower boundary.
For this test, a computational domain ranging from $0.0 \le x \le 3.0$
and $0.0 \le y \le 1.0$ in 2D Cartesian coordinates is used with a
step of height $0.2$ beginning at $x=0.6$.  Initially, the grid is
filled with gas of $\rho=1.4$ and $v=0.999$ with an adiabatic index of
$\Gamma=1.4$ moving with a Newtonian Mach number of $3.0$.  Using
eqn. \ref{cs} gives a pressure of $p=0.1534$.  Upper and lower
boundaries, as well as the face of the step, are reflective.  The left
boundary is a constant inflow with these initial parameters.  The
right boundary uses outflow boundary conditions.  
A CFL number of $0.5$ is used.
The test is run to a
time of $t=4$.  Figure~\ref{wind1pic} shows contours of density for
the results carried out on a uniform grid of $240\times80$ grid
points.  Our results have very little noise in the downstream region
and along the top of the step, as compared to the U-PPM results in
RAM, which employ a similar solver to FLASH (see RAM figure 7).  Our
results appear more comparable to the F-WENO method in RAM.

\begin{figure}
\epsscale{1.0}
%\plotone{windtunnel_amr_dens_vector.eps}
%\plotone{windtunnel_amr_pres_grid.eps}

\plotone{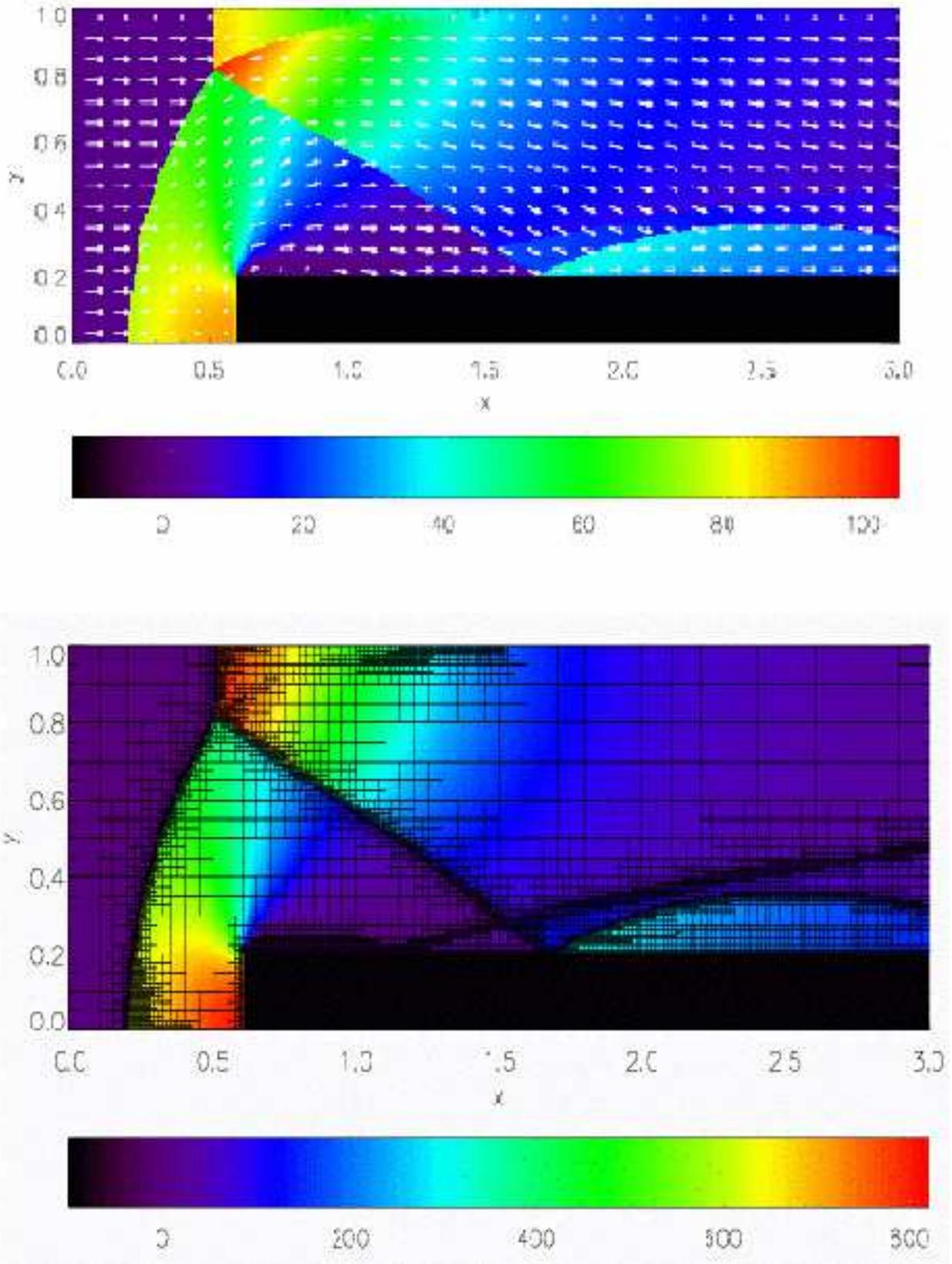}
\caption{\label{wind2pic} Emery step at $t=4$ for 5 level adaptive
mesh (maximum $3840\times1280$ grid points).  The upper panel shows
(linear) density along with the velocity field (white arrows).  The
lower panel shows pressure along with the block structure of the
adaptive grid (black lines).  Each block contains $8\times8$ grid
points.}
\end{figure}

This test has also been carried out on an adaptive mesh grid with 5
levels of refinement, for a maximum resolution of $3840 \times 1280$
grid points.  Other than the resolution, the setup is identical to
that of the uniform grid.  As this problem involves mostly smooth
flows, it provides a good test of the ability of FLASH to selectively
refine and de-refine while still capturing shocks and discontinuities.
Logarithmically spaced density contours of the results at $t=4$ are
shown in the lower frame of figure \ref{wind1pic}.  The density and
pressure are plotted in figure \ref{wind2pic} along with the velocity
field and block structure of the mesh.  The figure is comparable to
figure 15 from RAM, which shows the same results for an identical
setup and maximum resolution.  Both codes concentrate refinement
around the shocks that develop and the contact discontinuity that
originates from the bottom of the Mach stem.  Using our code,
Kelvin-Helmholtz instabilities are clearly seen along this contact
discontinuity.  This instability does not develop using RAM, which may
indicate that FLASH is less diffusive.

\subsection{Double Mach Reflection}

The double Mach reflection test presented here follows the same setup
as in RAM.  This test consists of a computational domain ranging from
$0.0 \le x \le 4.0$ and $0.0 \le y \le 1.0$ in 2D Cartesian
coordinates.  A shock is placed on the grid moving down and to the
right at a 60 degree angle to the x axis.  The lower boundary is
reflecting for $x > 1/6$.  At the initial time, the shock is just
making contact with the reflecting portion of the lower boundary.  The
lower boundary for $x \le 1/6$ is set to the post-shock conditions, as
is the left boundary.  The upper boundary is set to either the pre- or
post-shock conditions depending on the time of the simulation.  The
right boundary is always set to the pre-shock conditions for the test
considered here.  The unshocked conditions are $\rho_0=1.4$,
$p_0=0.0025$, and $v_0=0.0$ with an adiabatic index of $\Gamma=1.4$ in
both the pre- and post-shock state.  The shock is moving with a
classical Mach number of $M=v_s/c_s$ of $10$ where $v_s$ is the shock
speed and $c_s$ is the sound speed of the unshocked gas.  The
relativistic shock jump conditions can then be used to determine the
post-shock conditions and the shock speed.  Using equation \ref{cs},
we find that $v_s=10 \times c_s=0.4984$ in the observer's frame.  In
the shock frame, this is the velocity of the incoming unshocked gas.
Solving the relativistic shock jump conditions yields a post-shock
state of $\rho_1=8.564$, $p_1=0.3808$, $v_1=0.09358$ in the shock
frame.  In the observer frame this transforms to $v_1=0.4247$.  The
speed of the leading edge of the shock is the relativistic sum of the
shock speed, $v_s$, and the sound speed of the shocked gas, which is
$c_{s1}=0.2321$, giving a total speed of $0.5978$ for the leading edge
of the shock.  This is the velocity used to determine if a point on
the upper boundary is in the pre- or post-shock state.  A CFL number of $0.5$ is used.  
The test is run to a time of $t=4$.  

\begin{figure}
\epsscale{1.0}
%\plotone{doublemach_dens_contours.eps}
%\plotone{doublemach_dens_contours_amr.eps}

\plotone{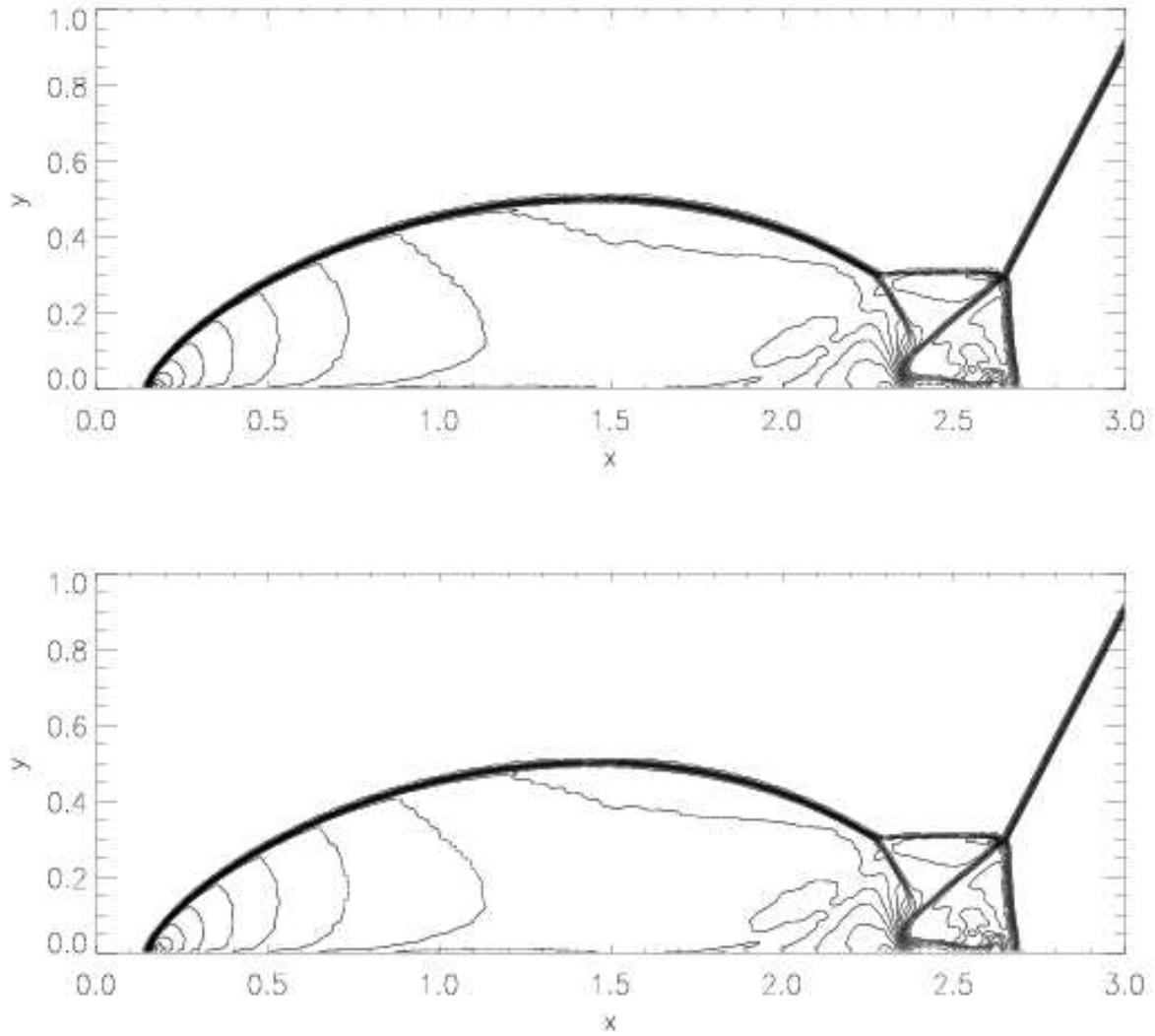}
\caption{\label{doublemachpic} Double Mach reflection at $t=4$.
Thirty equally spaced contours of density are plotted for a uniform
grid of $512\times128$ grid points (upper panel) and an adaptive mesh
grid with 4 levels of refinement with a maximum equivalent resolution
of $512\times128$ grid points (lower panel). Note that the entire
computational domain is not shown.}
\end{figure}

Figure~\ref{doublemachpic} shows density contours of this test run on
a uniform grid of $512\times128$ grid points and an adaptive mesh grid
with the same maximum resolution.  The contours of the two plots are
nearly identical whether or not adaptive mesh is used.  Contours
produced using FLASH do not appear to be as smooth as those from RAM
(see RAM figure 16), particularly to the left of the vertical shock at
$x = 2.7$.  This may indicate that RAM produces more accurate results
for this test.

%START 2D Cylindrical Coordinate Tests

\subsection{Spherical Implosion in Cylindrical Coordinates}

In order to test the behavior of the FLASH code in cylindrical
coordinates we have carried out tests with spherical implosions and
explosions.  The spherical implosion problem consists of a spherically
symmetric flow converging on a single point.  This test is carried out
on a computational domain of $0.0 \le r \le 1.0$ and $0.0 \le z \le
1.0$ in 2D cylindrical coordinates.  Initially, $\rho_0=1$ and $p=0$
everywhere and the material is flowing toward the origin at a fixed
speed $v$, and the adiabatic index is $\Gamma=4/3$.  The $r=0$ and
$z=0$ boundaries are reflecting and the $r=1$ and $z=1$ boundaries are
set to the analytic solution for a spherically converging flow for the
time in the simulation (Mart\'i\ et al. 1997).  The analytic solution
used at the boundaries is

\begin{equation}
\rho(r,t)=\rho_0 \left ( 1+\frac{|v|t}{r} \right )^2  
\end{equation}
where $r$ is the spherical radius, $t$ is the simulation time, and $v$
is the inflow velocity. Pressure at the boundaries is always $0$ and
$v$ is fixed.  The converging flow forms a spherical region of
shocked, stationary gas which increases in radius with time according
to

\begin{equation}
R_s=\frac{(\Gamma-1)\gamma|v|}{\gamma+1} t
\end{equation}
where $\gamma$ is the Lorentz factor of the inflowing gas.  The
density in the post-shock state is given by

\begin{equation}
\rho_s=\rho(R_s,t) \left ( \frac{\gamma\Gamma+1}{\Gamma-1} \right )
\end{equation}
This test was run for inflow velocities of $v=0.9$, $v=0.999$, and
$v=0.99999$ to a time of $t=2$.  All test were run with a CFL number
of $0.2$.  For these three tests, the average error in density in the
post-shock gas is $3.16\%$, $1.91\%$, and $8.37\%$, respectively.
Decreasing the CFL number will decrease the error, but as this is
generally a hard test problem, these errors are acceptable.

\subsection{Spherical Explosion in Cylindrical Coordinates}

\begin{figure}
\epsscale{1.0}
%\plotone{blast1fig.eps}
\plotone{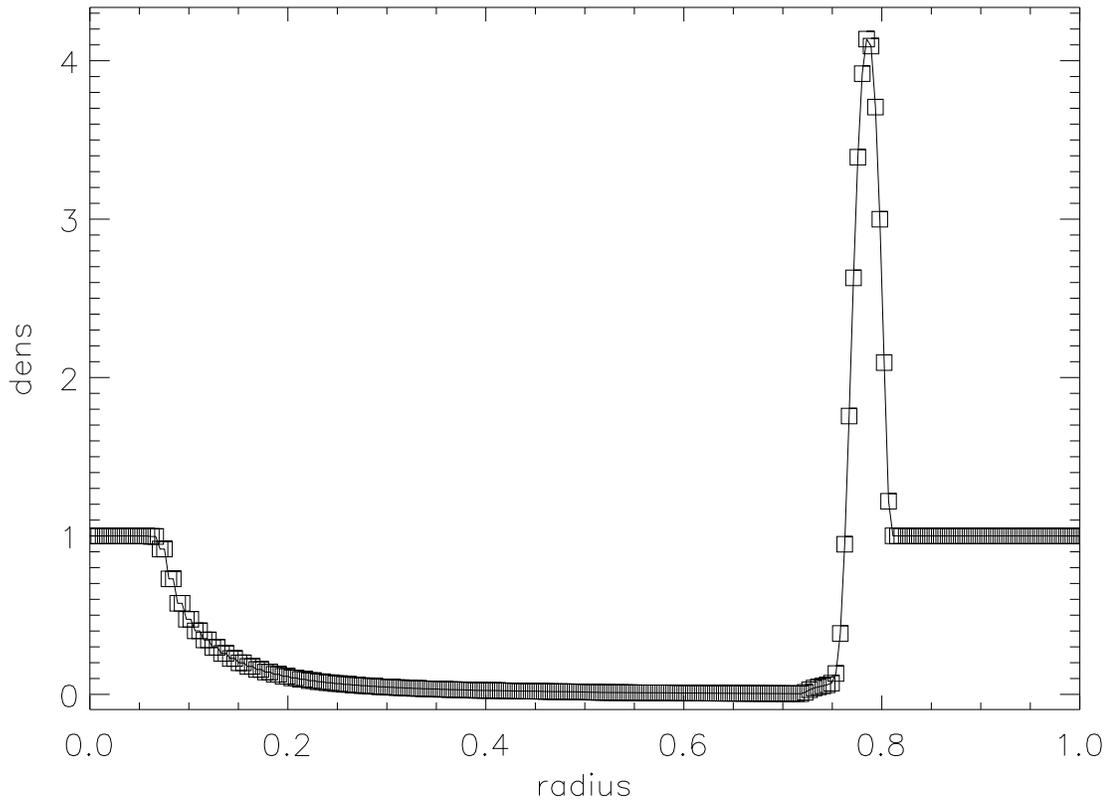}
\caption{\label{blastpic} Spherical explosion in 2D cylindrical
coordinates at $t=0.4$.  Density along the $r=z$ line is plotted for
results on an adaptive grid with a maximum resolution of
$320\times320$ grid points(solid line and squares).  The front of the
shock is captured by $\sim4$ points.}
\end{figure}

For the spherical explosion test, the computational domain ranges from
$0.0 \le r \le 1.0$ and $0.0 \le z \le 1.0$ in 2D cylindrical
coordinates.  Inside a radius of $R=0.4$ from the origin there is
initially a gas with $\rho=1$, $p=1000$, and $v=0$ while outside this
radius the gas has $\rho=1$, $p=1$, and $v=0$.  The adiabatic index is
$\Gamma=5/3$ in both states.  This pressure discontinuity gives rise
to a spherical shock traveling outward and a rarefaction wave
traveling inward.  Our test was carried out on a adaptive mesh grid
with 4 levels of refinement for a maximum resolution of $320\times320$
grid points.  The test was run with a CFL number of $0.2$ to a time of
$t=0.4$.  The $r=0$ and $z=0$ boundaries are reflecting and the $r=1$
and $z=1$ boundaries allow free outflow.  Figure~\ref{blastpic} shows
the density vs. (spherical) radius for points along the line $r=z$.
Our results are consistent with results from other codes.  The shock
front is resolved by $\sim4$ grid points.

\clearpage
\setlength{\hoffset}{-15mm}

%START TABLES

\begin{deluxetable}{ccccc}
\tablewidth{0pt}
\tablecaption{1D Riemann Problem 1 \label{riemann1table}}
\tablehead{
\colhead{Grid Points} & \colhead{FLASH L1 error} & \colhead{FLASH convergence rate} & \colhead{WENO L1 error} & \colhead{WENO convergence rate} }

\startdata
$100$  & $0.132$   & \nodata & $0.131$   & \nodata \\
$200$  & $0.0696$  & $0.92$  & $0.0725$  & $0.85$ \\
$400$  & $0.0357$  & $0.96$  & $0.0332$  & $1.1$ \\
$800$  & $0.0179$  & $1.0$   & $0.0208$  & $0.67$ \\
$1600$ & $0.00852$ & $1.1$   & $0.0100$  & $1.1$ \\
$3200$ & $0.00432$ & $0.98$  & $0.00507$ & $0.98$ \\
\enddata

\end{deluxetable}

\begin{deluxetable}{ccccc}
\tablewidth{0pt}
\tablecaption{1D Riemann Problem 2 \label{riemann2table}}
\tablehead{
\colhead{Grid Points} & \colhead{FLASH L1 error} & \colhead{FLASH convergence rate} & \colhead{WENO L1 error} & \colhead{WENO convergence rate} }

\startdata
100  & 0.206   & \nodata & 0.210   & \nodata \\
200  & 0.148   & 0.48       & 0.142   & 0.56 \\
400  & 0.0832  & 0.83       & 0.0929  & 0.61 \\
800  & 0.0461  & 0.85       & 0.0554  & 0.75 \\
1600 & 0.0249  & 0.89       & 0.0254  & 1.1  \\
3200 & 0.0130  & 0.94       & 0.0151  & 0.75 \\
\enddata

\end{deluxetable}

\begin{deluxetable}{ccccc}
\tablewidth{0pt}
\tablecaption{1D Riemann Problem 3 \label{riemann3table}}
\tablehead{
\colhead{Grid Points} & \colhead{FLASH L1 error} & \colhead{FLASH convergence rate} & \colhead{WENO L1 error} & \colhead{WENO convergence rate} }

\startdata
100  & 0.0587   & \nodata & 0.0997   & \nodata \\
200  & 0.0347   & 0.76       & 0.0629   & 0.67 \\
400  & 0.0214   & 0.70       & 0.0301   & 1.1 \\
800  & 0.0133   & 0.69       & 0.0169   & 0.83 \\
1600 & 0.00845  & 0.65       & 0.00948  & 0.83 \\
3200 & 0.00329  & 1.36       & 0.00524  & 0.86 \\
\enddata

\end{deluxetable}

\begin{deluxetable}{ccccc}
\tablewidth{0pt}
\tablecaption{1D Riemann Problem 4: "Easy" Shear Velocity Test \label{riemann4table}}
\tablehead{
\colhead{Grid Points} & \colhead{FLASH L1 error} & \colhead{FLASH convergence rate} & \colhead{WENO L1 error} & \colhead{WENO convergence rate} }

\startdata
100  & 0.627  & \nodata & 0.758   & \nodata \\
200  & 0.337   &  0.90      & 0.392  & 0.95 \\
400  & 0.172   &  0.97      & 0.231  & 0.76 \\
800  & 0.0843  &  1.0      & 0.118  & 0.97 \\
1600 & 0.0441  &  0.93      & 0.0658  & 0.84 \\
3200 & 0.0232  &  0.93      & 0.0344  & 0.94 \\
\enddata

\end{deluxetable}

\begin{deluxetable}{ccccc}
\tablewidth{0pt}
\tablecaption{1D Riemann Problem 5: "Hard" Shear Velocity Test \label{riemann5table}}
\tablehead{
\colhead{Grid Points} & \colhead{FLASH L1 error} & \colhead{FLASH convergence rate} & \colhead{WENO L1 error} & \colhead{WENO convergence rate} }

\startdata
100  & 0.512  & \nodata & \nodata   & \nodata \\
200  & 0.464  & 0.14       & \nodata  & \nodata \\
400  & 0.325  & 0.51        & 0.521   & \nodata \\
800  & 0.217  & 0.58        & 0.363   & 0.52 \\
1600 & 0.133  & 0.71        & 0.233   & 0.64 \\
3200 & 0.0833  & 0.68       & 0.126   & 0.89 \\
6400 & 0.0534  & 0.64       & 0.0649  & 0.96 \\
\enddata

\end{deluxetable}

\begin{deluxetable}{ccccc}
\tablewidth{0pt}
\tablecaption{1D Isentropic Flow \label{isen1Dtable}}
\tablehead{
\colhead{Grid Points} & \colhead{FLASH L1 error} & \colhead{FLASH convergence rate} & \colhead{WENO L1 error} & \colhead{WENO convergence rate} }

\startdata
$80$   & $5.48e-3$  & \nodata   & $2.07e-3$  & \nodata \\
$160$  & $1.55e-3$  & $1.8$     & $1.10e-4$  & $4.2$ \\
$320$  & $3.99e-4$  & $2.0$     & $1.70e-5$  & $2.7$ \\
$640$  & $1.00e-4$  & $2.0$     & $1.47e-6$  & $3.5$ \\
$1280$ & $2.50e-4$  & $2.0$     & $1.58e-7$  & $3.2$ \\
$2560$ & $5.35e-6$  & $2.2$     & $1.91e-8$  & $3.1$ \\
$5120$ & $1.56e-6$  & $1.8$     & $2.37e-9$  & $3.0$ \\
\enddata

\end{deluxetable}

\begin{deluxetable}{ccccc}
\tablewidth{0pt}
\tablecaption{2D Isentropic Flow \label{isen2Dtable}}
\tablehead{
\colhead{Grid Points} & \colhead{FLASH L1 error} & \colhead{FLASH convergence rate} & \colhead{WENO L1 error} & \colhead{WENO convergence rate} }

\startdata
$48\times64$    & $8.12e-3$  & \nodata   & $7.35e-2$  & \nodata \\
$96\times128$   & $2.23e-3$  & $1.9$     & $4.43e-3$  & $4.1$ \\
$192\times256$  & $5.87e-4$  & $1.9$     & $8.04e-4$  & $2.5$ \\
$384\times512$  & $1.48e-4$  & $2.0$     & $9.62e-5$  & $3.1$ \\
$768\times1024$ & $3.61e-5$  & $2.0$     & $1.12e-5$  & $3.1$ \\
\enddata

\end{deluxetable}

\clearpage
\setlength{\hoffset}{0mm}

% START APPENDIX FIGURES HERE

% START 1D FIGURES HERE

\clearpage

% START 2D FIGURES

\end{document}